\documentclass[print]{aa}
\usepackage{graphicx}
\usepackage[figuresright]{rotating}
\usepackage{txfonts}
\begin{document}
\title{Massive molecular outflows associated with UCHII/HII regions}
\author{Sheng-Li, Qin \inst{1,2}; Jun-Jie, Wang \inst{1}; Gang Zhao \inst{1}; Martin, Miller
\inst{3} \and Jun-Hui, Zhao\inst{2}} \offprints{S.-L., Qin,
\email{sqin@cfa.harvard.edu}} \institute{National Astronomical
Observatories, Chinese Academy of Sciences, Beijing
100012, P. R. China \\
\and Harvard-Smithsonian Center for Astrophysics,
60 Garden Street, Cambridge, MA 02138, USA \\
\and  I.Institute of Physics, University of Cologne, Cologne,
50937, Germany }

\date{Received date ~~~~~~~~~~~ ; Accepted date ~~~~~~~~~~~~~}

\abstract{Aims: We searched for the molecular outflows from
fifteen molecular clouds associated with ultra-compact and compact
HII (UCHII/HII) regions and discussed possible gas heating
mechanism. Methods: Mapping observations of CO $J=3-2$ and $J=2-1$
lines were carried out with the KOSMA 3m-Telescope towards the 15
HII regions/molecular cloud complexes. Results: Ten molecular
outflows were identified out of the fifteen HII region/molecular
cloud complexes. The higher outflow detection rate (67\%)
suggested that such outflows are as common in high mass star
forming regions as those in low mass star forming regions, which
is consistent with the results of other authors. The observations
also showed that the outflow might occur in the HII region. The
integrated CO line intensity ratios (${\rm
R_{I_{CO(3-2)}/I_{CO(2-1)}}}$) were determined from the core
component of the spectra as well as from both the blue and red
wings. Maximum line intensity ratios from the wings and core
components appeared to be related to the mid-infrared sources
imaged by Midcourse Space Experiment (MSX). The relationship
between the maximum line intensity ratios and MSX sources
indicates that the molecular gas could be heated by the emission
of dust associated with massive stars. Based on maser observations
reported in the literature, we found that ${\rm H_{2}O}$ masers
were only detected in seven regions. The ${\rm H_{2}O}$ masers in
these regions are located near the MSX sources and within the
maximum line intensity ratio regions, suggesting that ${\rm
H_{2}O}$ masers occur in relatively warm environments. 
}

\authorrunning{Qin et al.}   
\titlerunning{Massive Molecular Outflows}
\maketitle
\newpage
\section{Introduction}
Ultra-compact and compact HII regions provide information about
massive star formation within molecular clouds (Heyer et al. 1989;
Churchwell 2002). The dust surrounding the massive stars or HII
regions absorbs nearly all the stellar radiation and re-radiates
in the far-infrared and mid-infrared bands (e.g. Wood \&
Churchwell 1989; Churchwell 2002). The molecular clouds associated
with HII regions have higher ratio of CO $J=3-2$ to CO $J=2-1$ and
have higher gas temperature than those sources without HII regions
(Wilson, Walker \& Thornley 1997). The high gas temperature of
molecular clouds might be related to the warm dust components
associated with the HII regions (Kaufman et al. 1999; Goldreich \&
Kwan 1974). Various kinds of masers have been detected frequently
in massive star formation regions. Different masers maybe occur in
different astrophysical environments. For example, both
observations and theoretical models suggested that ${\rm H_{2}O}$
masers originate in hot cores, and are excited by the shocks
associated with outflows or accretion (Elitzur, Hollenbach \&
McKee 1989; Felli, Palagi \& Tofani 1992; Codella, Testi \&
Cesaroni 1997; Garay \& Lizano 1999). Molecular outflows have been
found in the regions associated with all kinds of protostars,
which are considered a good indicator of star formation (Beuther
et al. 2002; Shepherd \& Churchwell 1996a, 1996b; Bachiller 1996;
Bontemps et al. 1996; Wu et al. 2004, 2005). In the process of
massive star formation, the molecular outflows appear to be a
common phenomenon in the early stages prior to formation of UCHII
regions (e.g. Beuther et al. 2002) and of UCHII regions (e.g.
Shepherd \& Churchwell 1996b). However, it has not been fully
understood that how the outflows initially start from HII regions
and what kinds of links exist between HII regions and molecular
clouds in large scale structures. To understand the relationship
between massive stars and those associated with astrophysical
phenomena, we carried out observations of CO $J=3-2, J=2-1$ and
$^{13}$CO $J=2-1$ towards fifteen HII regions/molecular cloud
complexes. The sources were selected from the sample of HII
regions detected at 1.46 and 5.89 GHz by VLA (Fich 1993) and the
catalog of UCHII regions compiled by Kurtz, Churchwell \& Wood
(1994). Comparing to the sources of Beuther et al. (2001) and
Shepherd \& Churchwell (1996b), all fifteen sources in this sample
have H$_{\alpha}$ emission as seen from the red plate of the
Palomar Observatory Sky Survey (POSS)\footnote{Available at
http://archive.stsci.edu/dss/} and are in relatively evolved
stages. Although the sources in our sample are not statistically
complete, the observations illustrate the massive star forming and
molecular gas properties associated with HII regions.

The different transitions of CO trace different molecular
environments. For example, the CO $J=3-2$ and CO $J=2-1$
transitions have distinct upper energy level temperatures and
critical densities. The upper energy level temperature and
critical density in the CO $J=2-1$ line are 16.6 K and about
$10^4$ ${\rm cm^{-3}}$ while those for CO $J=3-2$ line are 33.2 K
and $5\times 10^{4}$ ${\rm cm^{-3}}$ (Kaufman et al. 1999).
Therefore, the ratio of line intensities between CO $J=3-2$ and CO
$J=2-1$ indicates the temperature and density structure of
molecular cloud environments if the molecular gas is optically
thin (Hasegawa et al. 1994; Wilson, Walker \& Thornley 1997).
Previous studies have shown that the emission of CO ($J=3-2$ and
$J=2-1$) in line wings and line center is optically thick in star
formation regions (Choi, Evans II \& Jaffe 1993; Correia, Griffin
\& Saraceno 1997; Langer \& Penzias 1990). According to Wilson et
al. (1997, 1999) and Kim \& Koo (2002), the line intensity ratios
based on the optically thick CO transitions indicate that the
temperature varies at different positions. The mid-infrared
emission at 8.3 ${\mu}$m is thought to be from small dust grains
and polycyclic aromatic hydrocarbons (PAHs), and be excited by the
UV radiation leaking from the HII regions (Leger \& Puget 1984;
Deharveng et al. 2003, 2005). The radio continuum at 1.4 GHz from
NRAO VLA Sky Survey (NVSS)\footnote{Available at
http://www.cv.nrao.edu/nvss/postage.shtml.} have an angular
resolution of 45$^{\prime\prime}$ and a sensitivity of 45 mJy
(Condon et al. 1998). The Midcourse Space Experiment (MSX) band A
(at 8.3 ${\mu }$m) emission was imaged at an angular resolution of
18.3$^{\prime\prime}$ with a sensitivity\footnote{Available at
http://iras.ipac.caltech.edu/applications/MSX/.} of 0.1 Jy. The
relationship of gas temperature traced by the line intensity
ratio, warm dust traced by 8.3 ${\mu }$m emission and the radio
continuum at 1.4 GHz will be discussed here.

In this paper, we report results from observations of fifteen HII
regions at the CO $J=3-2$ and CO $J=2-1$ transitions using KOSMA
(K$\ddot{\rm o}$lner Observatorium f$\ddot{\rm u}$r Sub-Millimeter
Astronomie )\footnote{The KOSMA 3m telescope is operated at
submillimeter wavelengths by the University of Cologne in
collaboration with Bonn University.} at mm/submillimter
wavelengths. We also observed $^{13}$CO $J=2-1$ for six sources.
In section 2, the observations are described. The results from the
observations are given in section 3. Possible mechanisms driving
the outflows and the gas heating are discussed in section 4.
Section 5 summarizes the results.

\section{Observations \& data reduction}

We carried out observations toward fifteen HII region/molecular
cloud complexes in CO $J=3-2$ and CO $J=2-1$ lines (345.796 and
230.538 GHz) using the KOSMA 3m telescope at Gornergrat,
Switzerland, between February and April 2004. The dual-channel
230/345 GHz SIS receivers (Graf et al. 1998) were used to
simultaneously observe the two transitions of CO. DSB receiver
noise temperatures were about 120 K. In addition, $^{13}$CO
$J=3-2$ and $2-1$ lines at the rest frequencies of 330.588 and
220.399 GHz were also observed toward 6 sources in this sample.
The integration time on the sources appeared not long enough to
derive significant results from $^{13}$CO $J=3-2$ data at 330.588
GHz.

The  medium and variable resolution acousto optical spectrometers
(Schieder et al. 1989) with 1501 and 1601 channels or bandwidth of
248 and 544 MHz at 230 and 345 GHz were used as backends. The
channel widths of 165 and 340 kHz correspond to velocity
resolutions of 0.21 and 0.29~${\rm km}~{\rm s}^{-1}$ at 230 and
345 GHz which result in different rms noise levels of ${\rm
T_{mb}}$ in each channel as shown in Table 1. The beam size and
main beam efficiency were determined using continuum cross scans
on Jupiter. The beam sizes at 230 and 345 GHz were
130$^{\prime\prime}$ and 80$^{\prime\prime}$, respectively. The
forward efficiency ${\rm F_{eff}}$ was 0.93 during our
observations. The corresponding main beam efficiencies ${\rm
B_{eff}}$ were 0.68 and 0.72 at 230 and 345 GHz. Pointing was
frequently checked on Jupiter and was better than
20$^{\prime\prime}$. Table 1 summarizes the observations. In
Table1, Table 2, Table 3 and Table 4, G139, G206, G189, G213, G192
and G70 represent G139.909+0.197, G206.543-16.347, G189.876+0.516,
G213.880-11.837, G192.584-0.041 and G70.293+1.600, respectively.
Six sources are from Kurtz, Churchwell \& Wood (1994), and the
other nine sources are from Fich (1993).

All the maps were made using the on-the-fly mode with a
$1^{\prime}\times1^{\prime}$ grid, except for G70.293 with a
$2^{\prime}\times2^{\prime}$ grid. The sources were scanned along
the right ascension (RA) at a fixed declination (Dec) with a
constant integration time of 4 second at each point. The sky level
was removed by observing the off-source position in real time.
Then, the scan continued by moving to the next row of the grids.
We repeated the observing procedure 6 times over the entire
source. The data were integrated at each point in order to obtain
a high S/N ratio. The data reduction was carried out using the
Continuum and Line Analysis Single-Disk Software (CLASS) and
Grenoble Graphic (GREG) software packages. A least-square fit to
baselines in the spectra was carried out with the first order
polynomial. The baseline slopes were removed for all the sources.
The correction for the line intensities to main beam temperature
scale was made using the formula ${\rm
T_{mb}=(F_{eff}/B_{eff})T_{A}^{\star}}$.

\section{Results and analysis}

We drew position-velocity (P-V) diagrams for those sources with
broad wings. According to the P-V diagrams, we selected the
integrated range of wings and determined the outflow intensities
of red and blue lobes. Ten molecular outflows were identified by
the contours of integrated intensities of CO $J=2-1$ line wings.
The molecular cloud cores in the CO $J=2-1$ line for the sources
with and without outflows were also imaged. The molecular cores in
the $^{13}$CO $J=2-1$ line for the six sources were imaged with
the same integrated velocity range as those in the CO $J=2-1$
line. The velocity intervals and the maximum integrated
intensities for the maps are summarized in Table 2. In order to
obtain the intensity ratio of the CO $J=3-2$ to CO $J=2-1$ lines,
we convolved the 80$^{\prime\prime}$ resolution of CO $J=3-2$ data
with an effective beam of size $\sqrt
{130^2-80^2}=102^{\prime\prime}$. The integrated intensities were
calculated for the CO $J=3-2$ line in the same velocity range as
for CO $J=2-1$. The ratio ${\rm R_{I_{CO(3-2)}/I_{CO(2-1)}}}$ of
the integrated line intensity in CO was determined from the core
component of the spectra as well as from both the blue and red
wings.

 Assuming the CO $J=2-1$ emission in the line wings is
optically thin, and taking the upper energy level temperature of
16.6 K as the excitation temperature, the column densities of line
wings within one beam were calculated using equation 1 (Garden et
al. 1991). If we assume the CO abundance ${\rm
[CO]/[H_{2}]=10^{-4}}$ and the mean atomic weight of the gas $\mu
_{g}=1.36$, the outflow mass, the momentum and the kinetic energy
within one beam were derived using equations 2-4 (cf. Goldsmith et
al. 1984),

\begin{equation}
{\rm N_{CO}=1.08 \times 10^{13}\frac{T_{ex}}{
exp(-16.6/T_{ex})}{\int}T_{mb}dv~~ (cm^{-2})},
\end{equation}
\begin{equation}
{\rm M_{i}=7.94 \times 10^{-4}D^{2}\frac{T_{ex}}{
exp(-16.6/T_{ex})}{\int}T_{mb}dv~~ (M_{\odot})},
\end{equation}
\begin{equation}
{\rm P_{i}=7.94 \times 10^{-4}D^{2}\frac{T_{ex}}{
exp(-16.6/T_{ex})}{\int}T_{mb}vdv~~ (M_{\odot}~ km~ s^{-1})},
\end{equation}

\begin{equation}
{\rm E_{i}=7.90 \times 10^{39}D^{2}\frac{T_{ex}}{
exp(-16.6/T_{ex})}{\int}T_{mb}v^{2}dv~~ (erg)},
\end{equation}
where v in km~s$^{-1}$ is the velocity of the gas with respect to
the cloud systemic velocity and T(v) in K is the brightness
temperature at v; D is the distance from the Sun in kpc; the
subscript i indicates the observed points within the outflow zone.
By summing the values at all the observed points in the red and
blue lobes, the outflow mass M, the momentum P and the kinetic
energy E were obtained.

The timescale t, the characteristic velocity V, the mass loss rate
${\rm \dot M_{loss}}$, the driving force F and mechanical
luminosity ${\rm L_{m}}$ were derived using the equations below
(cf. Goldsmith et al. 1984),
\begin{equation}
~~~~~~~~~~~{\rm V=\frac{P}{M}~ (km~s^{-1})},
\end{equation}

\begin{equation}
~~~~~~~~~~~{\rm t=9.78\times10^{5}\frac{R}{V}~ (yr)},
\end{equation}

\begin{equation}
~~~~~~~~~~~{\rm \dot {M}_{loss}= \frac{P}{tV_{w}}~(M_{\odot}~
yr^{-1})},
\end{equation}

\begin{equation}
~~~~~~~~~~~{\rm F= \frac{P}{t}~(M_{\odot}~ km~ s^{-1}~ yr^{-1})} ,
\end{equation}

\begin{equation}
~~~~~~~~~~~~{\rm L_{m}=
8.28\times10^{-42}\frac{E}{t}~(L_{\odot})},
\end{equation}
where R in pc is the outflow size defined by the average of the
the radius of the blue-shifted and red-shifted lobes, the wind
velocity ${\rm V_{w}}$ is assumed to be 500 km s$^{-1}$ (Marti,
Rodr¨ªguez \& Reipurth 1998). The physical parameters of the
outflows are summarized in Table 3. Our results give lower limits
on outflow parameters if line wings are optically thick and the
excitation temperature is larger than 16.6 K at the CO $J=2-1$
transition.

Using the formula of Casoli, Combes \& Dupraz (1986), the
far-infrared luminosities of IRAS sources associated with the HII
regions in our sample were derived from the infrared flux
densities at the 4 IRAS bands. The far-infrared luminosities of
IRAS sources are summarized in Table 4. In the images of the MSX
band A towards the sources in our sample, the brightest emission
at 8.3 $\mu$m is always centered on the MSX point-like sources and
located near or at the same position as the IRAS sources/massive
stars. G213.880-11.837 is outside of the MSX survey region. The
emission at 8.7 $\mu$m in G213.880-11.837 centered on IRAS
06084-0611 was detected by Persi \& Tapia (2003). The offset
positions of the MSX sources from the associated IRAS
source/massive stars are labeled in Figs.1-12 and presented in
Table 4.

\subsection{Outflows}
The blue and red lobes of the outflows were mapped for all the
outflow sources as shown in Figs.1-10 (see Appendix B). In each of
the figures, the dot symbol marks the mapped points. The spectra
averaged over the outflow zones are shown in the bottom-right
panel.

\noindent {\bf 3.1.1. S186}\\
There are two IRAS sources, IRAS 01056+6251 and IRAS 01053+6251,
within the HII region and molecular cloud complex S186. IRAS
01056+6251 is located near the center of the continuum at 1.4 GHz
(NVSS) and 6 GHz (Fich 1993), which is a possible exciting source
of the HII region S186. The derived far-infrared luminosity of
IRAS 01053+6251 corresponds to a B3-type star. There is no radio
continuum to be detected around IRAS 01053+6251. IRAS 01053+6251
appears to be a deeply embedded massive protostar.

In Fig. 1, the molecular core shows an elongated structure and
extends in the EW direction. The blue and red lobes of the outflow
have a similar extension to the molecular core. The two MSX
sources are located near the regions of the maximum line intensity
ratio (${\rm R_{I_{CO(3-2)}/I_{CO(2-1)}}}$) determined from the
core and wing components.

\noindent {\bf 3.1.2. G139.909+0.197}\\
G139.909+0.197 is located within the reflection nebula AFGL437,
which is associated with IRAS 03035+5819. IRAS 03037+5819 is
located about $2^{\prime}$ NE of IRAS 03035+5819. IRAS 03035+5819
is located near the center of the NVSS continuum at 1.4 GHz. A
bipolar outflow in the NS direction has been observed in the
$^{12}$CO (1-0) line (Arquilla \& Goldsmith 1984). Gomez et al.
(1998) suggested that the poor degree of collimation of the
outflow may be due to the wind sweeping up the inhomogeneous
molecular materials and creating a cavity or hollow in the bipolar
lobes.

From the core diagram of Fig. 2, the line intensity ratio of ${\rm
R_{I_{CO(3-2)}/I_{CO(2-1)}}}$ decreases gradually from SE to NW.
The MSX source near IRAS 03035+5819 is located near the region
with maximum value of the line intensity ratio.

From Fig. 2, the CO $J=2-1$ line wing map shows that the
morphology of the outflow is similar to that of CO $J=1-0$
(Arquilla \& Goldsmith 1984), but the size of the outflow is less
extended than that of Arquilla \& Goldsmith. The result indicates
that the outflow traced by CO  $J=2-1$ arises from warm layer
closer to the central exciting star.

From the line intensity ratio map of the blue wing, the SiO maser
1 (Harju et al. 1998), ${\rm H_{2}O}$ masers 2, 3, 4 (Cesaroni,
Palagi \& Felli 1988; Palagi et al. 1993; Wynn-Williams et al.
1986), IRAS 03035+5819 and its associated MSX source are located
within the maximum line intensity ratio region. SiO emission is
known to be a tracer of shocks. The SiO maser 1 and ${\rm H_{2}O}$
masers 2, 3, 4 are located within the maximum line intensity ratio
region, which is consistent with the model that ${\rm H_{2}O}$
masers are excited by shocks and associated with a warm molecular
environment (Elitzur, Hollenbach \& McKee 1989; Felli, Palagi \&
Tofani 1992; Codella, Testi \& Cesaroni 1997; Garay \& Lizano
1999).

The line intensity ratio from the red lobe of the outflow
correlates well with the MSX source near IRAS 03035+5819.

IRAS 03037+5819 and its associated MSX source are far away from
the maximum line intensity ratio regions of both the wings and
core components.

\noindent {\bf 3.1.3. G189.876+0.516}\\
G189.876+0.516 is associated with IRAS 06063+2040. There are three
B-type stars (ALS 8745, ALS 8748 and HD 252325) in this HII
region/molecular cloud complex. All of them are located south of
IRAS 06063+2040. Both IRAS 06063+2040 and HD 252325 are associated
with the NVSS continuum.

In core diagram of Fig. 3, the molecular cloud core has a compact
structure stretching in the NS direction, and bends toward the
west at $2^{\prime}$ north of IRAS 06063+2040. The location of the
massive stars and the distorted morphology of the molecular core
may be evidence that the molecular cloud are squeezed by the
stellar winds from stars ALS 8748 and HD 252325. The MSX source,
IRAS source and ALS 8745 appear to be associated with a maximum
${\rm R_{I_{CO(3-2)}/I_{CO(2-1)}}}$ region. However, a region with
the maximum line intensity ratio, 2 arcmin NE of IRAS 06063+2040,
has no identifications in existing observations of the
IR/FIR/optical and radio continuum emission.

In Fig. 3, the morphology of the blue-shifted lobe of the outflow
shows a similar structure to the NS extension of the molecular
core, while the red lobe corresponds to the bent structure of the
molecular core. In the blue-shifted lobe, the MSX source and IRAS
source are located within the maximum ${\rm
R_{I_{CO(3-2)}/I_{CO(2-1)}}}$ region. However, the maximum ${\rm
R_{I_{CO(3-2)}/I_{CO(2-1)}}}$ from the red-shifted wing appears to
be located along the NE edge of the red-shifted lobe.

\noindent {\bf 3.1.4. G213.880-11.837}\\
Cometary UCHII region G213.880-11.837 is located within the red
nebula GGD14 and is associated with IRAS 06084-0611. A ${\rm
H_{2}O}$ maser (Rodriguze et al. 1980; Tofani, Felli \& Faylor
1995) is located NE of IRAS 06084-0611. The NVSS continuum peaks
at IRAS 06084-0611. The radio source VLA7 (Gomez et al. 1998) has
a flux density of 0.17 mJy at 3.6 cm and is located in the ${\rm
H_{2}O}$ maser region. The 8.7 ${\mu }$m and 12.5 ${\mu }$m images
(Persi \& Tapia 2003) have a similar morphology to the radio
counterparts observed at 2 cm and 3.6 cm from G213.880-11.837
(Gomez et al. 1998, 2000). The similarity in morphology between
the Mid IR and radio continuum is evidence that ionized gas is
mixed with the dust in the cloud.

In the core diagram of Fig. 4, the core contours of CO $J=2-1$ and
$^{13}$CO $J=2-1$ have a similar morphology to that of HCO$^{+}$
$J=1-0$ (Heaton et al. 1988). The map of ${\rm
R_{I_{CO(3-2)}/I_{CO(2-1)}}}$ shows a large intensity gradient
from south to north. The ${\rm H_{2}O}$ maser and 8.7 $\mu$m
emission are associated with the region with maximum
line-intensity ratio.

From the outflow map of Fig. 4, the line intensity maps of the
wing components show a highly collimated bipolar outflow. VLA7 is
located on the axis of outflow and appears to be a sign of the
energy source of the outflow. The outflow extends from SE to NW
and is perpendicular to the core of high density HCO$^{+}$ $J=3-2$
gas (Heaton et al. 1988). It is possible that the accretion disk
traced by dense gas is responsible for the highly collimated
bipolar outflow.

The maximum ${\rm R_{I_{CO(3-2)}/I_{CO(2-1)}}}$ of the red wing is
located at the head of red lobe of the outflow and is associated
with the 8.7 ${\mu }$m emission and IRAS source. However, the
${\rm H_{2}O}$ maser is located within the maximum line intensity
ratio region of the blue-shifted component and the maximum line
intensity ratio is not associated with the 8.7 ${\mu }$m emission,
suggesting that shocks may be responsible for the heating of CO
gas in the blue lobe of the outflow.

\noindent {\bf 3.1.5. G192.584-0.041}\\
There are three IRAS sources (IRAS 06099+1800, IRAS 06096+1757 and
IRAS 06105+1756) and two massive stars (HD 253327 and HD 253247)
in the G192.584-0.041 HII region/molecular cloud complex. Both the
IRAS sources and massive stars are associated with the radio
continuum (NVSS).

The core contours of Fig. 5 show a dense core elongated NS
surrounded by the shell structure. The core structure in CO
$J=2-1$ is consistent with the column density contours of
$^{13}$CO $J=1-0$ (Heyer et al. 1989) and the molecular core of CS
$J=5-4$ (Shirley et al. 2003). The blue and red lobes of the
outflow have a similar morphology to the shell and dense core of
the molecular cloud, respectively.

From the outflow diagram of Fig. 5, along with Fig. 5 of Heyer et
al. (1989) and Fig. 16 of Zinchenko et al. (1994), the extension
of the outflow of CO $J=2-1$ is larger than that of CS $J=2-1$,
but smaller than that of CO $J=1-0$. This result suggests that the
outflows may be caused by a stellar wind sweeping up the
surrounding material in different intensity layers. The material
of the outflows come from the surrounding materials entrained by
the winds or jets from the central stars, and not from the
material ejected from the central stars.

In Fig. 5, the MSX source is centered at IRAS 06099+1800. ${\rm
H_{2}O}$ masers 1, 2, 3 (Henkel et al. 1986; Cemoretto et al.
1990; Henning et al. 1992) and the class II ${\rm CH_{3}OH}$ maser
4 (Caswell et al. 2000) are associated with IRAS 06099+1800 and
located near the maximum ${\rm R_{I_{CO(3-2)}/I_{CO(2-1)}}}$
region. The class II ${\rm CH_{3}OH}$ masers associated with UCHII
regions trace the earliest stage of massive stars and are excited
by infrared radiation (Cragg, Sobolev \& Godfrey 2002). ${\rm
H_{2}O}$ masers 5 and 6 (Henning et al. 1992; Palagi et al. 1993)
are located at the edge of molecular cloud.

\noindent {\bf 3.1.6. S288}\\
S288 is excited by a B0-type star ced92 and is associated with
IRAS 07061-0414. The NVSS continuum peaks at IRAS 07061-0414.

In Fig. 6, the molecular cloud core extends in the NE-SW direction
and the peak intensity of the molecular cloud core is centered at
IRAS 07061-0414. Another B8-type star HD 296489 lies NE of the
molecular cloud core.

The outflow contours have a similar morphology to that of the
molecular cloud core. The MSX source is associated with the region
with maximum ${\rm R_{I_{CO(3-2)}/I_{CO(2-1)}}}$ of the red wing.

\noindent {\bf 3.1.7. G70.293+1.600}\\
According to the continuum observations at 2 cm and 3.6 cm (Kurtz,
Churchwell \& Wood 1994), the UCHII region G70.293+1.600 consists
of a compact core extended EW and an elongated shell towards NS.
The NVSS continuum is associated with IRAS 19598+3324. IRAS
19598+3324 is at the same position as the MSX source.

From Fig. 7, the molecular cloud core and the blue lobe of the
outflow in the CO $J=2-1$ line have a similar morphology to that
of the compact core at 2 cm and 3.6 cm, and the morphology of the
red lobe is similar to the shell at 2 cm and 3.6 cm continuum
(Kurtz, Churchwell \& Wood 1994). The extension of the red lobe
exceeds the edge of the molecular core. It is possible that the
fast winds or jets entrain the surrounding material breaking
through the edge of molecular cloud and stretching out the
red-shifted CO gas.

From Fig. 7, the maximum line intensity ratios of ${\rm
R_{I_{CO(3-2)}/I_{CO(2-1)}}}$ determined from the wings are
elongated EW. ${\rm H_{2}O}$ masers 1, 3 and OH maser 2 (Benson et
al. 1990; Argon, Reid \& Menten 2000) are located in the NE-SW
direction of the outflow and within the maximum ${\rm
R_{I_{CO(3-2)}/I_{CO(2-1)}}}$ region determined from the core
component. OH masers are formed in the envelope of massive stars
and are indicators of newly formed O- and early B-type stars
(Argon, Reid \& Menten 2000). The association of the OH maser with
IRAS 19598+3324 suggests that IRAS 19598+3324 is a deeply embedded
young massive star. The MSX source is associated with the maximum
line intensity ratio determined from the core and wings.

\noindent {\bf  3.1.8. S127}\\
S127 is associated with IRAS 21270+5423. The NVSS continuum peaks
near IRAS 21270+5423. Two MSX sources are located NE and SW of
IRAS 21270+5423.

The molecular cloud core has an extended structure from NE to SW
(see Fig. 8). The maximum ${\rm R_{I_{CO(3-2)}/I_{CO(2-1)}}}$
correlates well with the MSX source.

In the outflow diagram of Fig. 8, the two lobes of the outflow
extend in the different direction. The maximum ${\rm
R_{I_{CO(3-2)}/I_{CO(2-1)}}}$ from the blue and red wings are
related to the MSX sources to the SW and NE, respectively.

\noindent {\bf 3.1.9. S138}\\
S138 is associated with IRAS 22308+5812. The O7.5-type star
GRS105.63-00.34 is located east of IRAS 22308+5812. The NVSS
continuum peaks at IRAS 22308+5812. The contours of CO $J=1-0$ and
$^{13}$CO $J=1-0$ show a structure elongated NS and the core of
C$^{18}$O $J=1-0$ is elongated EW (Johansson et al. 1994).

In the core diagram of Fig. 9, molecular cloud cores in CO $J=2-1$
and $^{13}$CO $J=2-1$ lines show a triangle structure. Massive
stars are located at the center of the molecular cloud. Compared
with the contours of CO $J=1-0$ and $^{13}$CO $J=1-0$ (Johansson
et al. 1994), the relatively warm gas of CO $J=2-1$ appears to be
surrounded by the cold CO $J=1-0$ and $^{13}$CO $J=1-0$ gas. The
line intensity ratio determined from the core extends from the
center to the west. The line intensity ratio map has a similar
morphology to that of the ${\rm H_{\alpha}}$ image (Deharveng et
al. 1999).  A possible explanation is that the expanding HII
region injects the hot ionized gas into the molecular core and
heats the dust and molecular gas in the molecular core.

In the outflow diagram of Fig. 9, the red lobe has a similar
morphology to that of the core contours of CO $J=1-0$ and
$^{13}$CO $J=1-0$ (Johansson et al. 1994), and the blue lobe has a
similar extension to that of the core contours of C$^{18}$O
$J=1-0$ (Johansson et al. 1994). The extension of the red lobe
exceeds the boundary of the molecular core. The strong stellar
winds or jets from central stars appear to be responsible for
driving the red lobe of the outflow out from the core region. The
MSX emission and IRAS source or massive stars are not associated
with the maximum line intensity ratio of the red lobe in the
outflow. It is possible that the shock is responsible for the
molecular gas heating of the red lobe. The ${\rm H_{2}O}$ maser
(Cesaroni, Palagi \& Felli 1988) is overlaid at the IRAS
22308+5812 position and is located east of the maximum ${\rm
R_{I_{CO(3-2)}/I_{CO(2-1)}}}$ region.

\noindent {\bf 3.1.10. S149}\\
There are three IRAS sources, IRAS 22542+5815, IRAS 22546+5814 and
IRAS 22543+5821 in the S149 HII region/molecular cloud complex.
IRAS 22542+5815 is located near the peak of the NVSS continuum.

Fig. 10 shows that there are two molecular cloud cores in the S149
HII region/molecular cloud complex. The northern one is associated
with IRAS 22543+5821 and the southern one is associated with IRAS
22542+5815 and IRAS 22546+5814.

The complex structure of red and blue lobes indicates
multi-outflows produced in this molecular/HII region complex.

\subsection {Relationship between NVSS continuum, MSX sources and line intensity ratios}
In addition to the ten sources with detection of outflows
discussed above, there are five sources with no detections of
outflows (see Figs. 11 and 12 in Appendix A). According to Figs.
1-12, all fifteen sources have NVSS continuum emission. The NVSS
continuum peaks near or at the IRAS sources with greatest
far-infrared luminosities (see also Table 4) in each of the HII
region/molecular cloud complexes. Except for S193, the MSX
point-like sources are located near the IRAS sources with NVSS
continuum emission. We consider an MSX source to be associated
with the maximum ratios if the MSX sources are located within the
range of 65$^{\prime\prime}$, half of the CO J=2-1 beam, from the
peak of the ratio. Those MSX point-like sources associated with
the maximum line intensity ratios of the blue and red wings and/or
the core components (${\rm R_{b}}$, ${\rm R_{r}}$ and ${\rm
R_{c}}$) are presented in Table 4.

As seen in Table 4, there are twenty-seven IRAS sources and twelve
massive stars without IRAS counterparts in the fifteen regions. Of
the twenty-seven IRAS sources, the twenty IRAS sources are located
at or near the MSX point-like sources. In the S217 region, IRAS
04547+4753 is located at the center of the MSX 8.3 $\mu$m image
although it has no MSX point-like source counterpart. Of the
twelve massive stars without IRAS counterparts, only ALS6206 in
the S175 region and ALS8745 in the S138 region are associated with
MSX point-like source counterparts.

From Figs. 1-12, there are nineteen maximum line intensity ratio
${\rm R_{I_{CO(3-2)}/I_{CO(2-1)}}}$ regions determined from core
components in the fifteen HII region/molecular cloud complexes.
Sixteen of the nineteen maximum line intensity ratio regions are
associated with MSX point-like sources. This result shows that all
the maximum line intensity ratio regions determined from core
components are related to the MSX 8.3 ${\mu}$m emission.

Among the ten sources associated with outflows, we found that nine
sources with a maximum line intensity ratio determined from the
blue wing appears to be related to the eleven MSX point-like
sources while seven sources with a maximum line intensity ratio
determined from red wings are related to the eight MSX point-like
sources. However, for G213.880-11.837 and S138, we have found no
evidence for the maximum line intensity ratios of the red-shifted
wings in these sources associated with either MSX sources or IRAS
sources or massive stars.

Based on our sample discussed in this paper, there are 78\% IRAS
sources found to be associated with MSX 8.3 $\mu$m emission. For
the core components, 87\% of the maximum line intensity ratio
regions appear to be associated with MSX sources. For both the
blue and red wings, about 90\% and 70\% of maximum line intensity
ratio regions are associated with MSX sources. Our analysis
suggests that there is a possible relationship between the regions
with maximum line intensity ratio and the MSX 8.3 ${\mu}$m as well
as FIR emission.

\section {Discussion}
\subsection {Outflows}

Ten molecular outflows were identified out of the fifteen HII
region/molecular cloud complexes observed in this program. All
outflows in our sample show poor collimation except for
G213.880-11.837. However, the angular resolution
(130$^{\prime\prime}$ in CO $J=2-1$) corresponds to a linear scale
range of 0.3--7.5 pc for the program sources at distances ranging
from 0.5 to 11.5 kpc. Our angular resolution appears to be not
adequate to image the highly collimated outflow on a smaller
scale. Thus, the outflow detection rate from our KOSMA
observations only imposes a lower limit. Nevertheless, the higher
outflow detection rate (67\%) suggests that in the process of
high-mass star formation, outflows appear to be common phenomena,
which is consistent with the results obtained from other groups
(Beuther et al. 2002; Shepherd \& Churchwell 1996b). Beuther et
al. (2002) selected twenty-six massive star forming regions prior
to the stage of UCHII regions and identified twenty-one massive
molecular outflows. Shepherd \& Churchwell (1996b) identified five
massive molecular outflows from ten massive star forming regions
associated with UCHII regions. All the fifteen sources in our
sample have H$_{\alpha}$ emission from the red plate of the POSS
survey and are in relatively evolved stages. When combined with
the results of Beuther et al. (2002) and Shepherd \& Churchwell
(1996b), the outflows appear to present in the evolutionary stages
from the high mass protostars and UCHII regions to the HII regions
in the process of massive star formation.

Many models for the origin of the outflows have been discussed
(e.g. Lada 1985; Shu et al. 1991 ). The molecular outflows were
commonly thought to be driven by winds or jets from the energy
sources located in the molecular clouds. The winds and jets likely
sweep up the ambient material around central stars and create the
outflow structure (Lada 1985; Shu et al. 1991). In the high mass
star forming regions, the outflow likely plays the same important
role in dissipating energy and angular momentum as those in
low-mass star formation regions.

Comparing the extension of the outflows with that of molecular
cores, the scale of the outflows except for S138 and G70.293+1.600
appears not to exceed the edge of molecular cloud cores. The peaks
of the integrated intensity in the red and blue lobes of the
outflows in our sample are close to the peaks of their associated
molecular cores. A possible driving mechanism of the outflows is
that the surrounding material is entrained by the winds or jets as
the winds or jets pass through the inhomogeneous molecular clouds.
As a result, the larger integrated intensities of red and blue
lobes in the outflows are close to the dense regions in the
molecular cloud cores. Comparing the far-infrared luminosities
with the masses of the outflows in Table 3, the sources with
higher far-infrared luminosities appear to have higher outflow
masses. This is consistent with the result found by Shepherd \&
Churchwell (1996b). From Table 3, the outflow masses (except for
G213.880-11.837 and S288 with 11 and 33 ${\rm M_{\odot}}$) in our
sample are greater than 83 ${\rm M_{\odot}}$, which is far greater
than the mass of an O4-type star. The higher outflows mass
suggests that the outflow masses are not likely to originate from
the stellar surface, but could be caused by the entrainment of
ambient gas (Shepherd \& Churchwell 1996b).

\subsection {Heating mechanism}

The molecular clouds associated with HII regions have higher line
intensity ratios of CO $J=3-2$ to CO $J=2-1$ lines and have higher
temperatures than those sources without HII regions (Wilson,
Walker \& Thornley 1997). The association of higher line intensity
ratios with IRAS sources implies that the molecular gas could be
heated by the thermal emission of dust (Goldreich \& Kwan 1974).
Cohen \& Green (2001) carried out a survey at 8.3 ${\mu }$m and
834 MHz toward thermal and non-thermal radio sources covering an
area of $310^{\circ}.4 \leq l \leq 313^{\circ}.7$ and
$-0^{\circ}.9 \leq b \leq 1^{\circ}.4$. There is a complete
absence in this field of any detected MSX 8.3 ${\mu}$m
counterparts to non-thermal radio sources. This result suggests
that MSX 8.3 ${\mu}$m emission is dominated by the thermal
emission of dust and that the dust is heated by the radiation from
massive stars (Cohen \& Green 2001).

We found that about 87\% of maximum line intensity ratio regions
from core components in our sample are associated with MSX
point-like sources. All the IRAS sources with NVSS continuum
(except for S193 region) have MSX 8.3 ${\mu}$m emission. A
possible explanation is that massive stars with IRAS counterparts
are still deeply embedded in their parental gas/dust envelope. The
dust acts as a buffer which absorbs the UV radiation leaking from
the HII region associated with massive stars and reradiates at the
far-infrared and mid-infrared bands so that the molecules in the
core are not destroyed in the heating process. The straightforward
statistics in our analysis favor the possibility that the
radiation from far-infrared and mid-infrared regions is the main
heating source of the gas in the molecular cores.

Alternatively, for G213.880-11.837, the 12.5 $\mu$m image (Persi
\& Tapia 2003) has a similar morphology to that of the continuum
at 2 cm (Gomez et al. 1998, 2000). The line intensity ratio map
from the core of S138 has a similar morphology to that of the
${\rm H_{\alpha}}$ image (Deharveng et al. 1999). It is possible
that the expanding HII regions inject hot ionized gas into the
molecular cores and lead to the increase in temperature of both
dust and molecular gas in the two molecular cores in  the
G213.880-11.837 and S138 regions.

For the ten sources associated with the outflows, the higher rate
of detection of MSX emission in maximum line intensity ratio
regions associated with the blue and red wings suggests that
molecular gas in the outflows is mainly heated by the emission of
dust. However, heating through shocks cannot be excluded. For
example, the maximum line intensity ratio determined from
red-shifted lobes in G213.880-11.837 and S138 regions are far from
any radiative heating sources including IRAS, MSX and HII
emission. Shocks produced by the interaction between the outflows
and molecular clouds might be a plausible heating mechanism for
the red lobe in the outflows of G213.880-11.837 and S138.

Furthermore, in the literature, ${\rm H_{2}O}$ masers are only
detected in seven regions in our sample (S201, G206.543-16.347,
G139.909+0.197, G192.584-0.041, G213.880-11.837, G70.293+1.600 and
S138). The model based on the shock structure and maser pumping
scheme (Elitzur, Hollenbach \& McKee 1989) suggests that in star
forming regions, ${\rm H_{2}O}$ maser emission results from dense
regions behind the shock. Observations showed that most ${\rm
H_{2}O}$ masers were found along the axes of the outflows (e.g.
Garay \& Lizano 1999). The CO outflows with high mechanical
luminosities were associated with ${\rm H_{2}O}$ masers (Felli,
Palagi \& Tofani 1992). The fact that ${\rm H_{2}O}$ masers were
associated with outflows with high mechanical luminosities and
observed along the axes of the outflows suggests that only winds
with high mechanical luminosities can produce shocks to excite
${\rm H_{2}O}$ masers. Observations also showed that ${\rm
H_{2}O}$ masers occur in relatively warm regions (e.g. Codella,
Testi \& Cesaroni 1997). Comparing with the images of CO lines,
the masers are located within the maximum line intensity ratio
regions and are associated with the outflows with high mechanical
luminosities and MSX 8.3 ${\mu }$m emission. The results are
consistent with the model that ${\rm H_{2}O}$ masers are related
to the outflows and occur in relatively warm environments
(Elitzur, Hollenbach \& McKee 1989; Felli, Palagi \& Tofani 1992;
Codella, Testi \& Cesaroni 1997; Garay \& Lizano 1999).

\section{Conclusions}
From our observations and analysis, we found ten molecular
outflows out of the fifteen HII region/molecular cloud complexes
in our sample. The higher outflow detection rate (67\%) appears to
indicate that the outflow plays an important role in the formation
process of high-mass stars. The HII regions in our sample are
relatively evolved. When combined with the detections of outflow
toward protostar cores (Beuther et al. 2002) and UCHII regions
(Shepherd \& Churchwell 1996b), massive molecular outflows might
occur in all stages of massive star formation.

Comparing the morphology of outflows with that of molecular cores,
the outflows appear to result from the surrounding materials that
are entrained by the winds or jets from central stars.

The maximum line intensity ratios determined from the molecular
cores and outflows are highly associated with the MSX 8.3 ${\mu}$m
emission. The molecular gas in the outflows and cores appears to
be mainly heated by the emission of dust. In a few cases of
outflows, shock heating might be inevitable.

Based on previous observations in the literatures, ${\rm H_{2}O}$
masers are located near the massive stars of the seven HII
region/molecular cloud complexes. The correlation between ${\rm
H_{2}O}$ masers, the outflows and the maximum line intensity
ratios is consistent with the result that ${\rm H_{2}O}$ masers
are related to the outflows and occur in relatively warm
environments.

\begin{acknowledgements}
We thank the anonymous referee for his/her constructive
suggestions. This work is supported by the National Natural
Science Foundation of China under Grant Nos.\, 10473014, 10328306,
10521001 and 10433010. This work is also supported by National
Basic Research Program of China (973 program) under grant No.
2007CB815103 and Chinese Academy Sciences under Grant No.
KJCX2-YW-T01. Sheng-Li Qin thanks Ms. Ni-Mei Chen for her help
during the observations.
\end{acknowledgements}

\begin{appendix}
\section{Sources without outflow}
\noindent {\bf 1. S175}\\
The optical HII region S175 is excited by an O9.5-type star ALS
6206 (White \& Gee 1986). IRAS 00244+6425 is located near the
O9.5-type star ALS 6206. The NVSS continuum peaks at IRAS
00244+6425.

In the left panel of Fig. 11, the contours of CO $J=2-1$ have a
morphology similar to that of $^{13}$CO $J=2-1$. The MSX source
and IRAS 00243+6427 are located within the maximum ${\rm
R_{I_{CO(3-2)}/I_{CO(2-1)}}}$ region.

\noindent {\bf 2. S193}\\
The molecular cloud/HII region complex S193 harbors three HII
regions, S192, S193 and S194. IRAS 02437+6145 is associated with
the NVSS continuum. Additional two IRAS sources of IRAS 02435+6144
and IRAS 02439+6143 are also located in this region.

The molecular cloud core has two peaks (see Fig. 11); the south
one is associated with IRAS 02435+6144. The MSX source and IRAS
02435+6144 are located within the maximum ${\rm
R_{I_{CO(3-2)}/I_{CO(2-1)}}}$ region.

 \noindent {\bf 3. S201}\\
IRAS 02593+6016 is associated with the HII region/molecular cloud
complex. The continuum at 2 and 6 cm are excited by IRAS
02593+6016, and the ${\rm H_{2}O}$ masers 1 and 2 (Cesaroni,
Palagi \& Felli 1988; Palagi, Cesaroni \& Comoretto 1993) are
close to the continuum peak at 1.4 GHz (NVSS), 2 and 6 cm (Felli,
Hjellming \& Cesaroni 1987). IRAS 02593+6016 is overlaid at the
MSX source position.

From the S201 diagram of Fig. 12, the molecular cloud extends in
the EW direction. The maximum ${\rm R_{I_{CO(3-2)}/I_{CO(2-1)}}}$
is far away from IRAS 02589+6014.

 \noindent {\bf 4. S217}\\
There are two massive stars, the O9.5-type star LSV 47$^{\circ}$24
and the B8-type star BD+47 1079 and two IRAS sources, IRAS
04551+4755 and IRAS 04547+4753 located within the S217 HII region
/molecular cloud complex.

From the right panel of Fig. 12, we see that the molecular cloud
core associated with S217 has an elongated structure. The
strongest emission is located to the SW of the molecular cloud,
centered at IRAS 04547+4753. The intensity falls off gradually
from the SW to NE. Comparing the NVSS continuum and Fig. 1 of
Roger \& Leahy (1993), the peak of the 21 cm continuum is located
at the same position as the CO peak; the O9.5-type star LSV
47$^{\circ}$24 (Roger \& Leahy 1993) is located about $3^{\prime}$
NE of IRAS 04547+4753; the B8-type star BD+47 1079 is close to the
northern weaker molecular core.

The line intensity ratio map shows that the maximum value of ${\rm
R_{I_{CO(3-2)}/I_{CO(2-1)}}}$ is close to IRAS 04547+4753. The
extended MSX 8.3 $\mu$m emission is centered at IRAS 04547+4753
but there is no MSX point source in this region.

\noindent {\bf 5. G206.543-16.347}\\
G206.543-16.347 is associated with IRAS 05393-0156. The molecular
cloud core in $^{13}$CO $J=2-1$ contours shows an elongated NS
structure, but the core contours in CO $J=2-1$ show an amorphous
structure. An ${\rm H_{2}O}$ maser (Palagi, Cesaroni \& Comoretto
1993) is located south of IRAS 05393-0156 and within the maximum
${\rm R_{I_{CO(3-2)}/I_{CO(2-1)}}}$ region. The brightest emission
at MSX 8.3 $\mu$m is consistent with the maximum ${\rm
R_{I_{CO(3-2)}/I_{CO(2-1)}}}$ region.

We fail to detect the highly collimated outflow that had been
observed in CO $J=2-1$ with an angular resolution of
23$^{\prime\prime}$ by Sanders \& Willner (1985), possibly because
of the low resolution of our observations.

\section{Figures}
\onecolumn
\newpage
\begin{figure}[h]

\includegraphics[width=80mm]{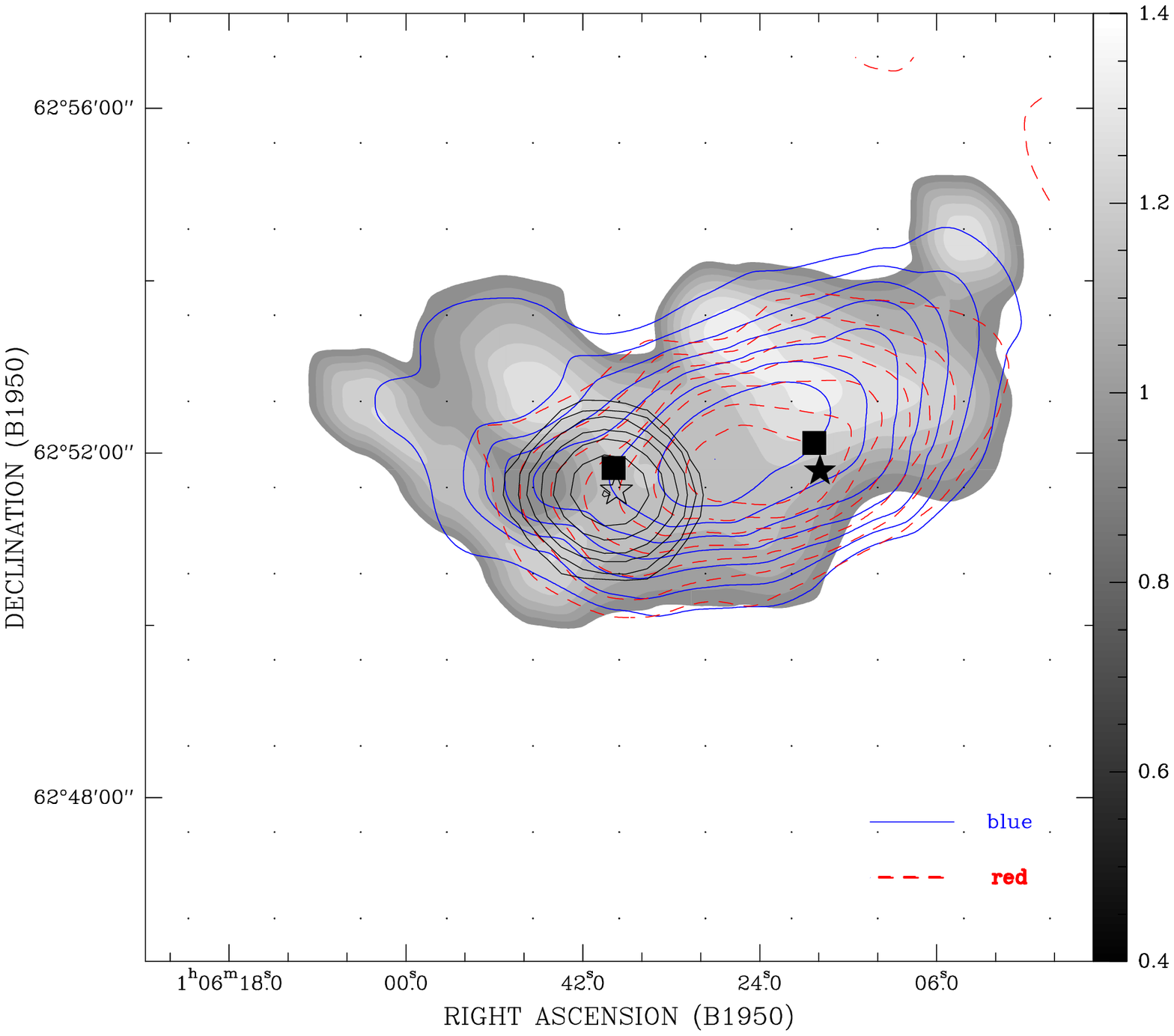}
\vspace{10mm}
\hspace{5mm}
\includegraphics[width=80mm]{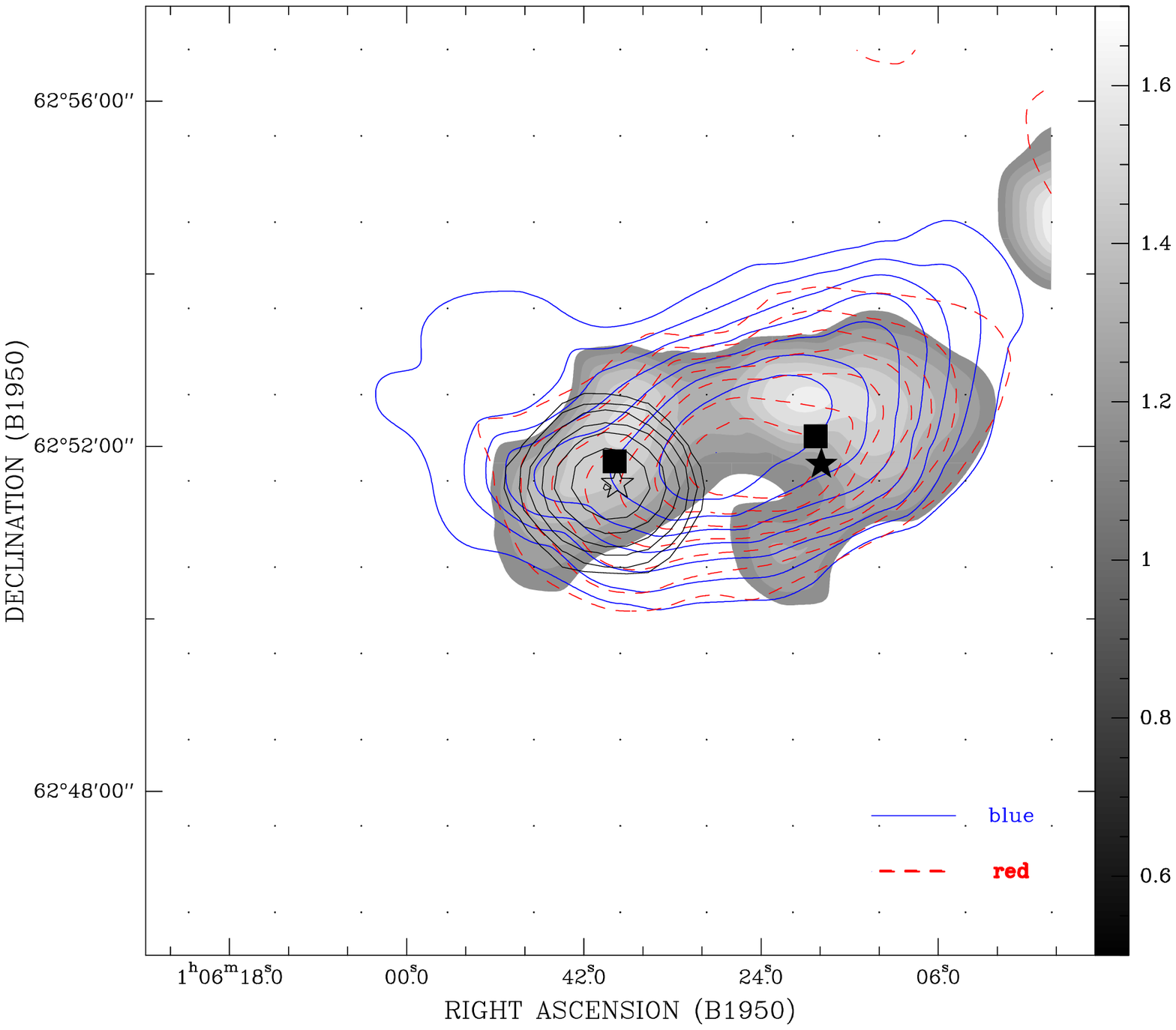}

\includegraphics[width=80mm]{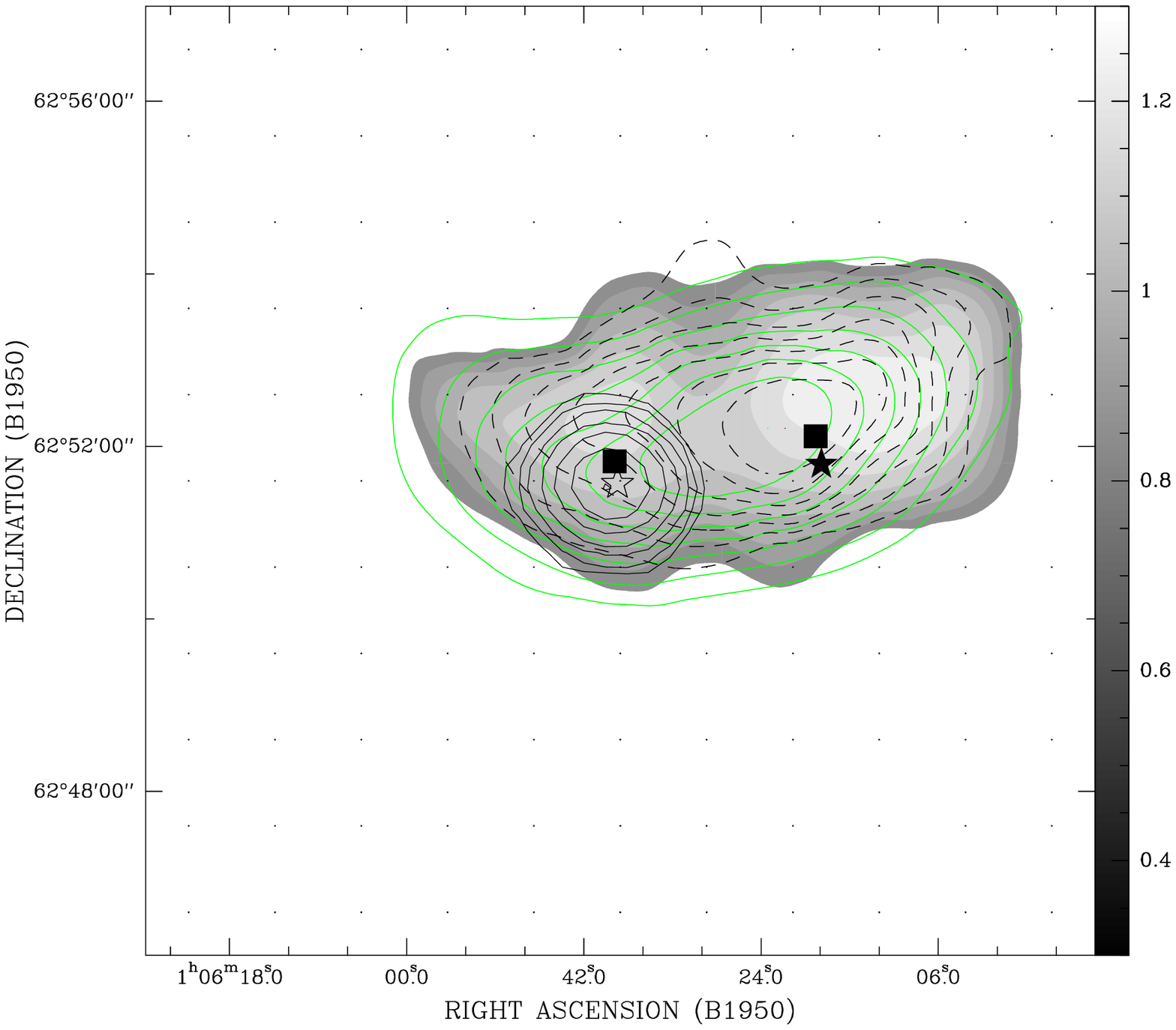}

\includegraphics[width=80mm]{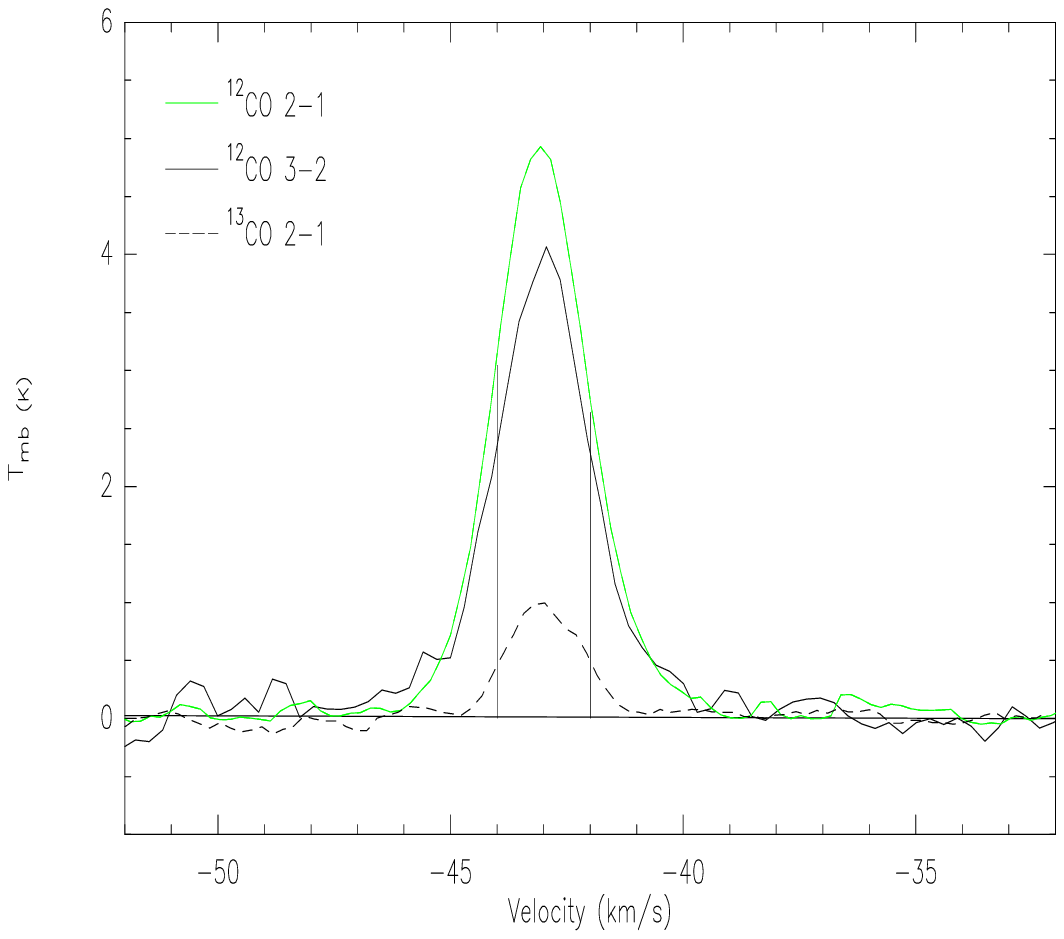}
\vspace{0.05cm}
 \caption {S186. Top left: outflow contours
(blue and red-shifted components are shown as blue solid and red
dashed lines) are superimposed on the line intensity ratio map
(grey scale) of the blue wing. Top right: outflow contours are
superimposed on the intensity ratio map (grey scale) of the red
wing. Bottom left: the integrated intensity map of the core in CO
$J=2-1$ (green solid contours) are superimposed on the grey map of
the line intensity ratio from the core component, and the dashed
contours are the integrated intensity map of the core in $^{13}$CO
$J=2-1$ line. The thin solid lines show the continuum at 1.4 GHz
from NVSS. The contours for the cores begin at 30\% of the maximum
integrated intensity ${\rm I_{max}}$ and increase with a step of
10\% of ${\rm I_{max}}$, and the integrated line intensity ratios
range from 30\% of the maximum ${\rm R_{I_{CO(3-2)}/I_{CO(2-1)}}}$
to the maximum ${\rm R_{I_{CO(3-2)}/I_{CO(2-1)}}}$. The intensity
ratios indicate the gas temperature variations at different
positions. The unfilled and filled stars mark IRAS 01056+6251 and
IRAS 01053+6251, respectively. The filled square symbols indicate
the MSX source. The dot symbols mark the mapped points. Bottom
right: the spectra averaged over the outflow zones (the CO $J=3-2$
data were smoothed to the resolution of the $J=2-1$ observations);
the vertical lines indicate the beginning of the blue and red
wings.}

  \end{figure}

\clearpage
\newpage
\begin{figure}[h]

\includegraphics[width=80mm]{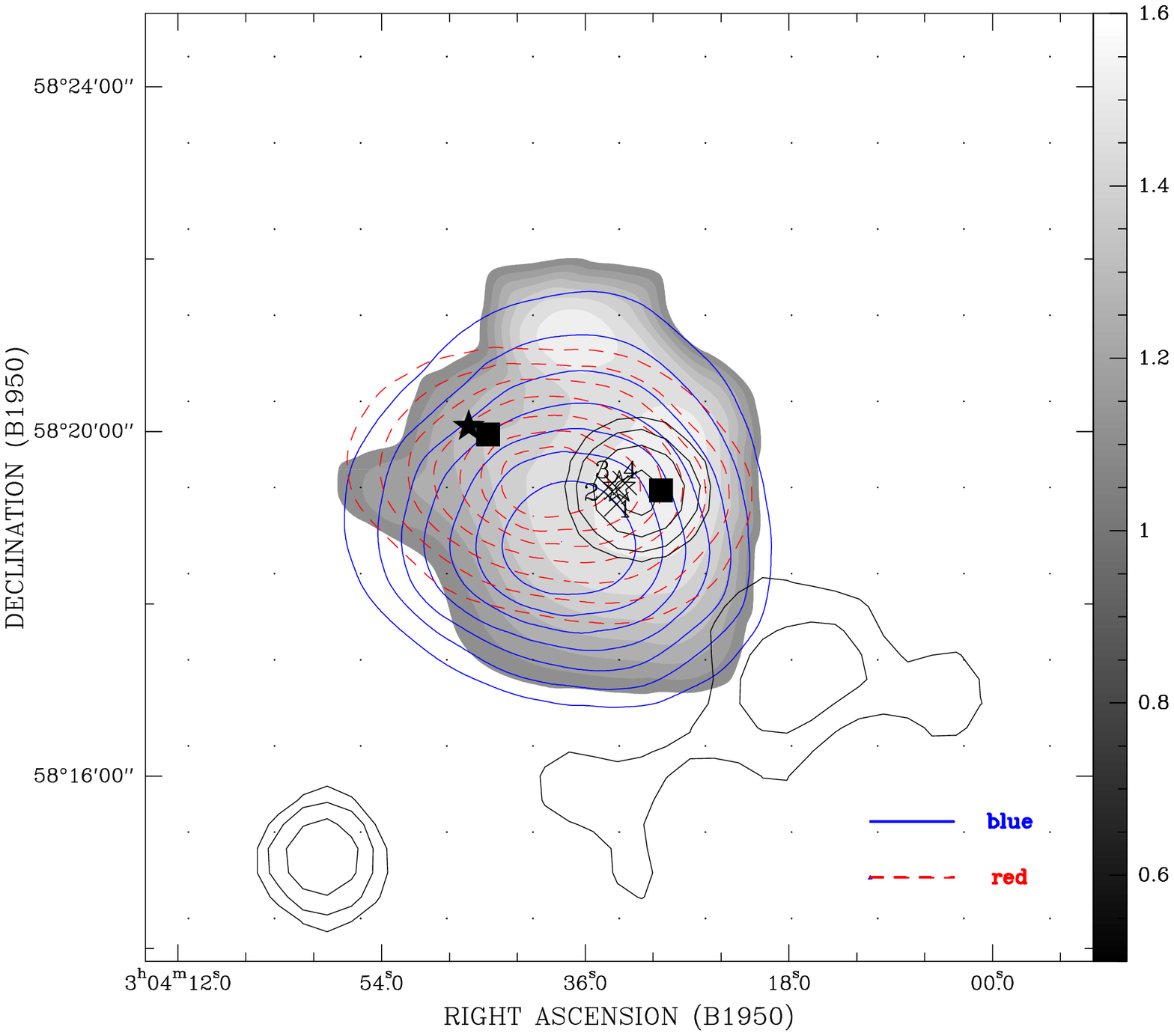}
\vspace{10mm} \hspace{5mm}
\includegraphics[width=80mm]{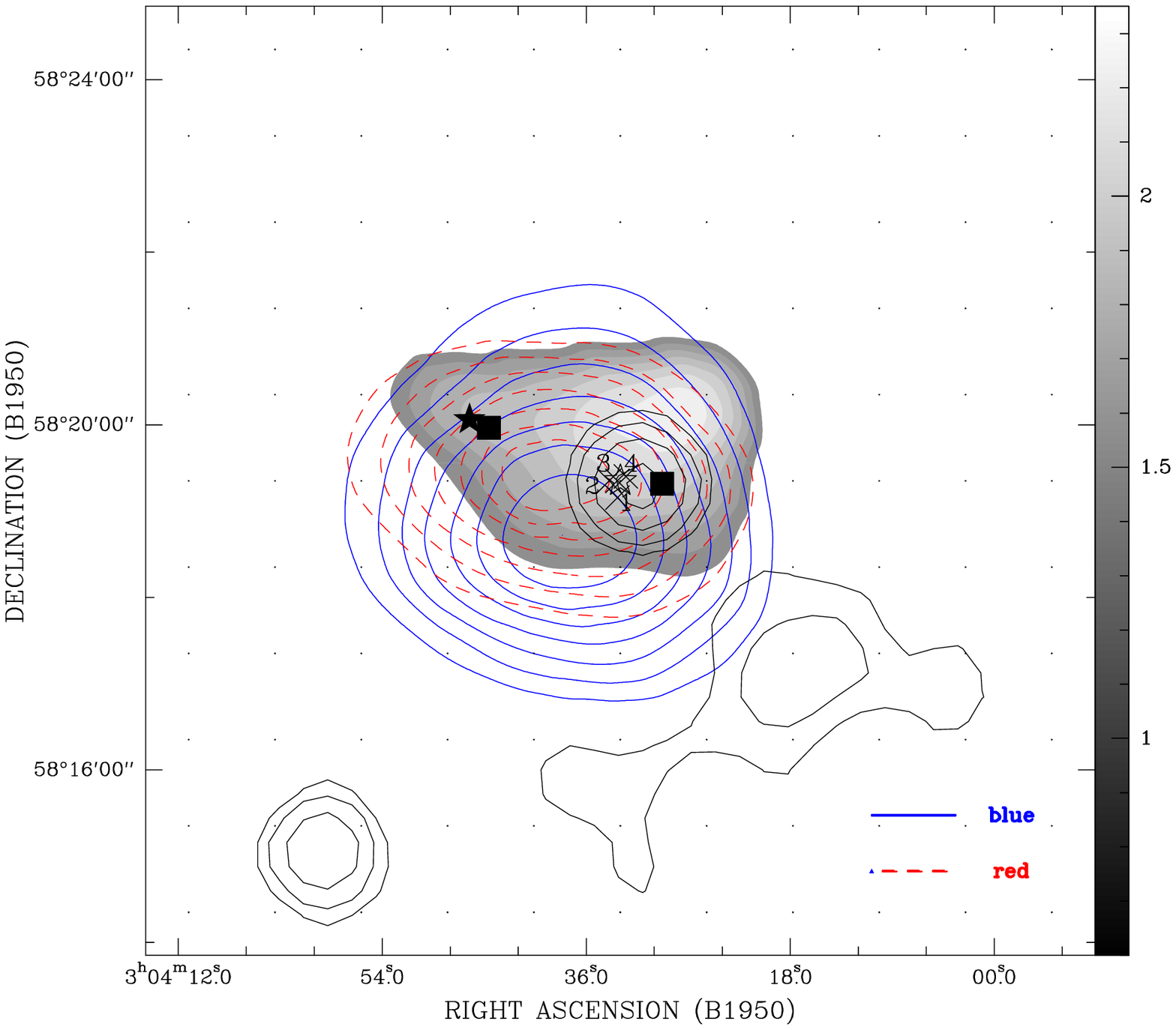}

\includegraphics[width=80mm]{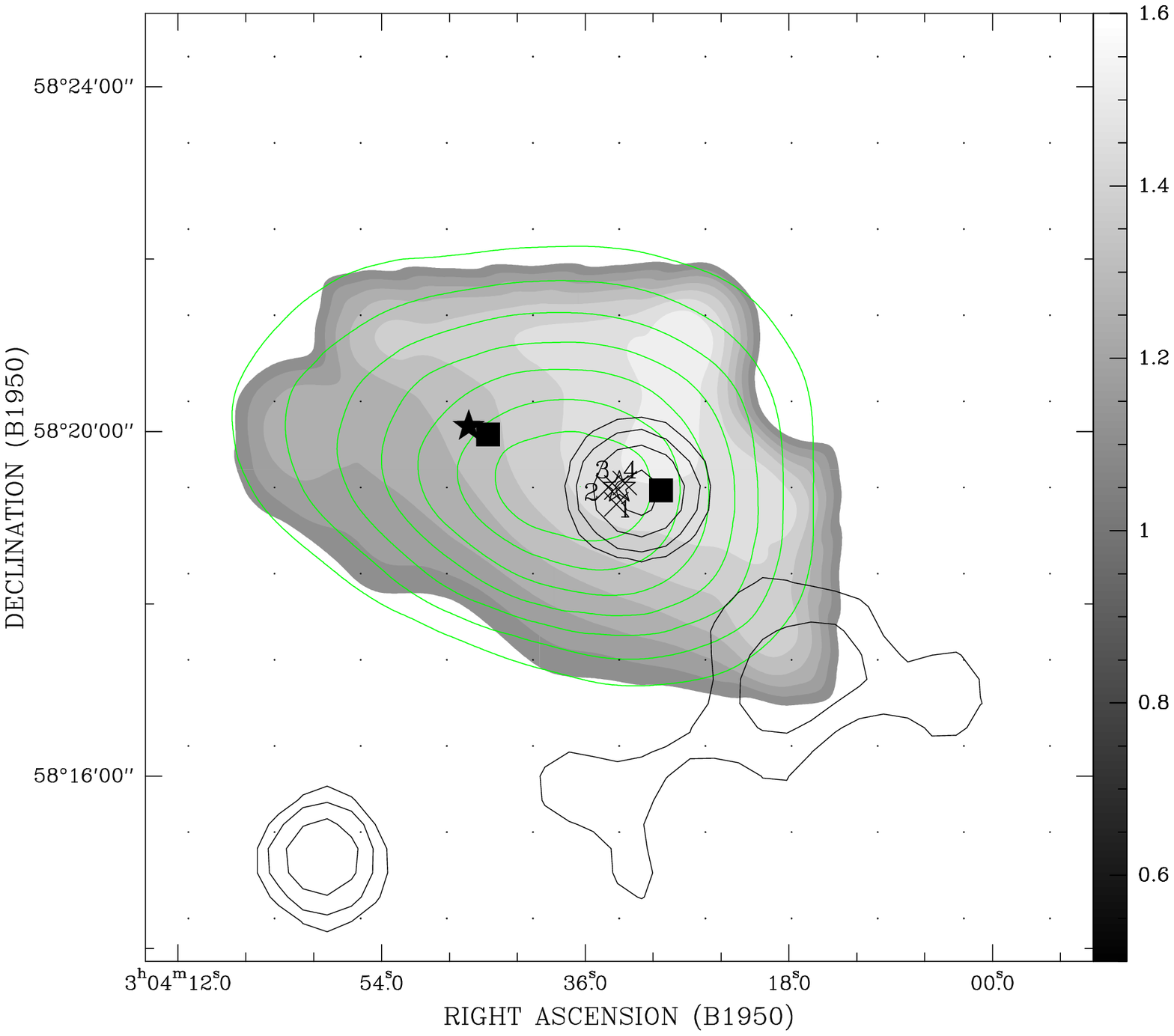}

\includegraphics[width=80mm]{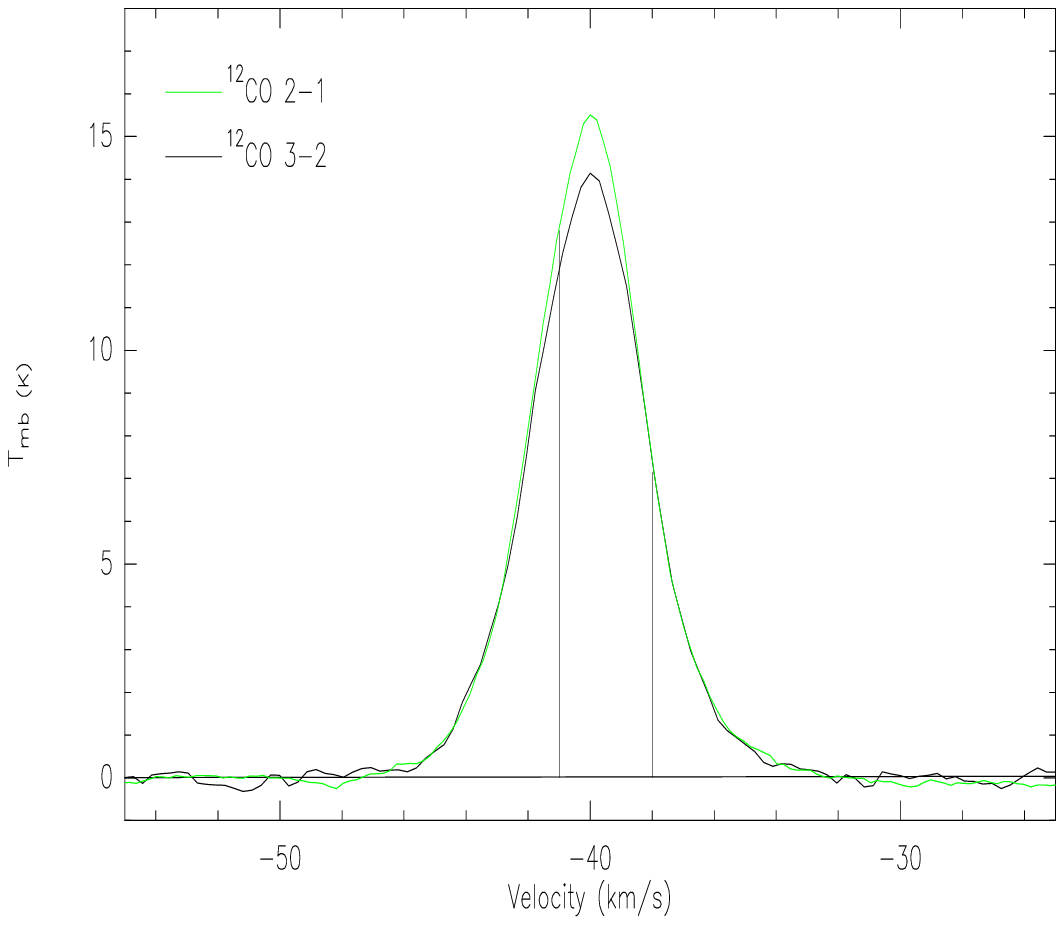}
\vspace{0.1cm}

\caption {G139.909+0.197. Caption as in Fig.B.1. The filled star
is IRAS 03037+5819, and unfilled star marks IRAS 03035+5819. The
cross symbol 1 is SiO maser, and the rest crosses 2, 3 and 4
indicate ${\rm H_{2}O}$ masers. }

\end{figure}
\clearpage
\newpage
\begin{figure}[h]

 \includegraphics[width=80mm]{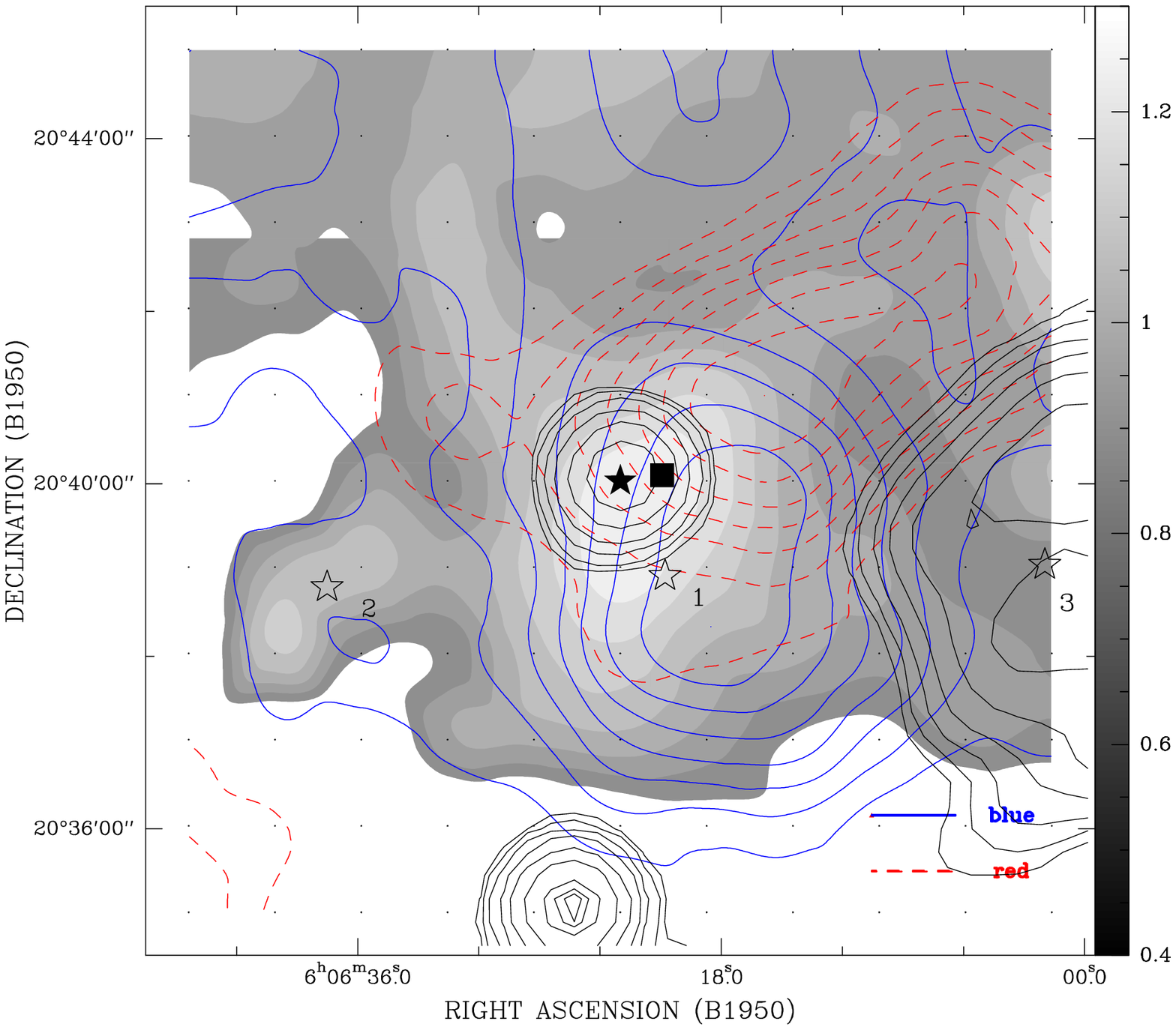}
\vspace{10mm} \hspace{5mm}
\includegraphics[width=80mm]{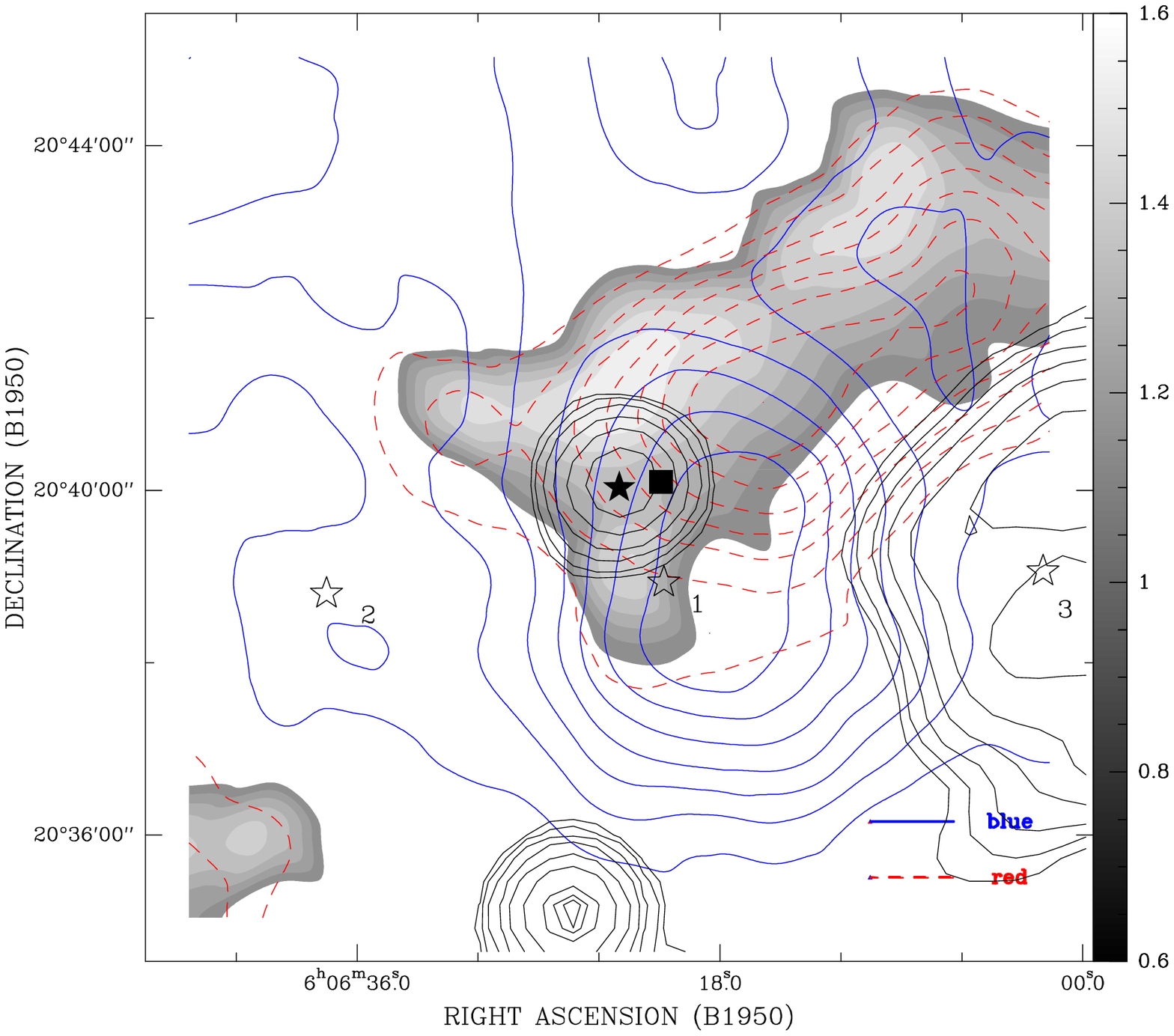}

\includegraphics[width=80mm]{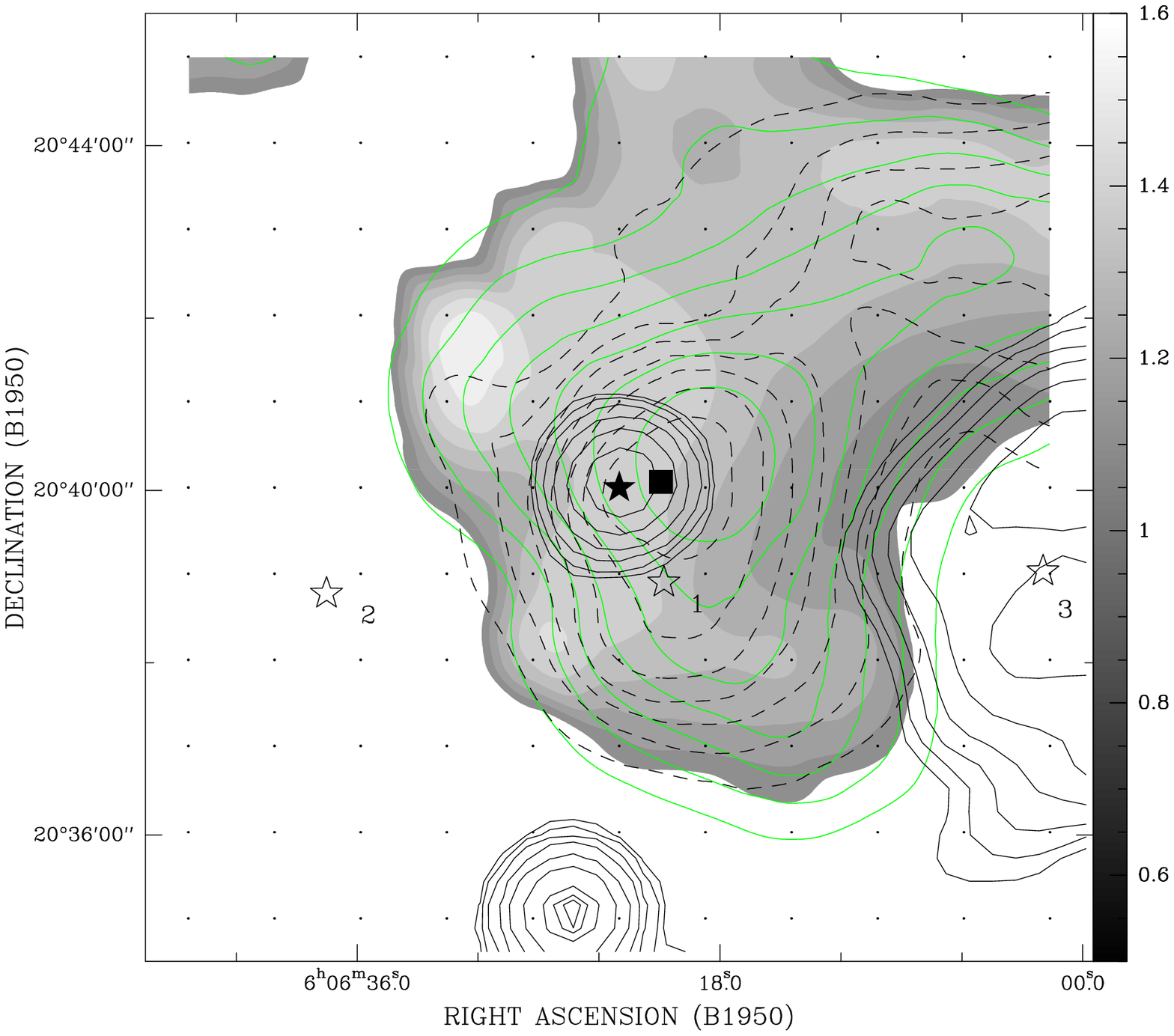}

\includegraphics[width=80mm]{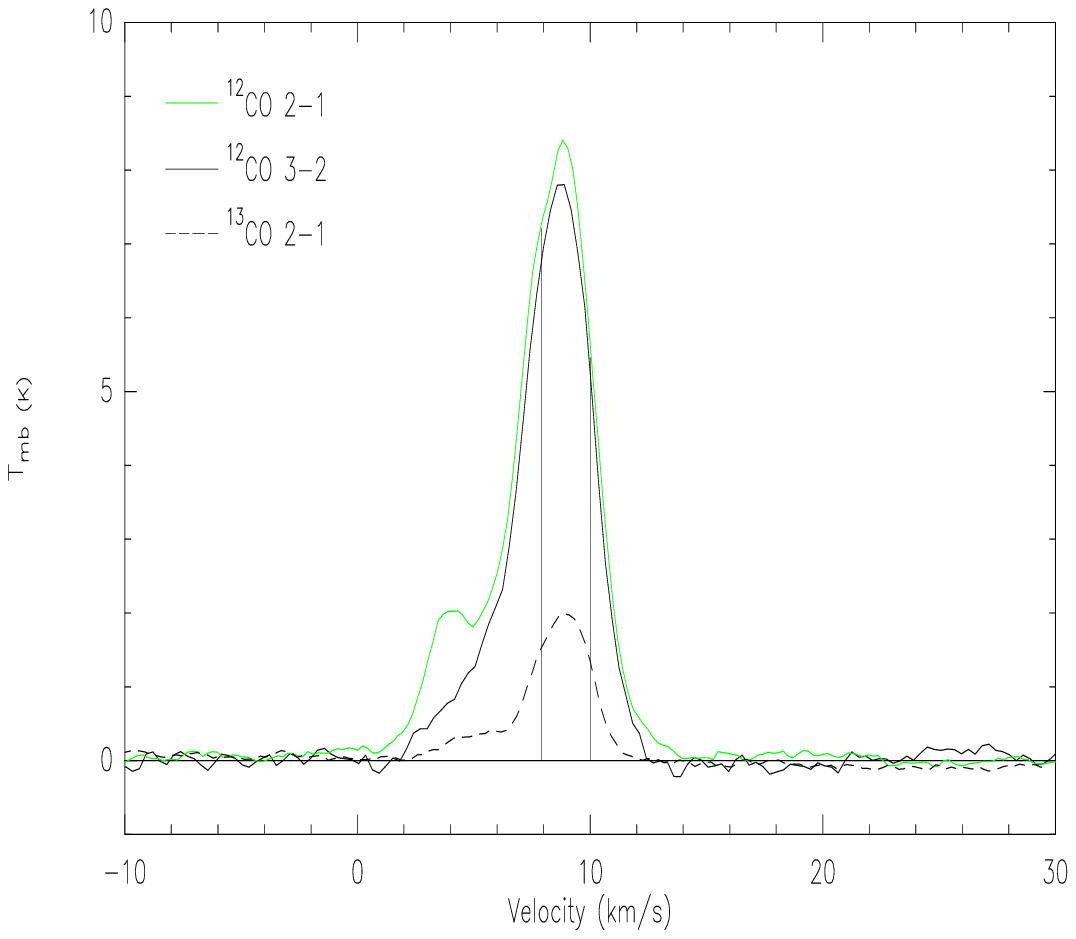}
\vspace{0.1cm}
 \caption {G189.876+0.516. Caption as in Fig.B.1. The unfilled
stars 1, 2, 3 are B-type stars ALS 8745, ALS 8748 and B1-type star
HD 252325, respectively; the filled star marks IRAS 06063+2040.}
\end{figure}
\clearpage
\newpage
\begin{figure}[h]

  \includegraphics[width=80mm]{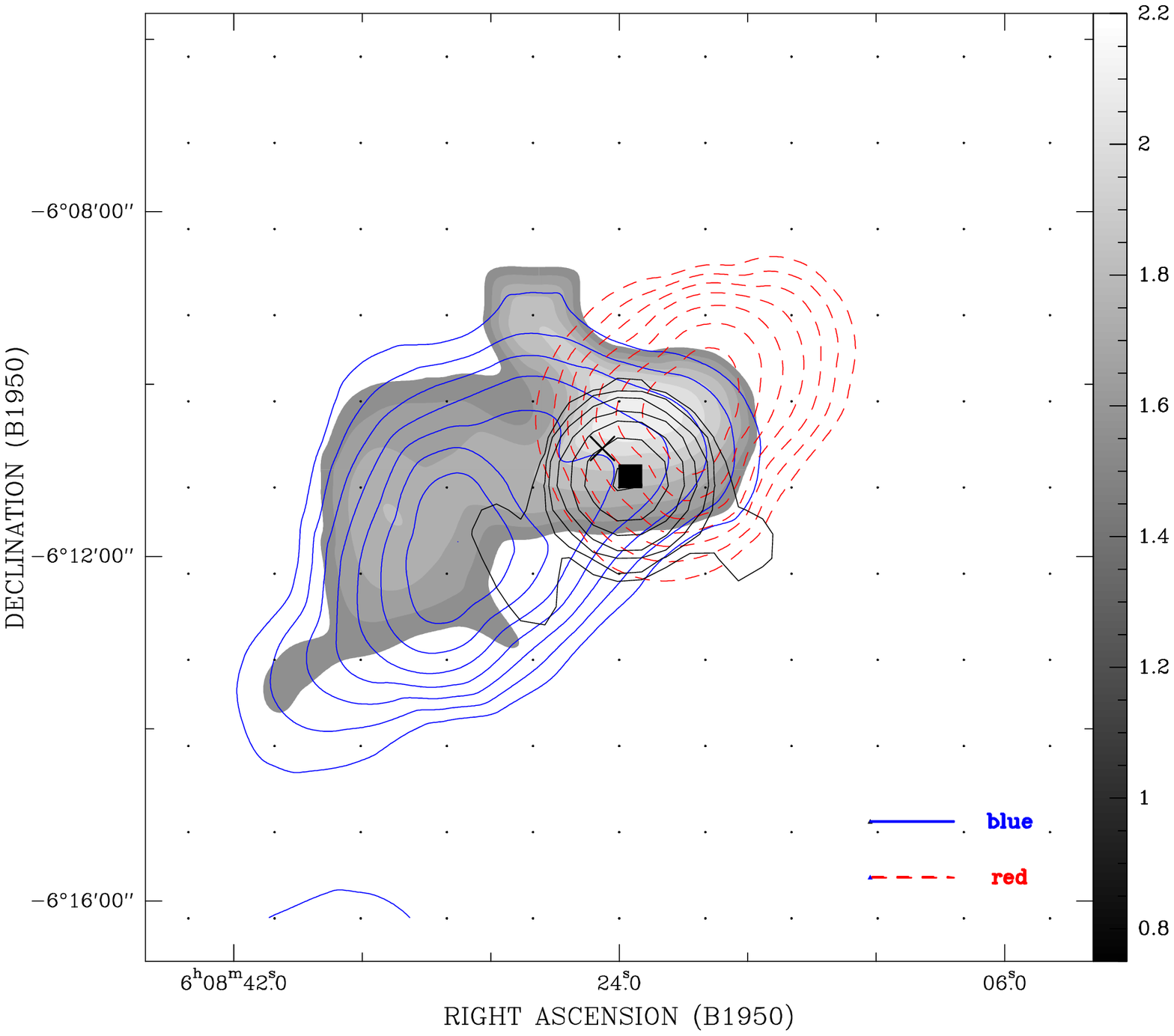}
\vspace{10mm} \hspace{5mm}
\includegraphics[width=80mm]{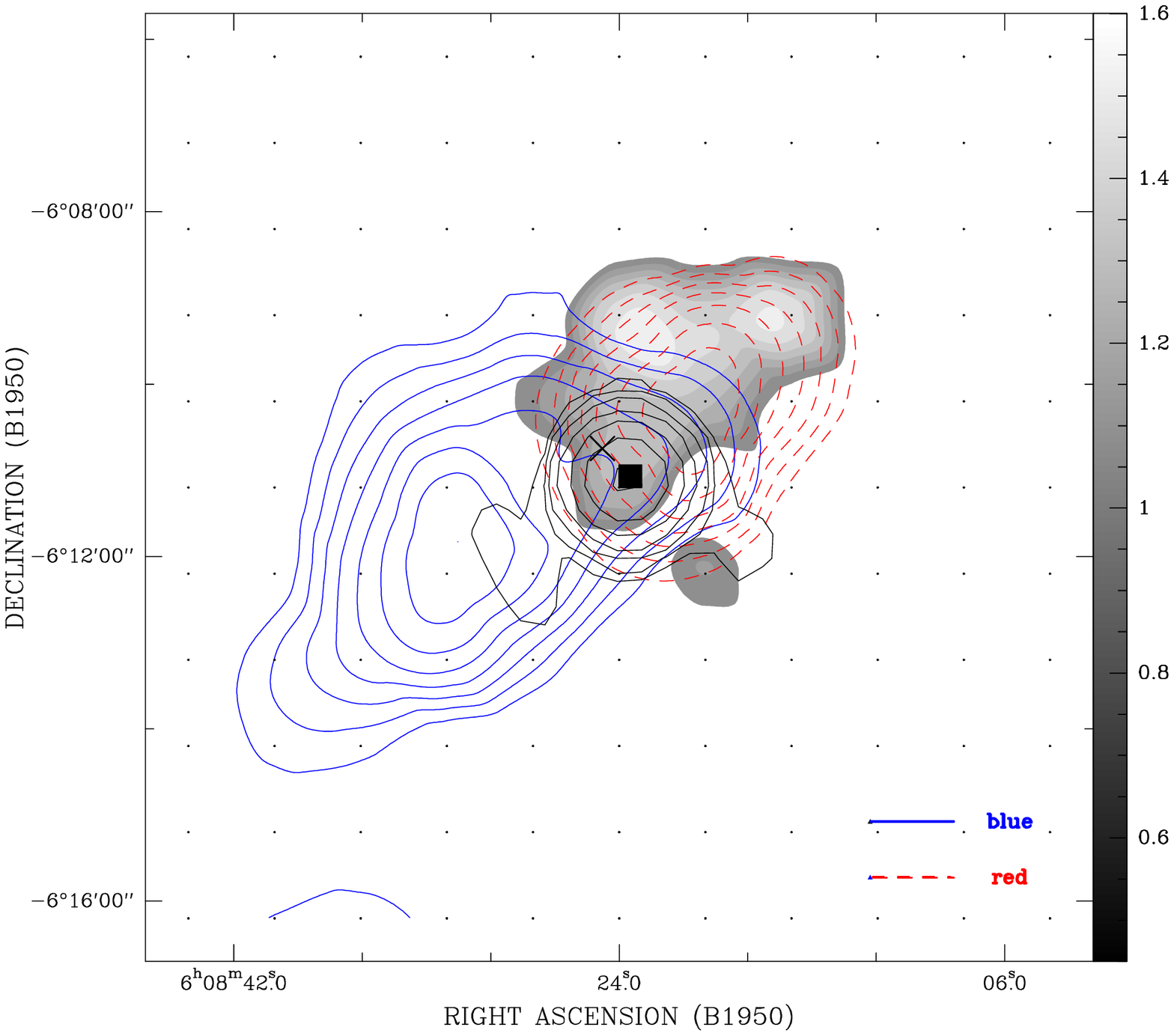}

\includegraphics[width=80mm]{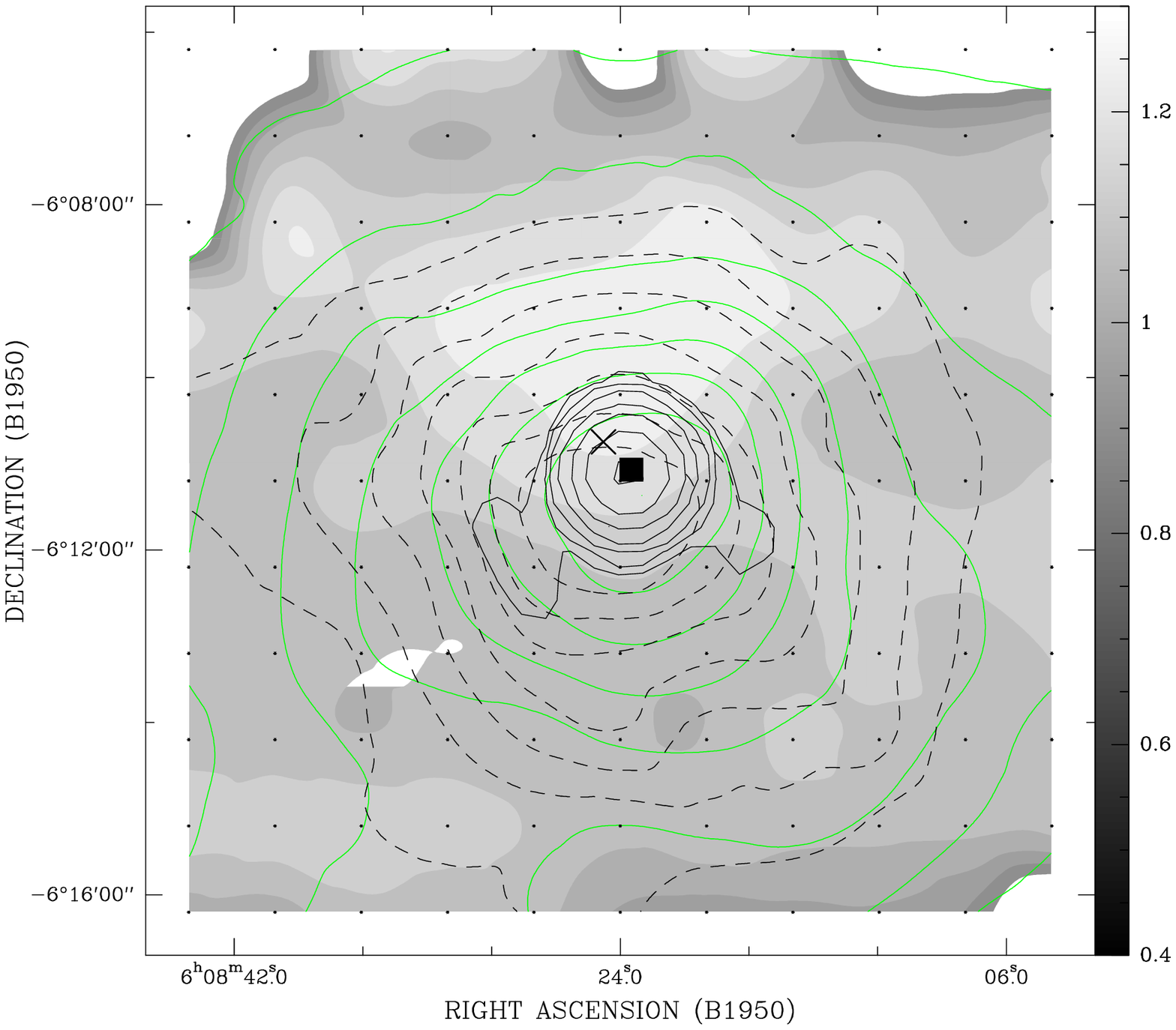}

\includegraphics[width=80mm]{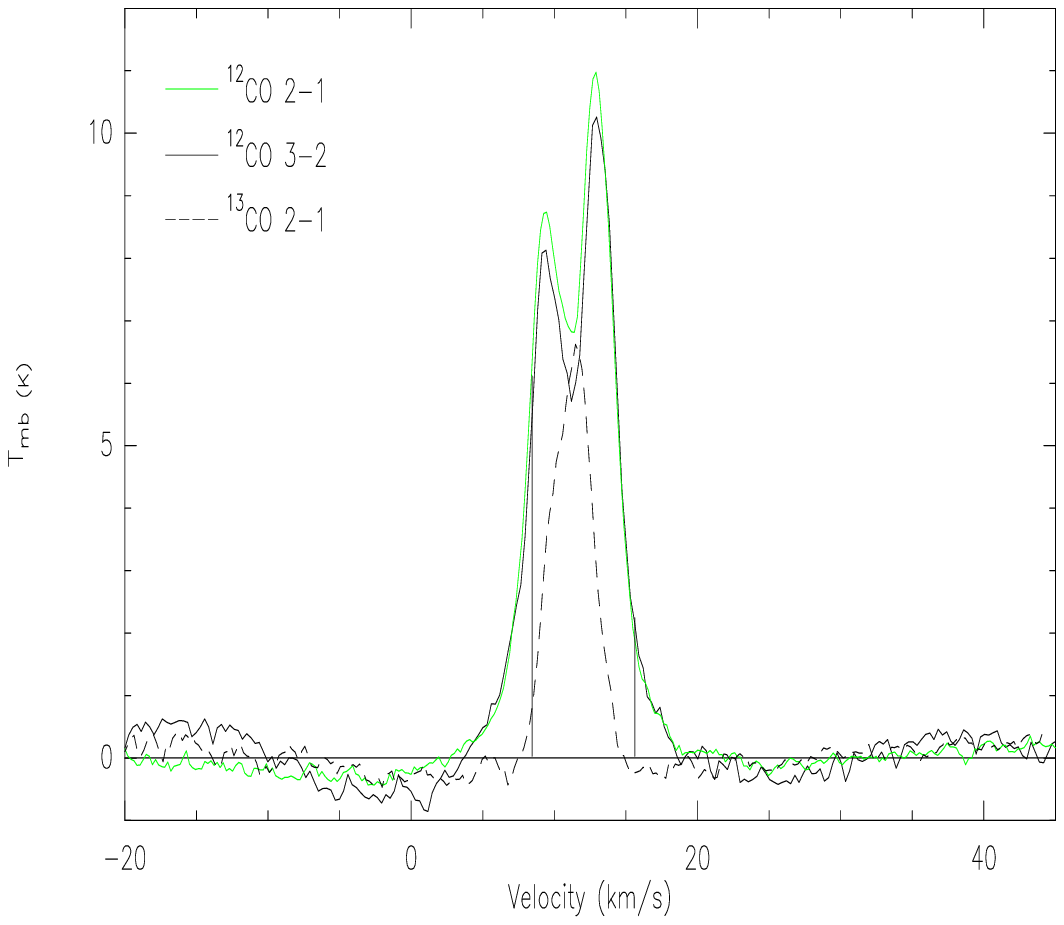}
\vspace{0.1cm} \caption {G213.880-11.837. Caption as in Fig.B.1.
The filled square symbol marks the peak of 8.7 $\mu$m emission.
IRAS 06084-0611 is located at the same position as the 8.7 $\mu$m
emission. The cross symbol marks ${\rm H_{2}O}$ maser.}
\end{figure}
\clearpage
\newpage
\begin{figure}[h]

  \includegraphics[width=80mm]{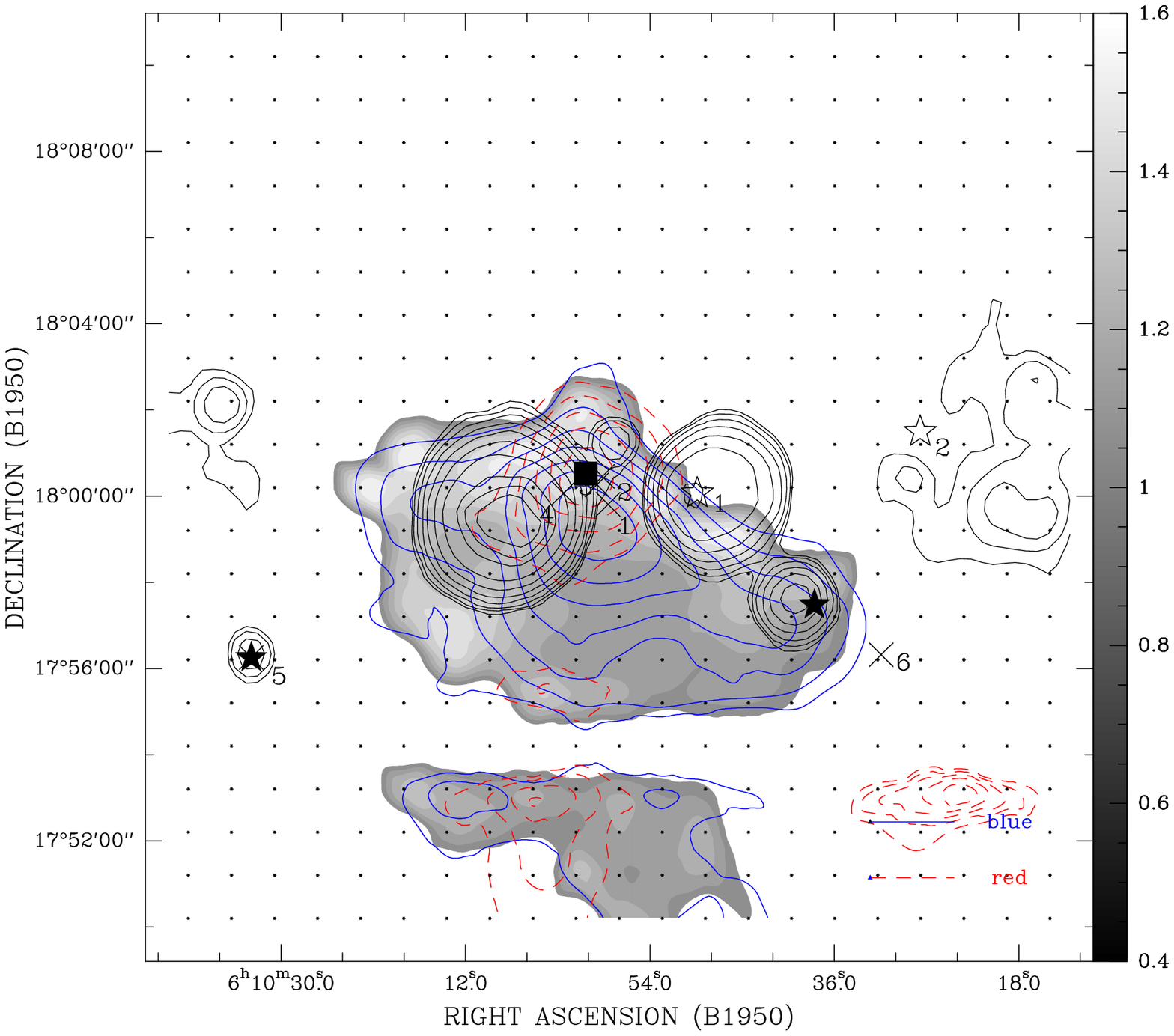}
\vspace{10mm} \hspace{5mm}
\includegraphics[width=80mm]{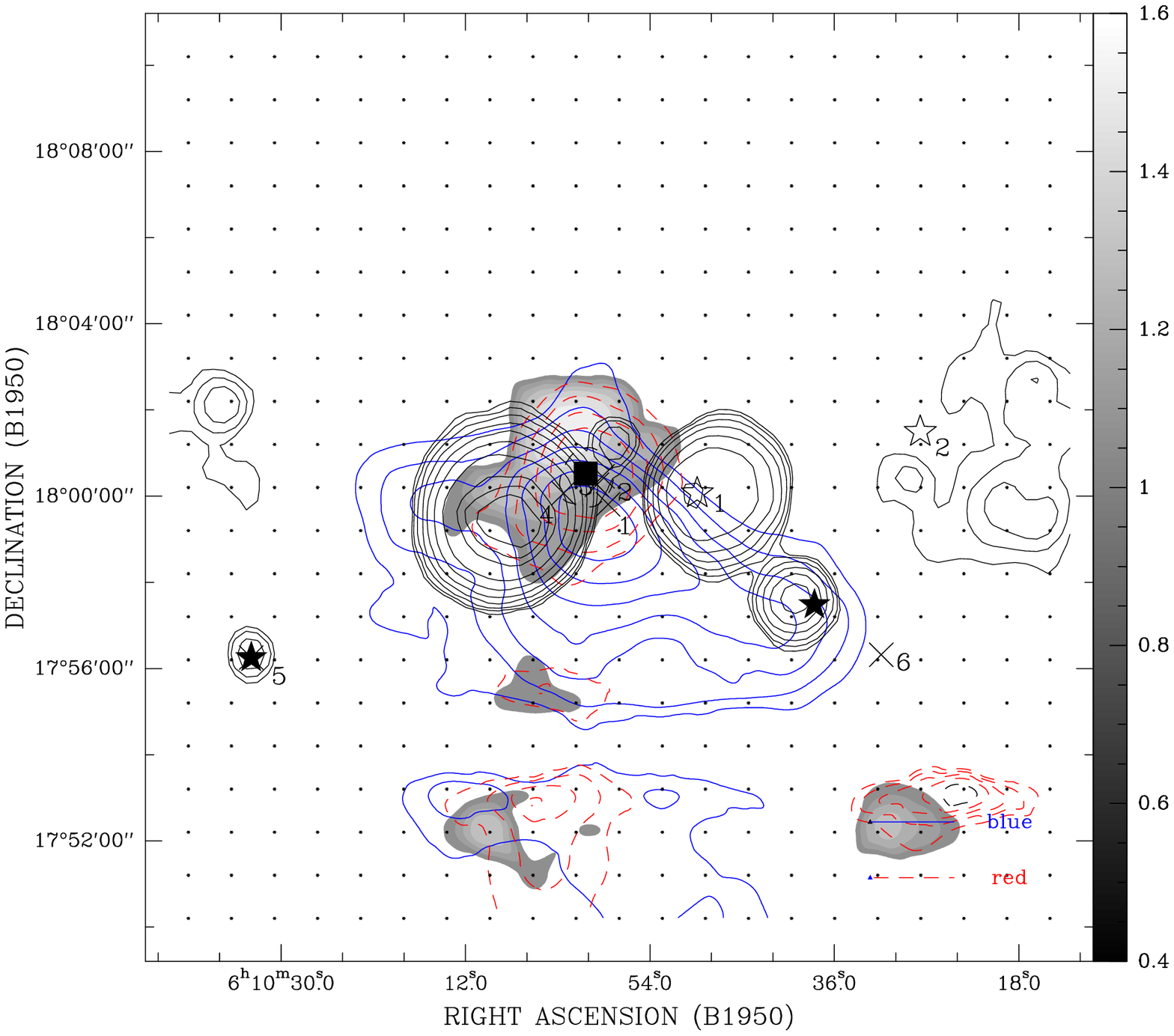}

\includegraphics[width=80mm]{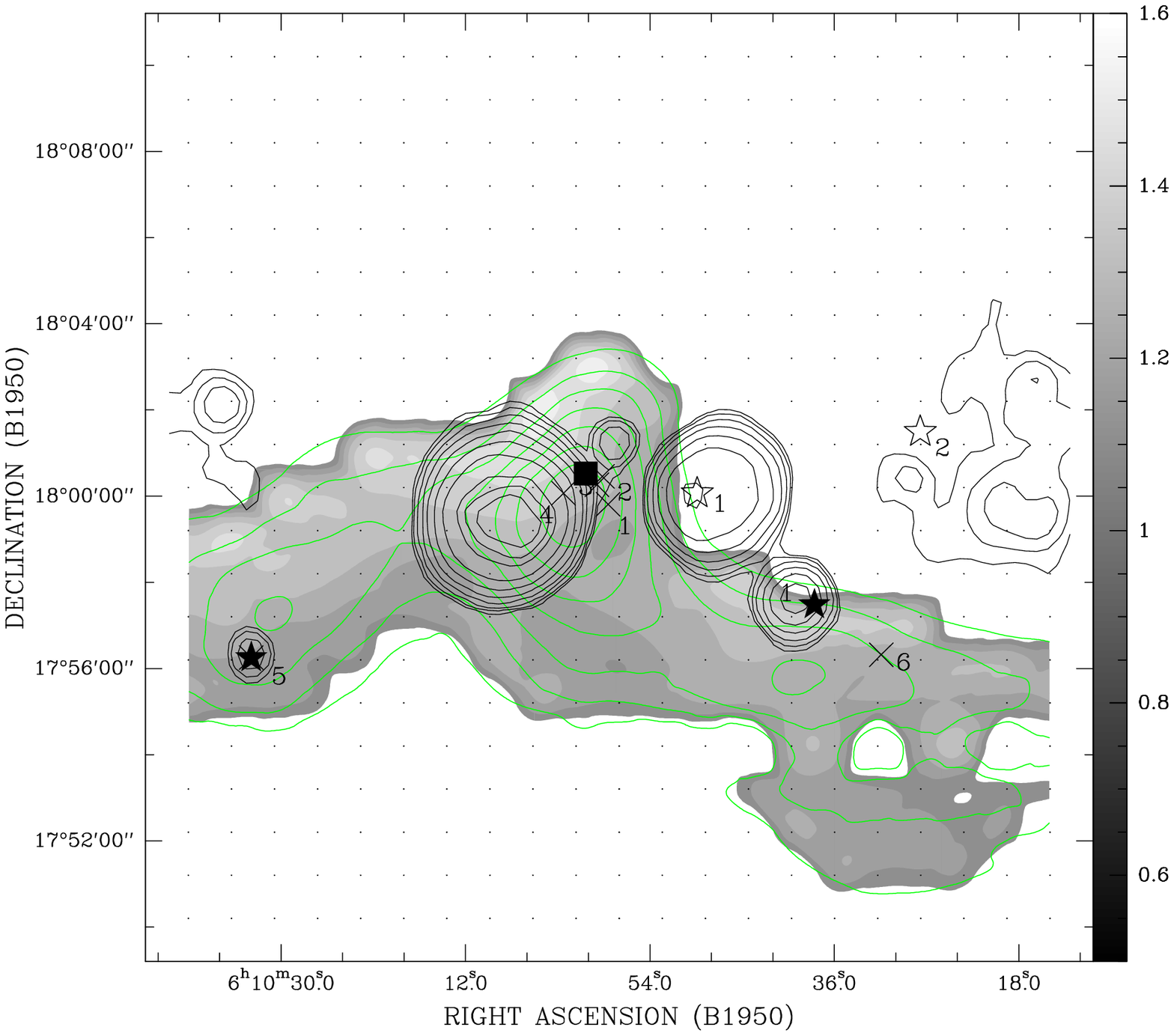}

\includegraphics[width=80mm]{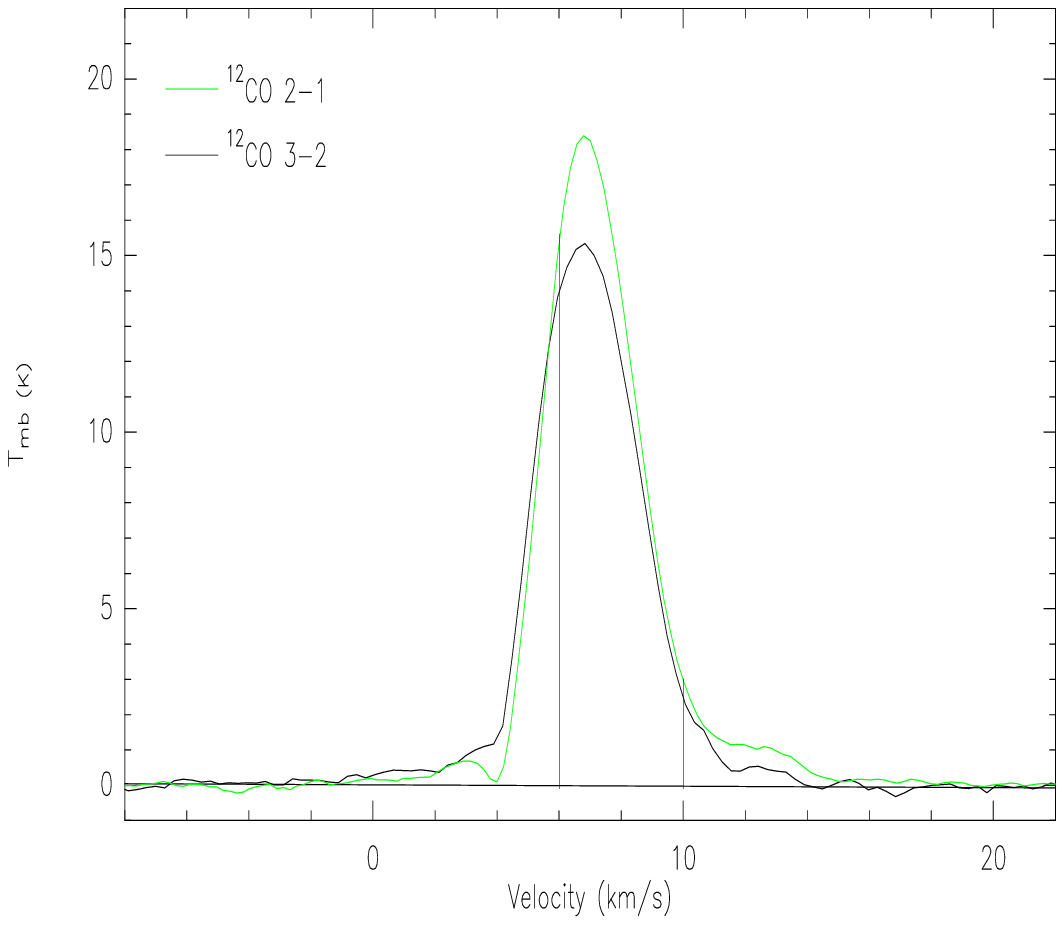}
\vspace{0.1cm} \caption {G192.584-0.041. Caption as in Fig.B.1.
The unfilled stars 1 and 2 are B0.5-type star HD 253327 and
B1-star HD 253247; the filled stars are IRAS 06105+1756 and IRAS
06096+1757. The cross symbols 1, 2, 3, 5 and 6 indicate ${\rm
H_{2}O}$ masers, the cross symbol 4 is class II ${\rm CH_{3}OH}$
maser. IRAS 06099+1800 is located at the MSX source position. }

 \label{fig:fig.5.}
\end{figure}
\clearpage
\newpage
\begin{figure}[h]

 \includegraphics[width=80mm]{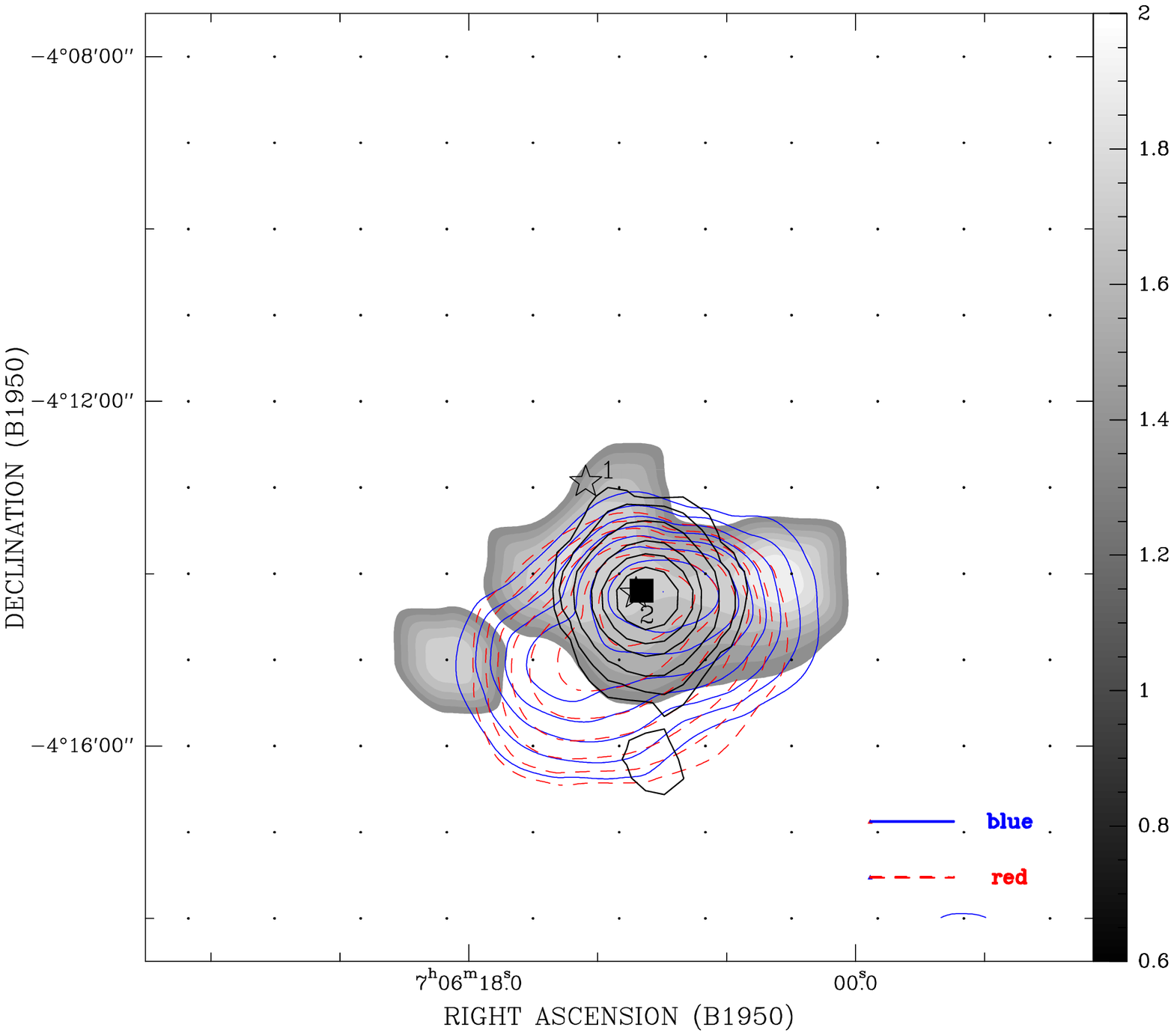}
\vspace{10mm} \hspace{5mm}
\includegraphics[width=80mm]{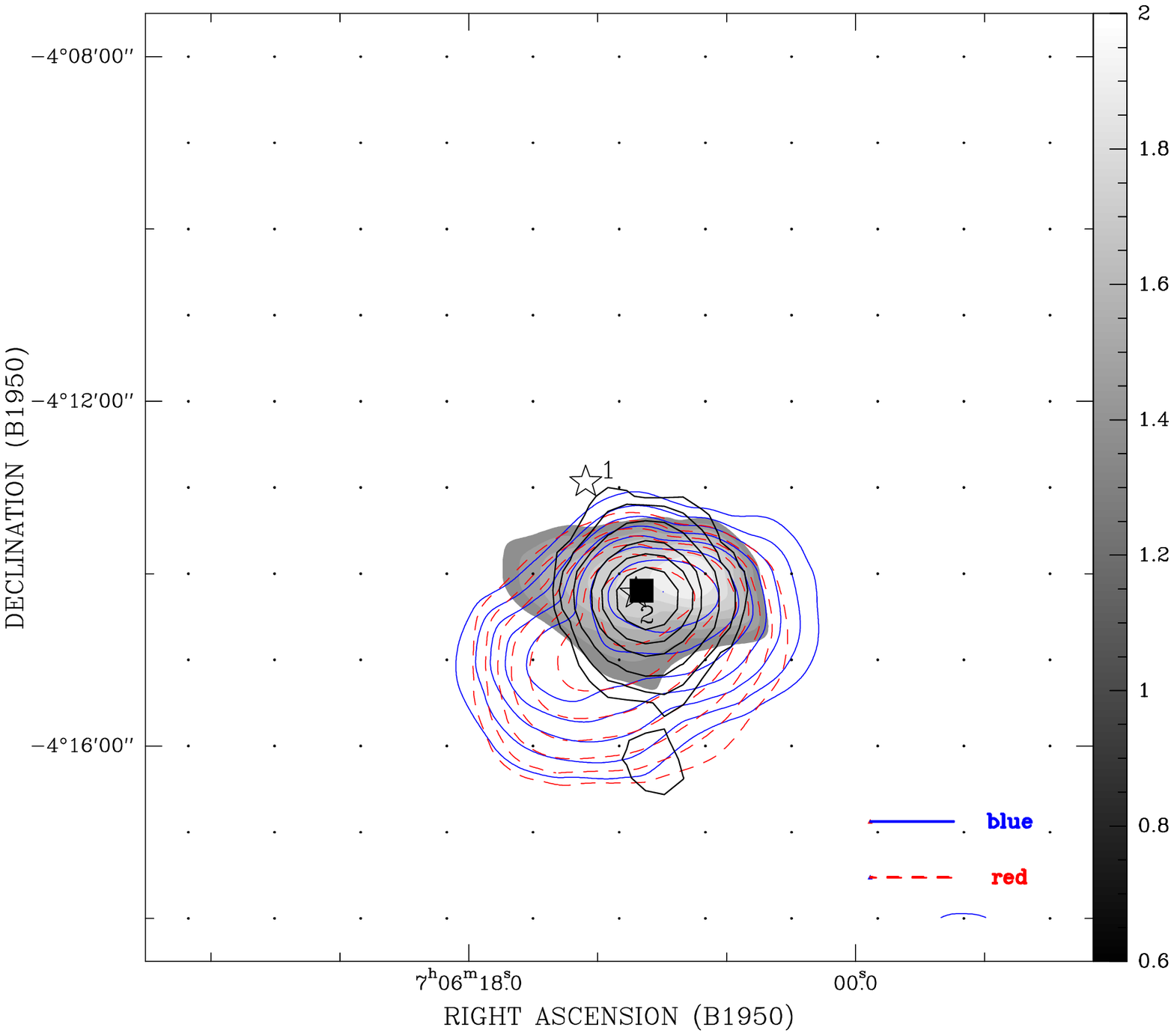}

\includegraphics[width=80mm]{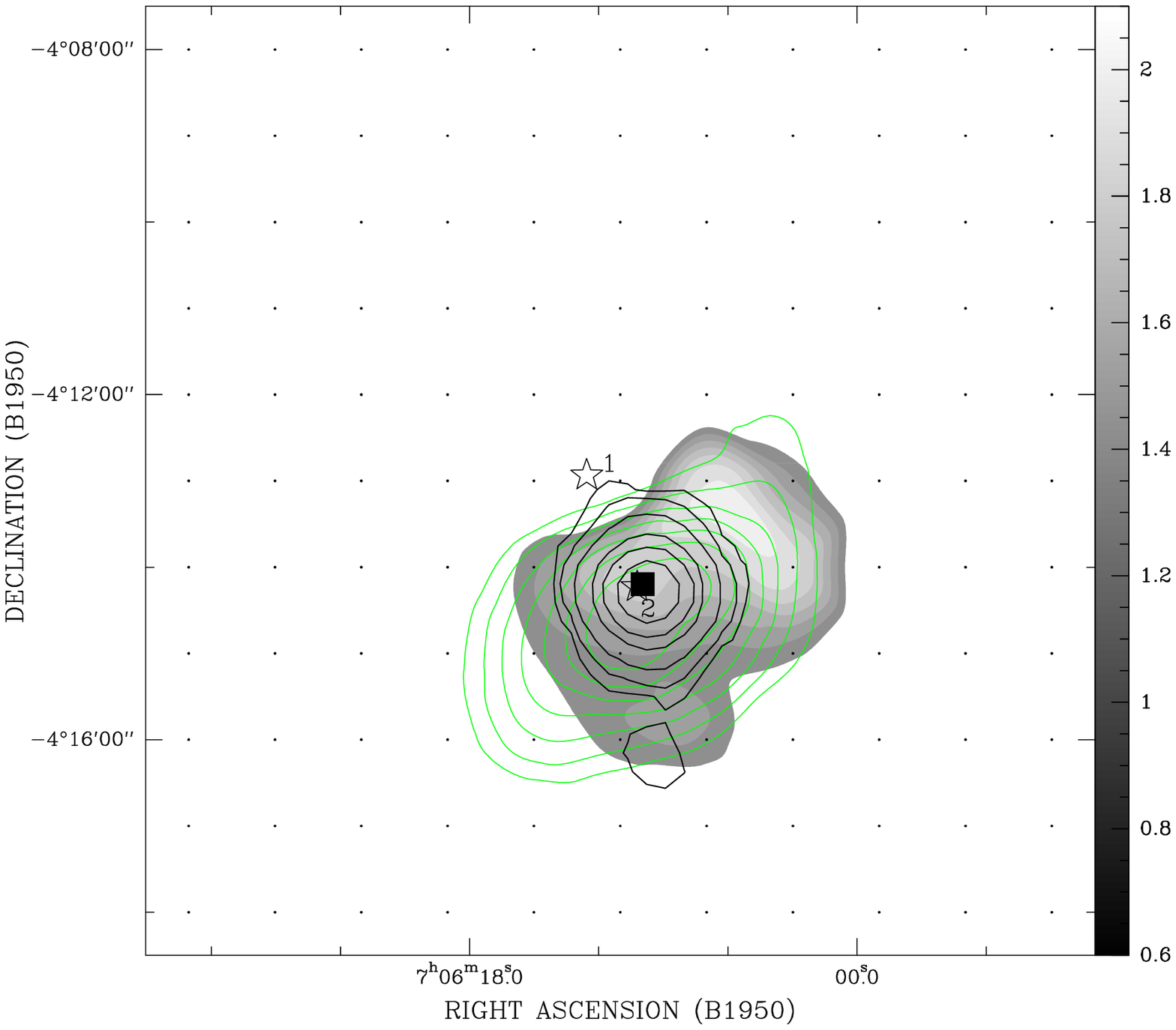}

\includegraphics[width=80mm]{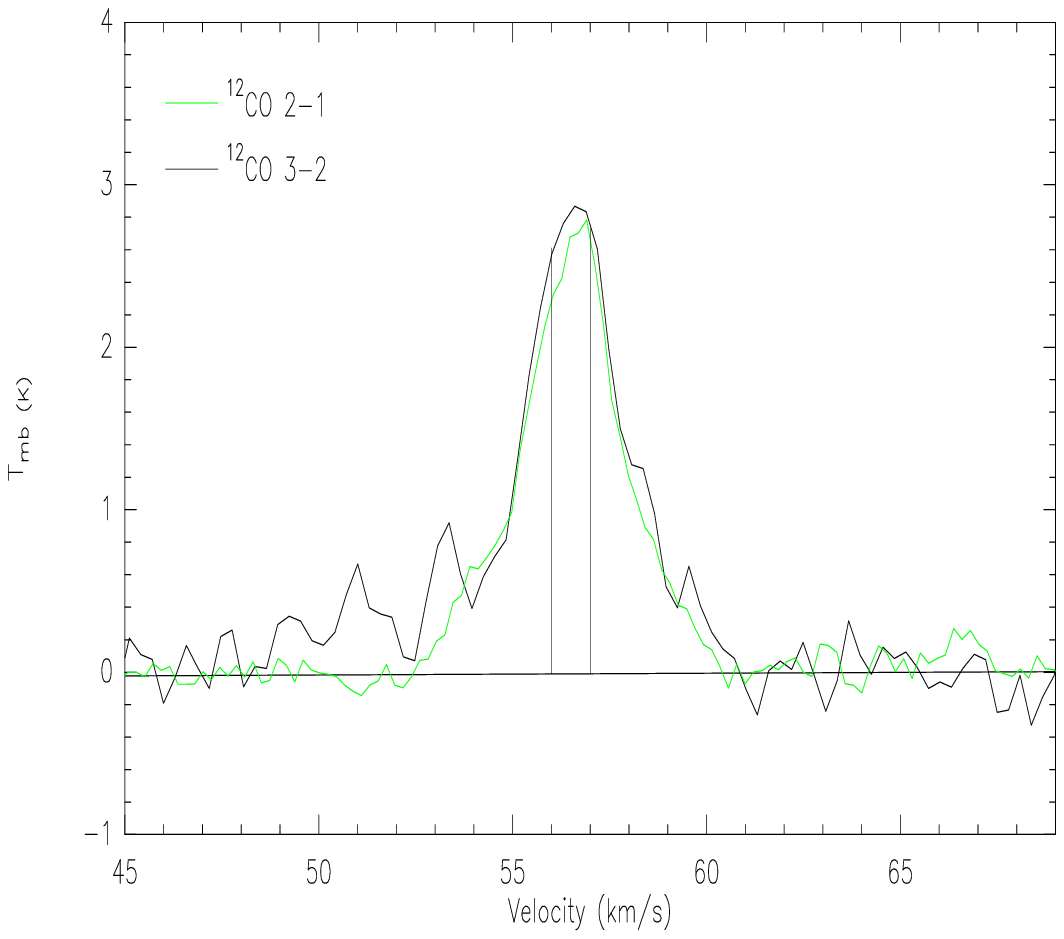}
\vspace{0.1cm} \caption {S288. Caption as in Fig.B.1. The star 1
is B8-type star HD 296489. The star 2 is B0-type star Ced 92
associated with IRAS 07061-0414. }

  \label{fig:fig.6.}
\end{figure}
\clearpage
\newpage
\begin{figure}[h]

 \includegraphics[width=80mm]{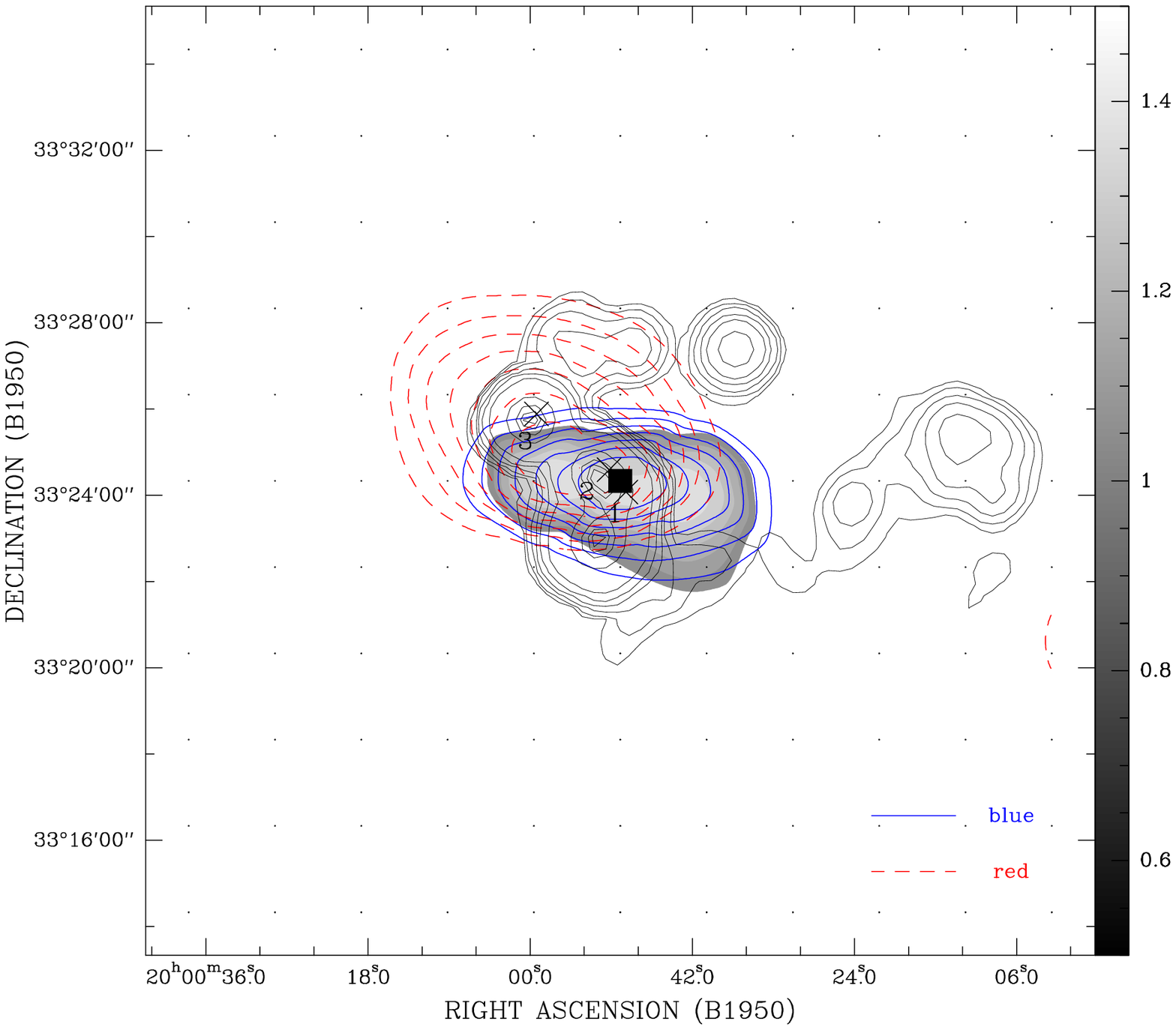}
\vspace{10mm} \hspace{5mm}
\includegraphics[width=80mm]{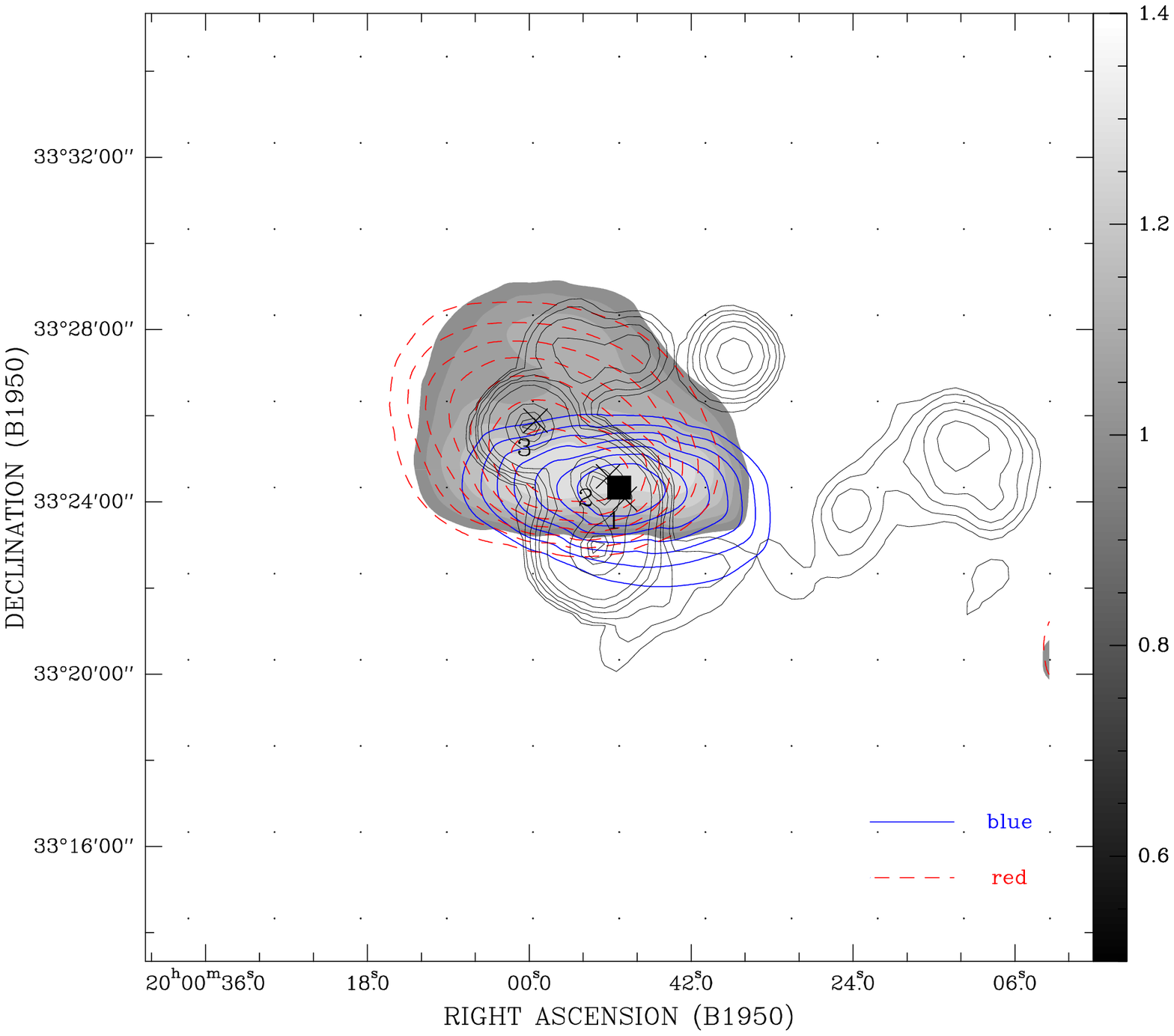}

\includegraphics[width=80mm]{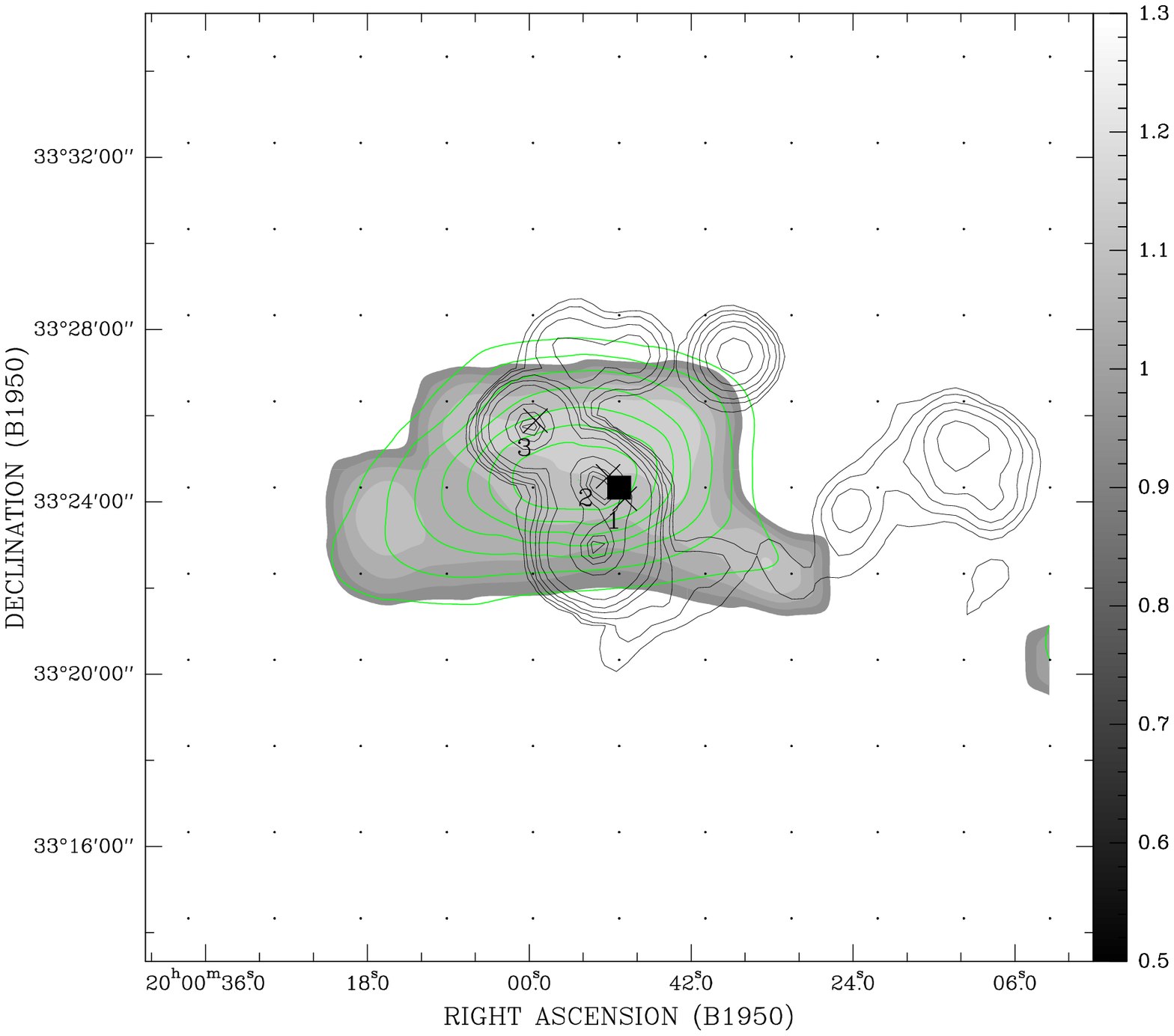}

\includegraphics[width=80mm]{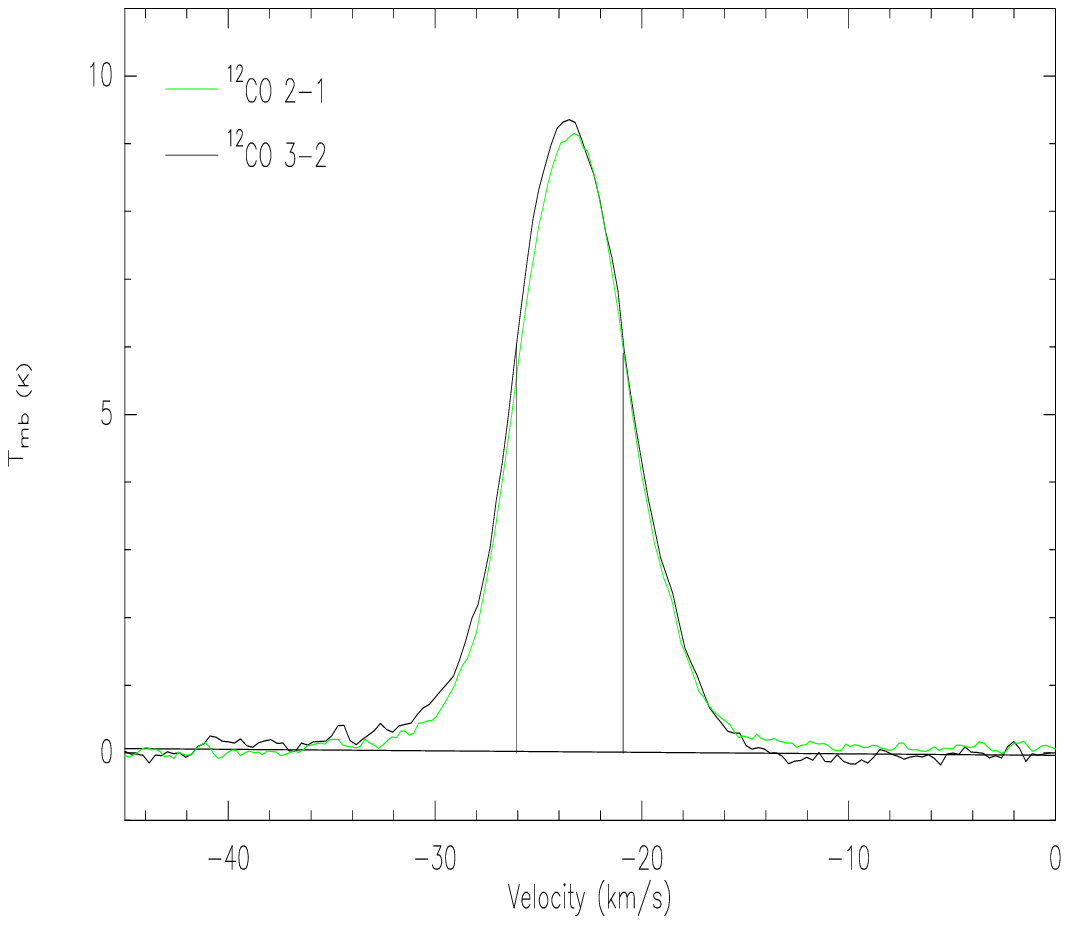}
\vspace{0.1cm}

\caption {G70.293+1.600. Caption as in Fig.B.1. The cross symbols
1, 3 mark ${\rm H_{2}O}$ masers and 2 is OH maser.}

 \label{fig:fig.7.}
\end{figure}
\clearpage
\newpage

\begin{figure}[h]

\includegraphics[width=80mm]{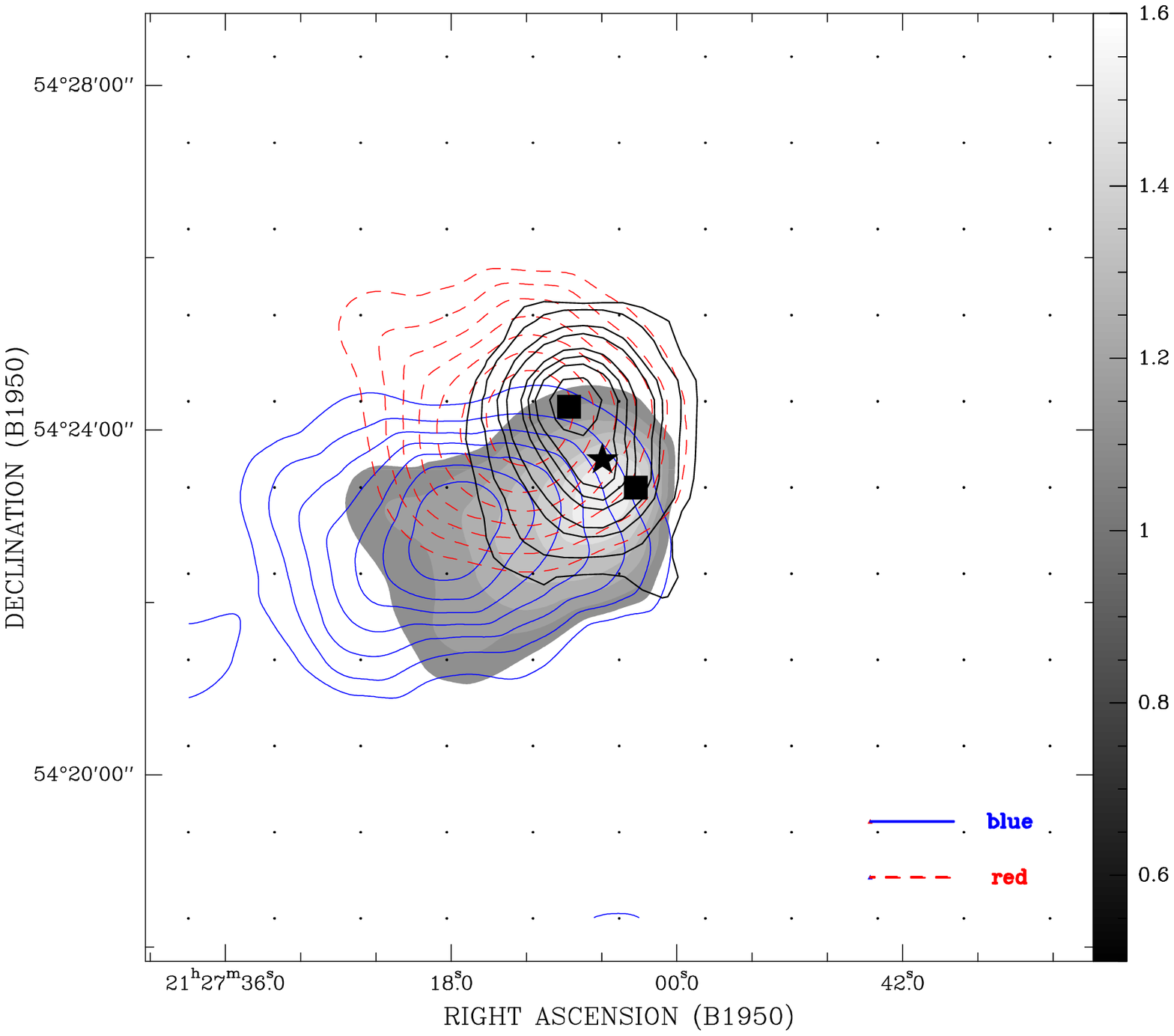}
\vspace{10mm} \hspace{5mm}
\includegraphics[width=80mm]{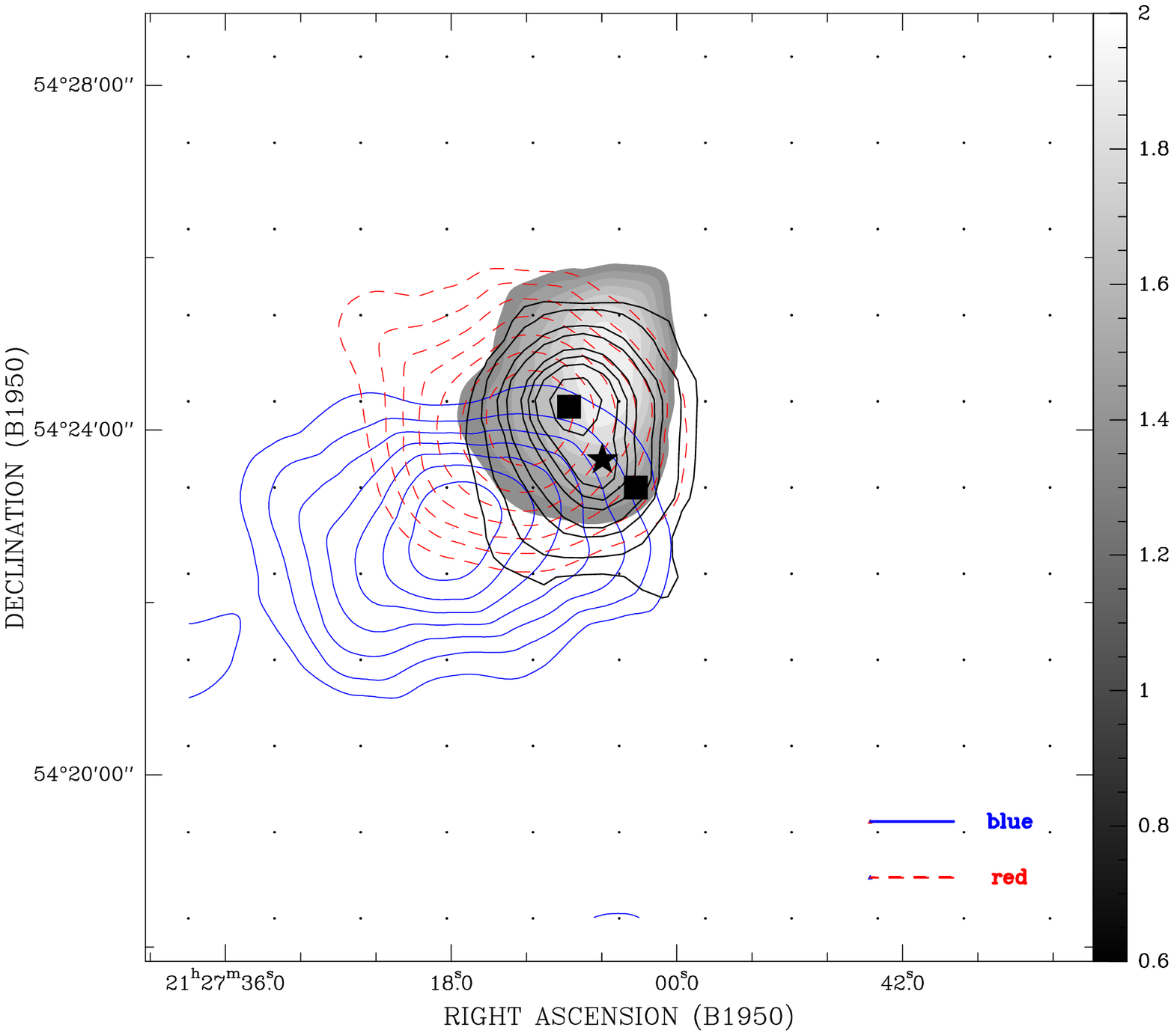}

\includegraphics[width=80mm]{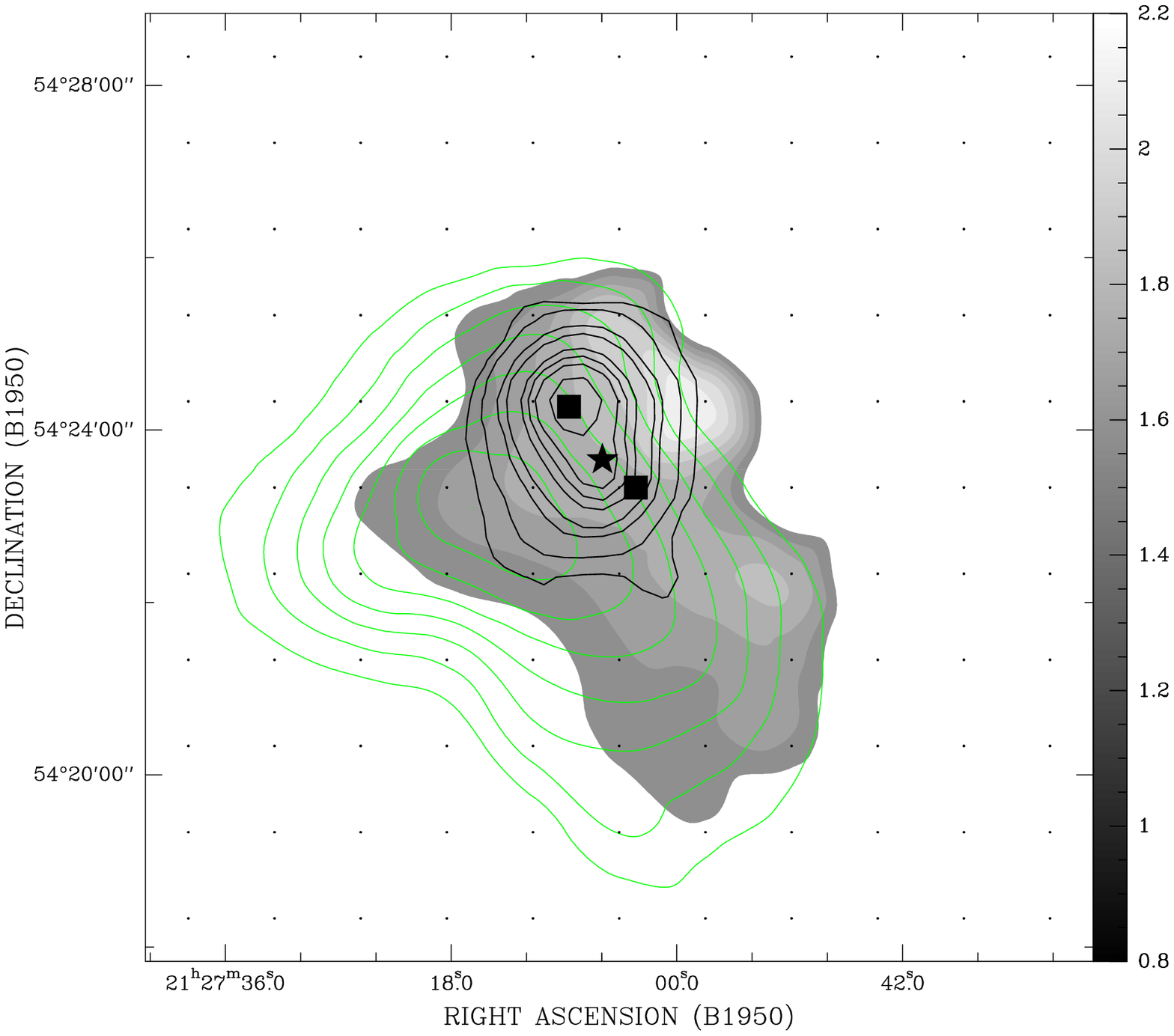}

\includegraphics[width=80mm]{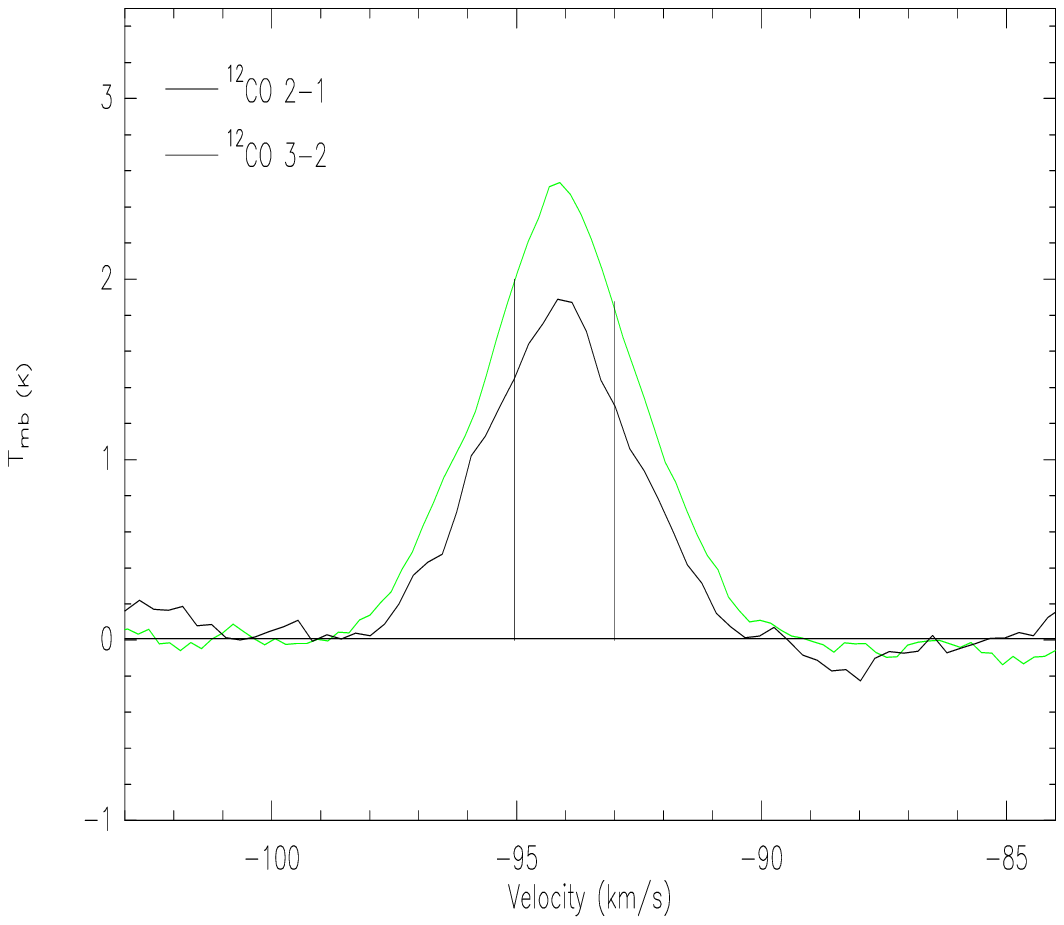}
\vspace{0.1cm} \caption{S127. Caption as in Fig.B.1. The star
symbol marks IRAS 21270+5423. }
 \label{fig:fig.8.}
\end{figure}
\clearpage
\newpage

\begin{figure}[h]

 \includegraphics[width=80mm]{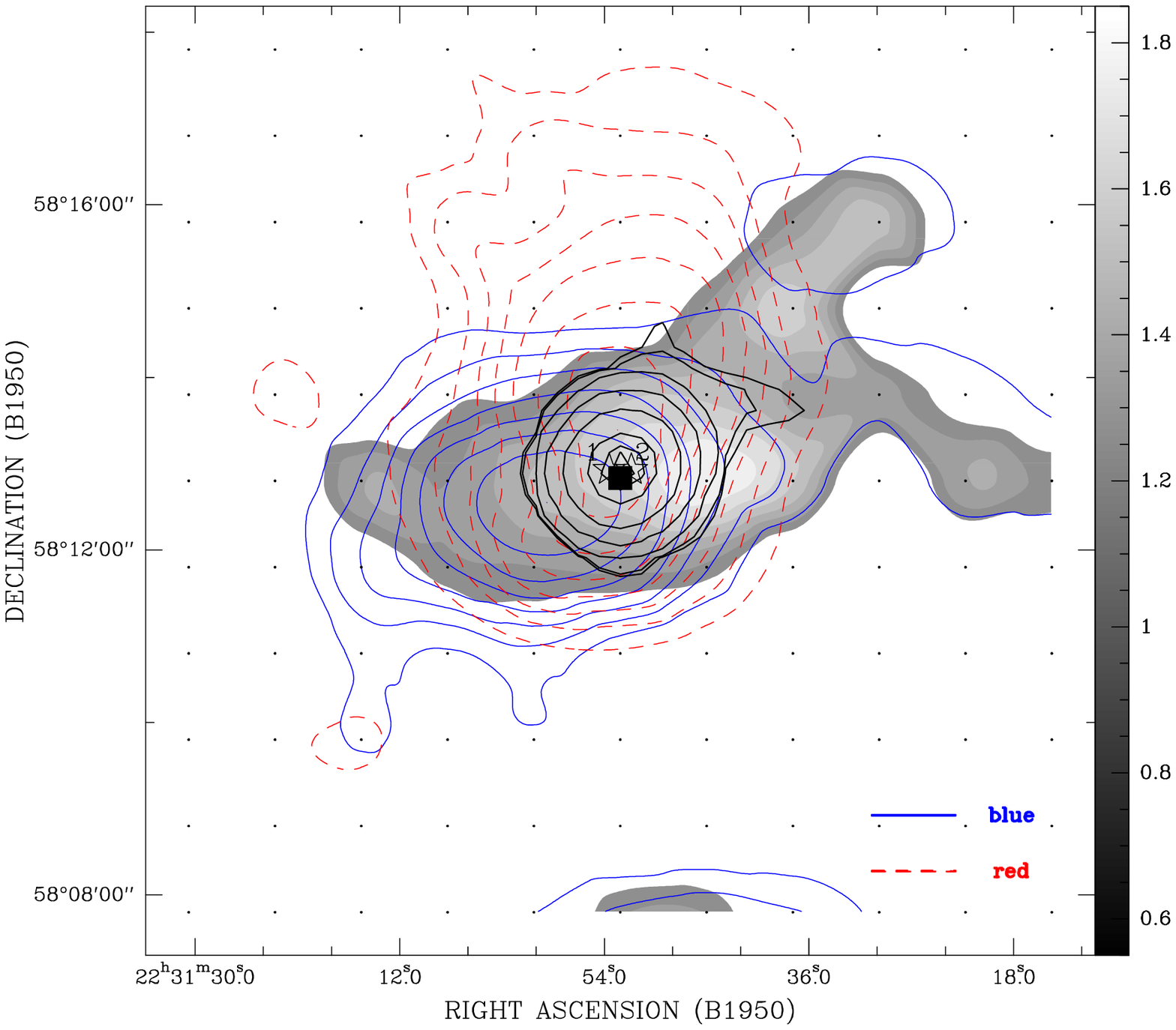}
\vspace{10mm} \hspace{5mm}
\includegraphics[width=80mm]{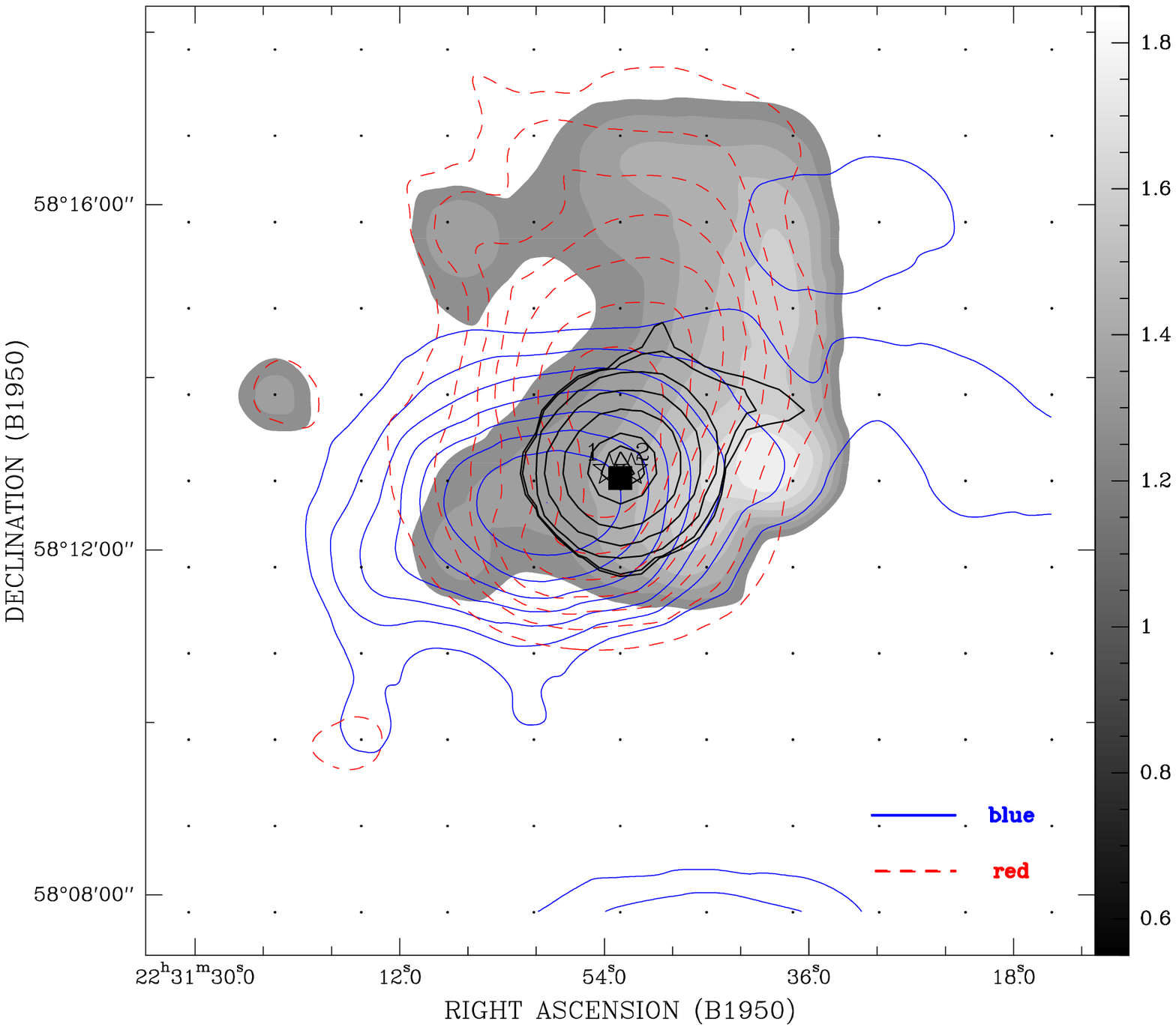}

\includegraphics[width=80mm]{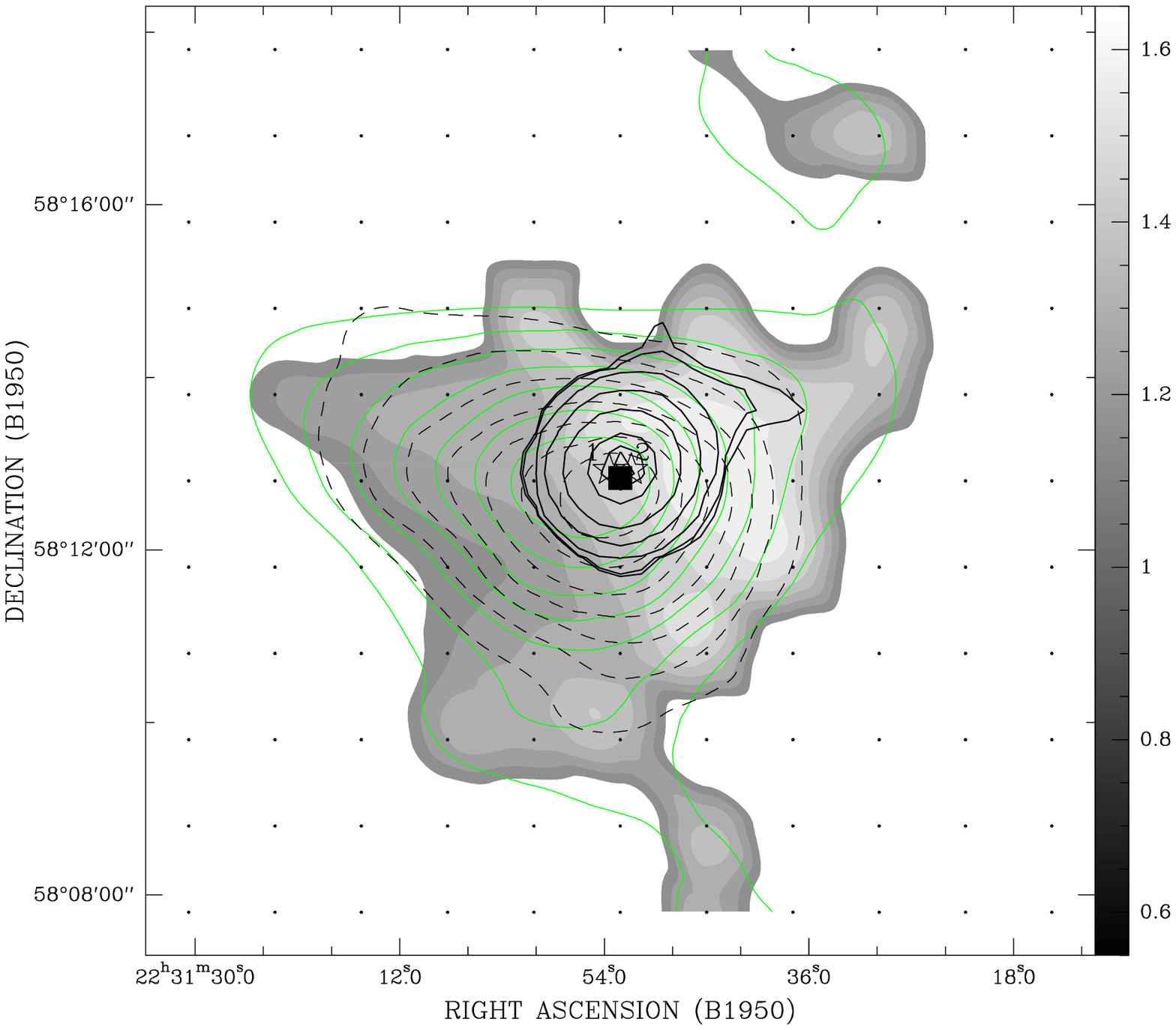}

\includegraphics[width=80mm]{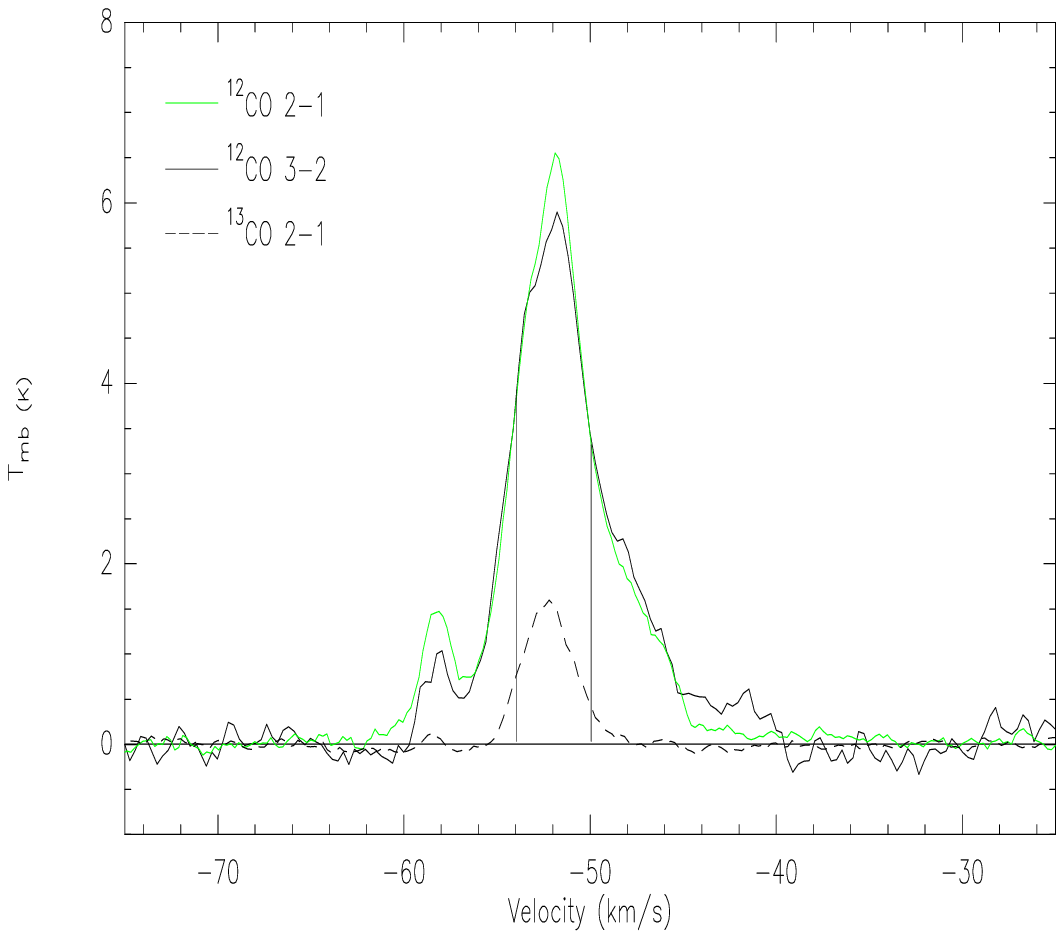}
\vspace{0.1cm} \caption {S138. Caption as in Fig.B.1. The Stars 1
and 2 mark IRAS 22308+5812 and O7.5-type star GRS 105.63-00.34.
${\rm H_{2}O}$ maser is overlaid at the IRAS 22308+5812 position.
}

 \label {fig:fig.9.}
\end{figure}
\clearpage
\newpage
\begin{figure}[h]

 \includegraphics[width=80mm]{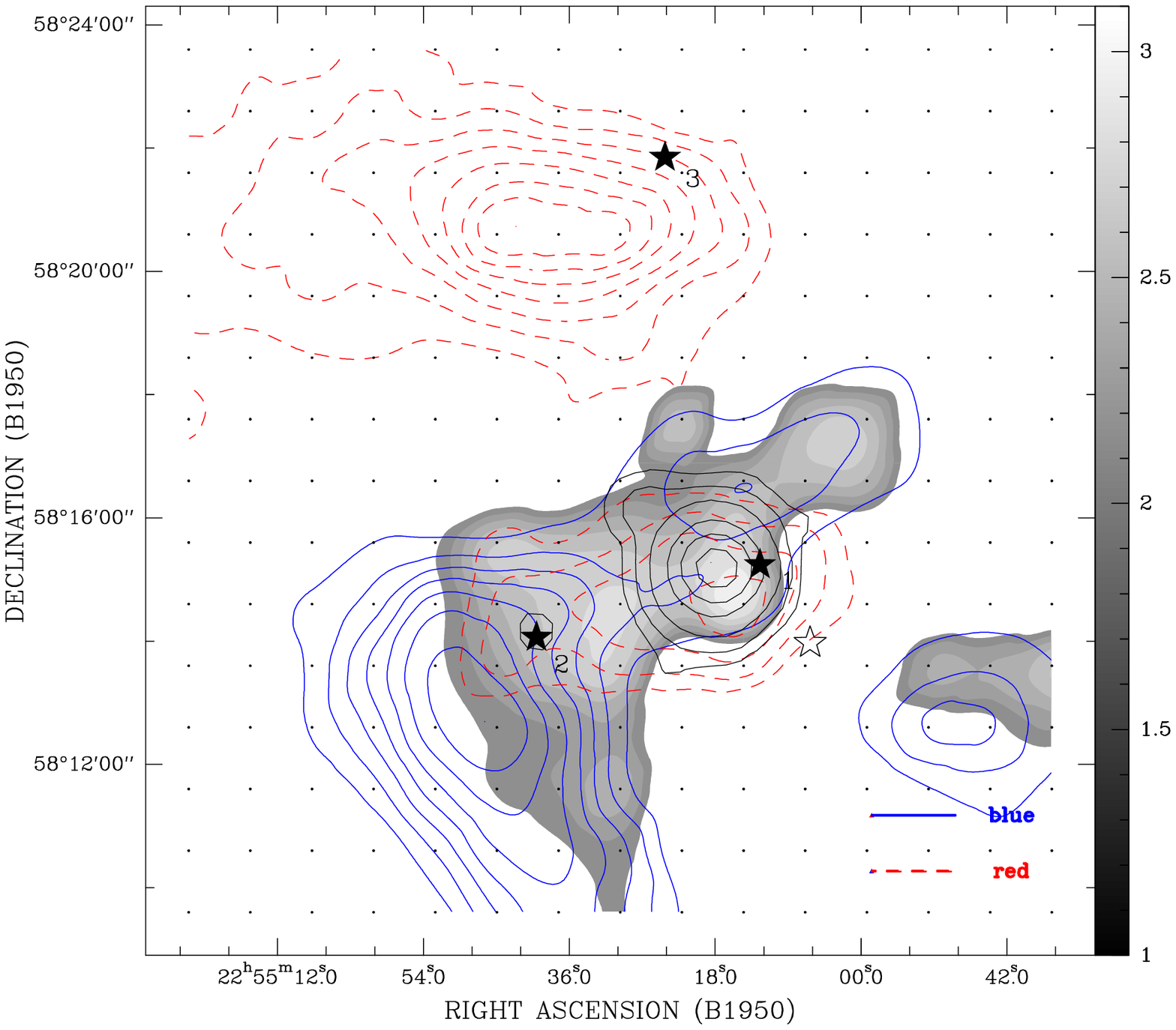}
\vspace{10mm} \hspace{5mm}
\includegraphics[width=80mm]{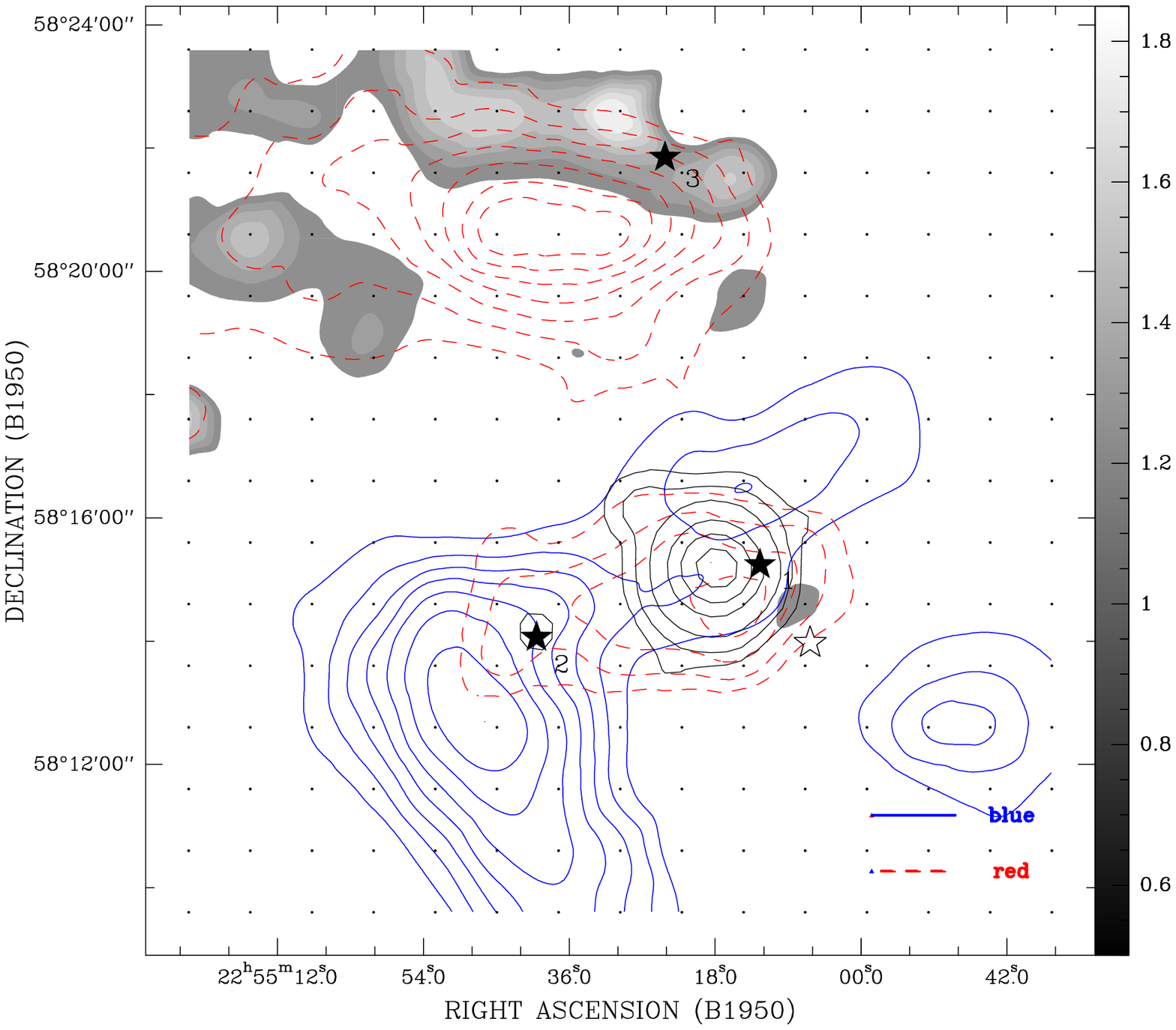}

\includegraphics[width=80mm]{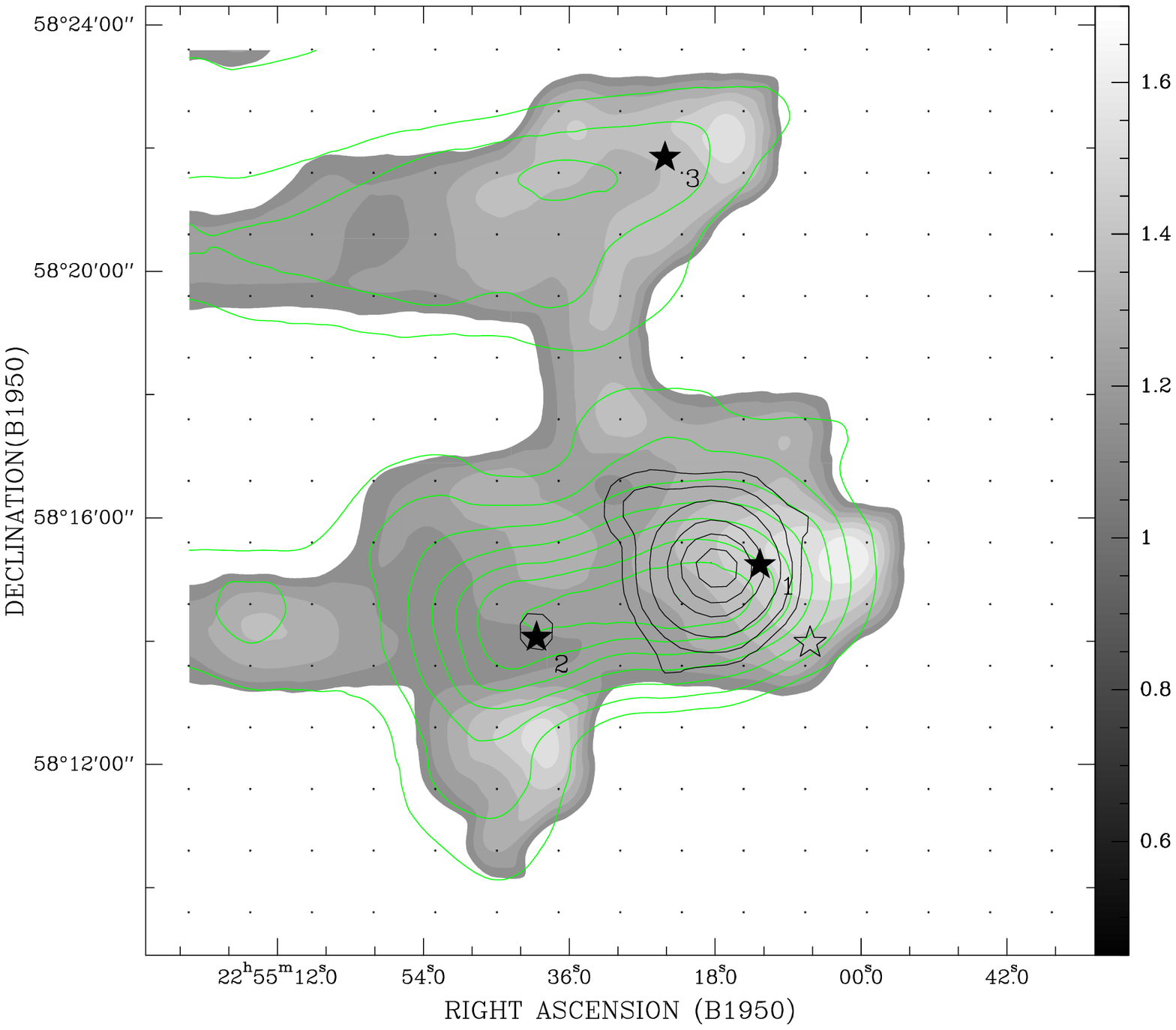}

\includegraphics[width=80mm]{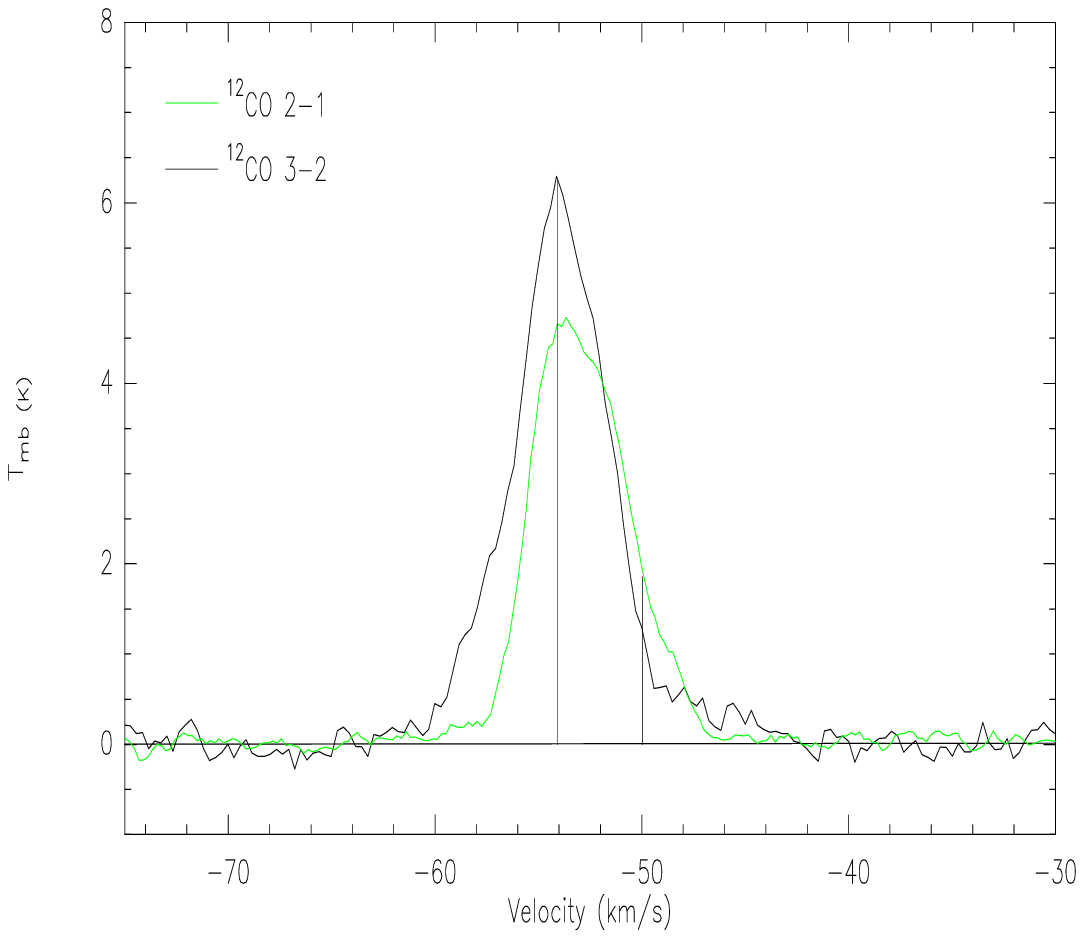}
\vspace{0.1cm}

\caption {S149. Caption as in Fig.B.1. The filled star 1, 2, 3 are
IRAS 22542+5815, IRAS 22546+5814 and IRAS 22543+5821; the unfilled
star is O8-type star SH2-148. }

\end{figure}
\clearpage
\newpage
\begin{figure}[h]

\begin{minipage}[t]{0.475\linewidth}
 \includegraphics[width=80mm]{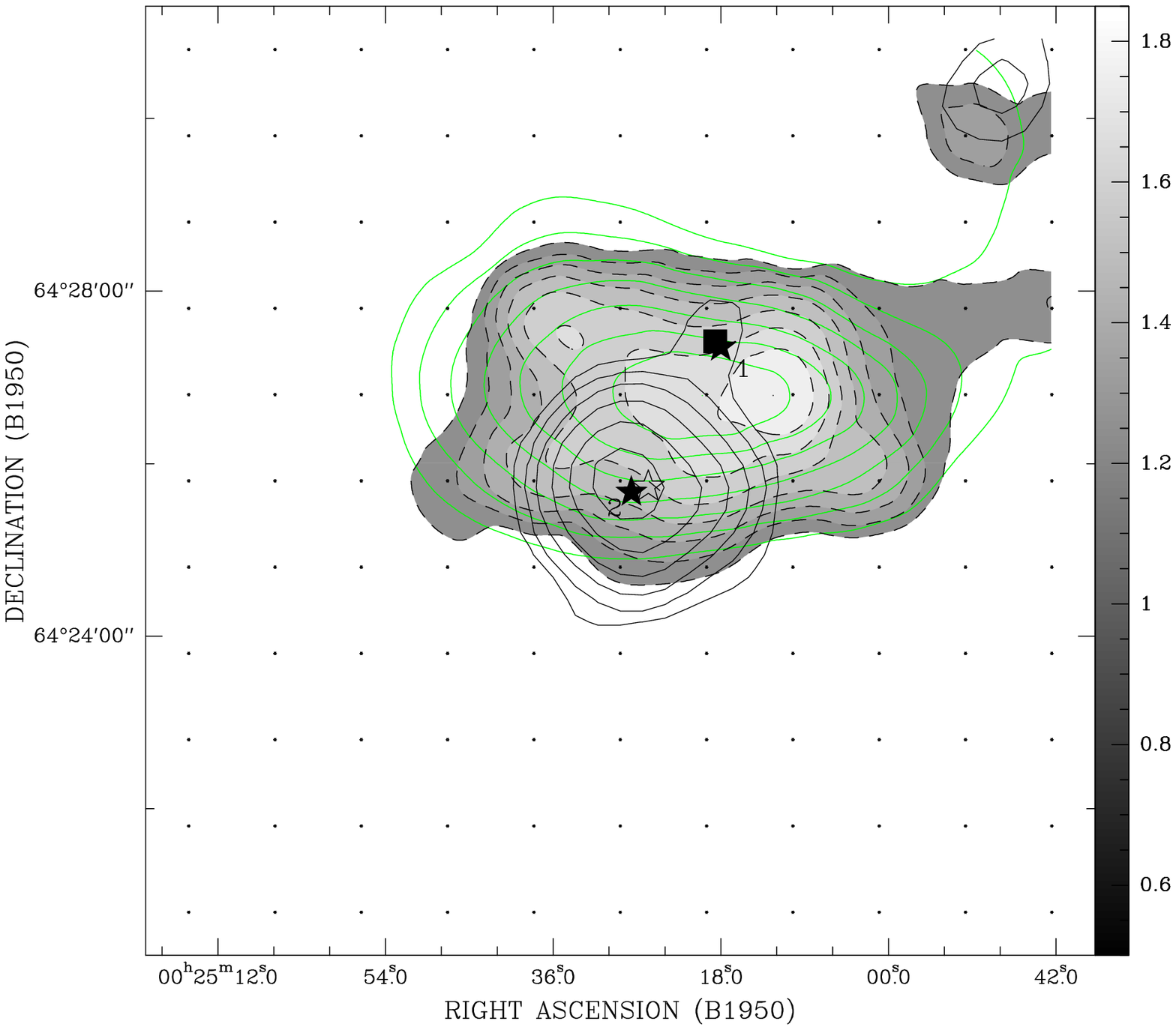}
 \end{minipage}
\begin{minipage}[t]{0.475\linewidth}
\includegraphics[width=80mm]{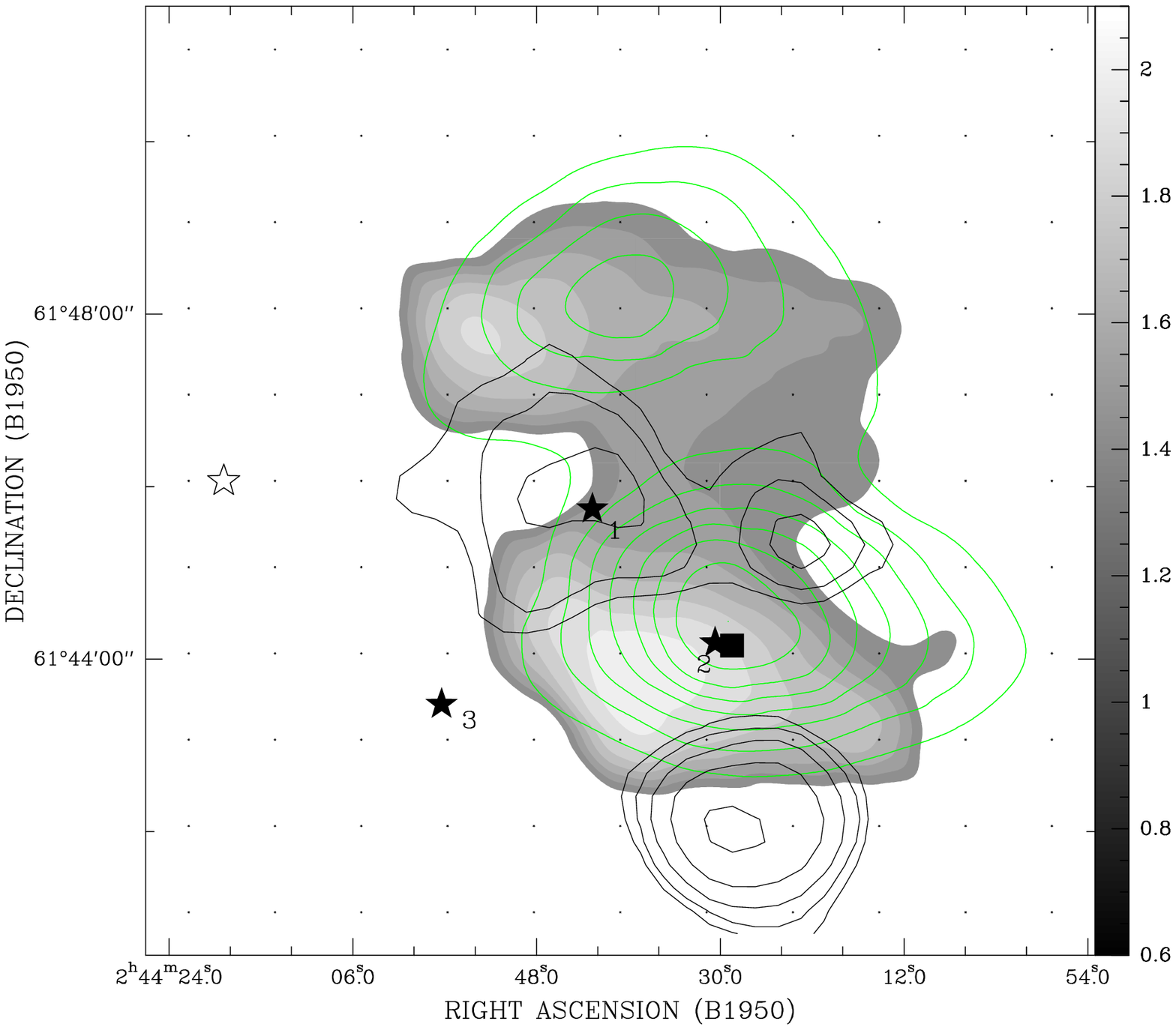}
\end{minipage}

\vspace{0.2cm}

\caption {Left panel for S175: the integrated intensity map of the
core in CO $J=2-1$ (green solid contours) are superimposed on the
grey map of the line intensity ratio from the core component, and
the dashed contours are the integrated intensity map of the core
in $^{13}$CO $J=2-1$ line. The thin solid lines show the continuum
at 1.4 GHz from NVSS. The filled stars 1 and 2 are IRAS 00243+6427
and IRAS 00244+6425, respectively. The unfilled star is O9.5 star
ALS 6206.
 Right panel for S193: the integrated intensity map of the core in CO
$J=2-1$ (green solid contours) are superimposed on the grey map of
the line intensity ratio from the core component. The thin solid
lines show the continuum at 1.4 GHz from NVSS. The contours for
the cores begin at 30\% of the maximum integrated intensity ${\rm
I_{max}}$ and increase at a step of 10\% of ${\rm I_{max}}$, and
the integrated line intensity ratios range from 30\% of the
maximum ${\rm R_{I_{CO(3-2)}/I_{CO(2-1)}}}$ to the maximum ${\rm
R_{I_{CO(3-2)}/I_{CO(2-1)}}}$. The intensity ratios indicate the
gas temperature variations at different positions. The filled
stars 1, 2 and 3 are IRAS 02437+6145, IRAS 02435+6144 and IRAS
02439+6143, respectively. The unfilled star is B4-type star GSC
04051-01677. }

 \label {fig:fig.11.}

  \end{figure}
\clearpage
\newpage
\begin{figure}[h]

\begin{minipage}[t]{0.455\linewidth}
 \includegraphics[width=80mm]{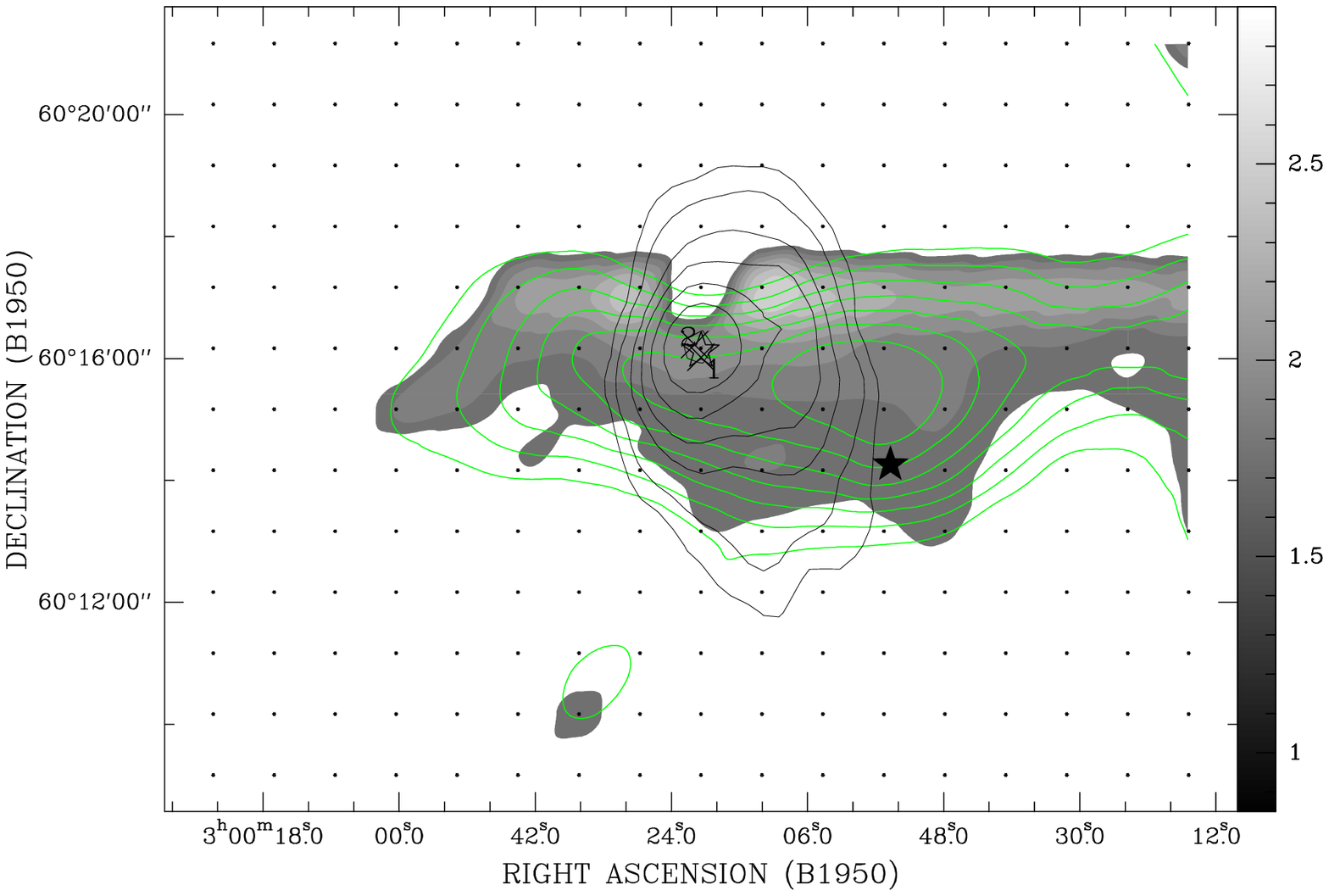}
 \end{minipage}
\begin{minipage}[t]{0.455\linewidth}
\includegraphics[width=80mm]{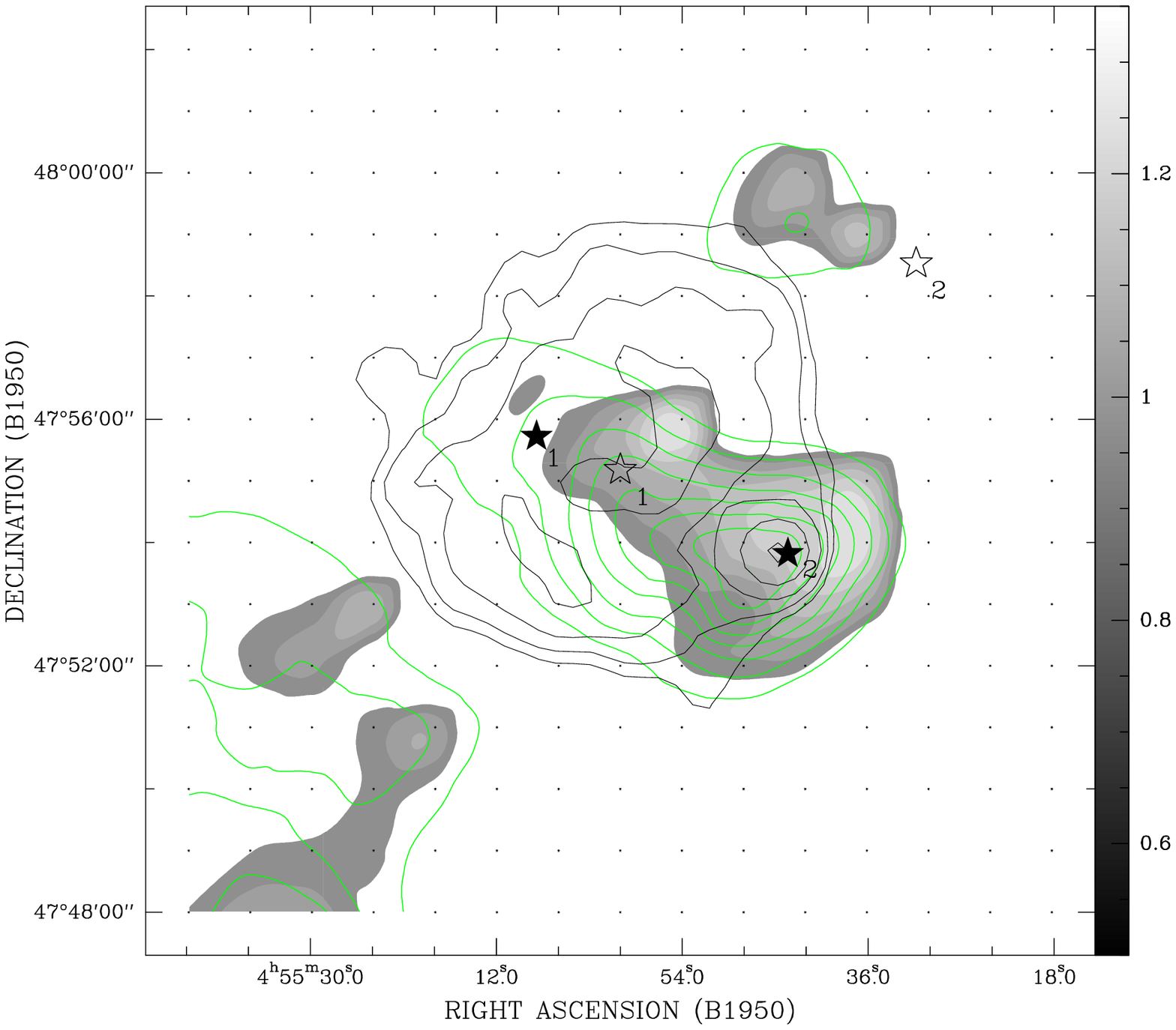}
\end{minipage}

\begin{center}
\includegraphics[width=80mm]{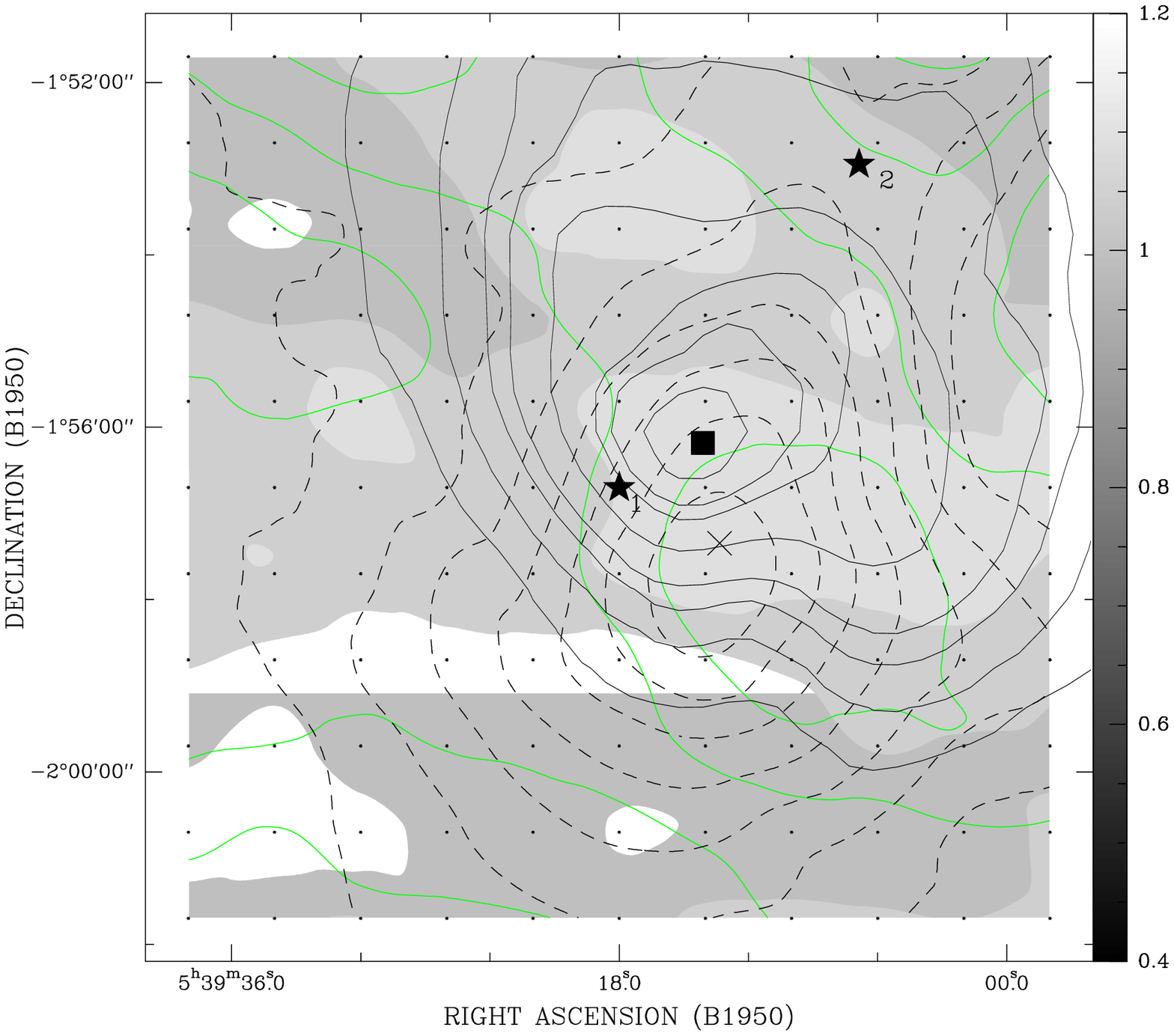}
\end{center}

\vspace{0.1cm} \caption {Caption as in Fig.B.11. Left panel for
S201: The star is IRAS 02589+6014, the cross symbols 1 and 2
indicate the ${\rm H_{2}O}$ masers. IRAS 02593+6016 is overlaid at
the maser 1 position. Right panel for S217: The filled stars 1 and
2 are IRAS 04551+4755 and IRAS 04547+4753; the unfilled stars 1
and 2 are O9.5-type star LSV 47$^{\circ}$24 and B8-type star BD+47
1079. Bottom panel for G206.543-16.347: the dashed contours are
the integrated intensity map of the core in $^{13}$CO $J=2-1$
line. The stars 1 and 2 are IRAS 05393-0156 and IRAS 05391-0152,
the cross symbol mark ${\rm H_{2}O}$ maser.}

 \label{fig:fig.12.}
\end{figure}
\end{appendix}
\twocolumn

\clearpage
\newpage
\onecolumn

\begin{table*}

\caption[1] {Observation Log. The columns give source name,coordinates$^a$, \\
transitions, number of mapped points, rms of spectra, system
temperatures.}

          \[
     \begin{array}{lcccccccl}
            \hline
{\rm Source }& \alpha (1950) & \delta (1950)& {\rm Transitions}&{\rm Pos.} &{\rm T_{mb}^{rms}} & {\rm T_{sys}}\\
& ({\rm h\>m\>s}) & ({{\circ}\>{\prime}\>{\prime\prime}}) &  & & ({\rm K}) & ({\rm K})   \\
          \hline
{\rm S175} & 00\> 24\> 28.8 & 64\> 25\> 48
&         {\rm CO (2-1)}& 121  &0.18 & 160 \\
& & &     {\rm CO (3-2)}& 121  &0.37 & 285 \\
& & &^{13}{\rm CO (2-1)}& 121  &0.18 & 221 \\

{\rm S186} & 01\> 05\> 38.3 & 62\> 51\> 36
&         {\rm CO (2-1)}& 121  &0.18 & 180 \\
& & &     {\rm CO (3-2)}& 121  &0.41 & 306 \\
& & &^{13}{\rm CO (2-1)}& 121  &0.08 & 173 \\

{\rm S193} & 02\> 43\> 39.8 & 61\> 46\> 04
&         {\rm CO (2-1)}& 121  &0.16 & 180 \\
& & &     {\rm CO (3-2)}& 121  &0.28 & 282 \\

{\rm S201} & 02\> 59\> 20.1 & 60\> 16\> 10
&     {\rm CO (2-1)}& 221  &0.24 & 195 \\
& & &     {\rm CO (3-2)}& 221  &0.71 & 280 \\

{\rm G139} & 03\> 03\> 33.0 & 58\> 19\> 21
&         {\rm CO (2-1)}& 121 &0.25 & 158 \\
& & &     {\rm CO (3-2)}& 121  &0.48 & 282 \\

{\rm S217} & 04\> 55\> 00.0 & 47\> 55\> 00
&         {\rm CO (2-1)}& 225  & 0.2 & 171 \\
& & &     {\rm CO (3-2)}& 225  &0.46 & 287 \\

{\rm G206} & 05\> 39\> 18.0 & -01\> 56\> 42
&     {\rm CO (2-1)}& 121  &0.53 & 175 \\
& & &     {\rm CO (3-2)}& 121  &1.34 & 295 \\
& & &^{13}{\rm CO (2-1)}& 121  &0.28 & 194 \\

{\rm G189} & 06\> 06\> 23.0 & 20\> 40\> 02
&         {\rm CO (2-1)}& 121  &0.21 & 167 \\
& & &     {\rm CO (3-2)}& 121  &0.43 & 273 \\
& & &^{13}{\rm CO (2-1)}& 121  &0.09 & 172 \\
{\rm G213} & 06\> 08\> 24.5 & -06\> 11\> 12
&     {\rm CO (2-1)}& 121  &0.30 & 173 \\
& & &     {\rm CO (3-2)}& 121  &0.70 & 288 \\
& & &^{13}{\rm CO (2-1)}& 121  &0.26 & 201 \\

{\rm G192} & 06\> 09\> 57.9 & 18\> 00\> 12
&     {\rm CO (2-1)}& 441  &0.26 & 181 \\
& & &     {\rm CO (3-2)}& 441  &0.56 & 296 \\

{\rm S288} & 07\> 06\> 11.0 & -04\> 13\> 00
&     {\rm CO (2-1)}& 121  &0.26 & 189 \\
& & &     {\rm CO (3-2)}& 121  &0.73 & 321 \\

{\rm G70}  & 19\> 59\> 50.0 & 33\> 24\> 20
&     {\rm CO (2-1)}& 121  &0.16 & 139 \\
& & &     {\rm CO (3-2)}& 121  &0.25 & 237 \\

{\rm S127} & 21\> 27\> 04.6 & 54\> 23\> 20
&         {\rm CO (2-1)}&  121  &0.13 & 149 \\
& & &     {\rm CO (3-2)}&  121  &0.27 & 246 \\

{\rm S138} & 22\> 30\> 52.6 & 58\> 12\> 48
&     {\rm CO (2-1)}& 121 &0.22 & 168 \\
& & &     {\rm CO (3-2)}& 121  &0.55 & 299  \\
& & &^{13}{\rm CO (2-1)}& 121  &0.09 & 163  \\

{\rm S149} & 22\> 54\> 14.5 & 58\> 16\> 36
&         {\rm CO (2-1)}& 225  &0.24 & 173  \\
& & &     {\rm CO (3-2)}& 225  &0.47 & 286  \\
           \hline
         \end{array}
      \]
\renewcommand{\footnoterule}{note}{$^a$Central coordinates of the maps.}
       \end{table*}
\clearpage
\newpage
\begin{table*}
 \caption[1]{Velocity intervals and maximum integrated intensities for the maps presented in Figs.1-12$^a$.}
  \label{Tab:publ-works}
  \begin{center}

  \begin{tabular}{lcccccccll}
  \hline\noalign{\smallskip}

{\rm Source}& $\triangle {\rm v}_{b}$ & ${\rm I_{max}(b)}$&
$\triangle {\rm v}_{r}$
&${\rm I_{max}(r)}$ & $\triangle {\rm v}_{c}$&${\rm I_{max}(c)}$&${\rm I_{max}(13)}$\\
& (${\rm km}~{\rm s}^{-1}$) & (${\rm K}~{\rm km}~{\rm s}^{-1}$)
&(${\rm km}~{\rm s}^{-1}$)& (${\rm K}~{\rm km}~{\rm s}^{-1}$)
&(${\rm km}~{\rm s}^{-1}$)&
$({\rm K}~{\rm km}~{\rm s}^{-1}$)&(${\rm K}~{\rm km}~{\rm s}^{-1}$)  \\
          \hline
{\rm S175} &           &   &           &    &$(-52,-46)$& 28 &6 \\
{\rm S186} & $(-46,-44)$ &4  & $(-42,-38)$ &5   &$(-44,-42)$& 15 &5 \\
{\rm S193} &           &   &           &    &$(-51,-42)$& 23    \\
{\rm S201} &           &   &           &    &$(-52,-46)$& 28    \\
{\rm G139} & $(-46,-41)$ &38 & $(-38,-33)$ &25  &$(-41,-38)$& 65    \\
{\rm S217} &           &   &           &    &$(-24,-14)$& 41    \\
{\rm G206} &           &   &           &    &  (0, 21) & 217&91\\
{\rm G189} &   (0, 8)   &32 &  (10, 15)  &15.7&  (8, 10) & 33 &16 \\
{\rm G213} &    (2, 8)  &11 &  (16, 22)  &10  &  (8, 16) & 81 &37 \\
{\rm G192} &    (3, 6)  &27 &  (10, 13)  &19  &  (6, 10) & 95     \\
{\rm S288} &  (52, 56)  &5.6&  (57, 60)  &6.2 &  (56, 57)& 4.8     \\
{\rm G70}  & $(-35,-26)$ &34 & $(-21,-12)$ &28  &$(-26,-21)$& 83      \\
{\rm S127} & $(-100,-95)$&6.4& $(-93,-88)$ &4   &$(-95,-93)$& 7.5     \\
{\rm S138} & $(-65,-54)$ &20 & $(-50,-39)$ &19  &$(-54,-50)$& 19 &15 \\
{\rm S149} &$(-60,-54)$  &25 & $(-50,-44)$ &14  &$(-54,-50)$& 41      \\
    \noalign{\smallskip}
    \hline

  \end{tabular}
\end{center}
\renewcommand{\footnoterule}{}{$^a$The maximum integrated intensities (${\rm
I_{max}}$) in $^{12}$CO $J=2-1$ and their corresponding integrated
velocity ranges ($\triangle {\rm v}$); b, r and c indicate the
blue, red-shifted outflow and core components, respectively. ${\rm
I_{max}(13)}$ indicate the maximum integrated intensities of the
cores in $^{13}$CO $J=2-1$, the cores in $^{13}$CO $J=2-1$ have
same integrated velocity range as those in $^{12}$CO $J=2-1$.}
\end{table*}
\clearpage
\newpage
\begin{table}[]
\centering
 \caption[] {Physical Parameters of Outflows. }
  \label {Tab:publ-works}
  \begin{center}\begin{tabular}{lccccccc}
  \hline\noalign{\smallskip}
{\rm Source}& {\rm M} & {\rm P}&{\rm E} & {\rm t}& ${\rm \dot
{M}_{loss}}$
& ${\rm F}_{m}$& ${\rm L}_{m}$\\
&(${\rm M}_{\odot}$)&(${\rm M}_{\odot}$~${\rm km}~{\rm
s}^{-1}$)&($10^{46}$ erg)&($10^{4}$ yr)&($10^{-4}$~${\rm
M}_{\odot}$~${\rm yr^{-1}}$)&($10^{-3}$~${\rm M}_{\odot}$~${\rm
km}~{\rm s}^{-1}$~${\rm
yr^{-1}}$)&(${\rm L}_{\odot}$)\\
  \hline\noalign{\smallskip}
{\rm S186}  & 83  &255  &7   & 1.7 &0.3 & 14&3\\
{\rm G139}  & 481 &2575 &12  & 0.7&7.4 & 366 &138 \\
{\rm G189}  & 384 &2024 &11  & 0.3&13.5  & 660 &288 \\
{\rm G213}  & 11  &123  &11  & 0.1&2.5 & 90 &684\\
{\rm G192}  & 170 &1172 &7   & 1.2&1.9 & 1200  &48\\
{\rm S288}  & 33  &94   &2.7 & 0.9&0.2  & 9&2\\
{\rm G70}   & 1045&8778 &65  & 1.8 &9.7   & 490 &300 \\
{\rm S127}  & 1021&2329 &9.2 & 6.1 &0.7 & 126&12 \\
{\rm S138}  & 840 &7875 &68  & 0.7 &22.5  & 1272 &870\\
{\rm S149}  & 764 &5265 &31  & 2.5 &4.2 & 210 &102\\
  \noalign{\smallskip}\hline
  \end{tabular}
\end{center}
\end{table}

\clearpage
\newpage
\begin{table*}
\centering
 \caption[1] {The Identifications$^a$ of the associated stellar objects,
NVSS continuum, 8.3 $\mu$m emission and line intensity ratio with
the HII complexes. }

 \begin{center}
  \begin{tabular}{lccccccccl}
  \hline\noalign{\smallskip}

{\rm Source}& {\rm D} & {IRAS/stars}&{\rm L} &ST & NVSS&${\rm
P_{MSX}}$ &
${\rm R_{b}}$ & ${\rm R_{r}}$ &${\rm R_{c}}$ \\
 &({\rm kpc})& &($10^{3}$ ${\rm L_{\odot}}$)&& &($^{\prime\prime}$,
 $^{\prime\prime}$) & \\
          \hline
{\rm S175} &1.7 & IRAS 00244+6425 &1.9  &B3  &Y &(7.9,6.5)   &  & &$-$\\
           &    & IRAS 00243+6427 &0.21 &ML  &N &(0.0,0.0)   &  & &+\\
           &    & ALS6206         &     &O9.5&N&            &  & &\\
{\rm S186} &4.0 & IRAS 01056+6251 &11.4 &B0.5&Y&(26.6,17.6) &+ &+&+\\
           &    & IRAS 01053+6251 &1.49 &B3  &N&(6.5,18.7)  &+ &+&+\\
{\rm S193} &5.2 & IRAS 02435+6144 &1.2  &B3  &N&($-$11.1,$-$0.4)&  & &+\\
           &    & IRAS 02437+6145 &5.4  &B2  &Y&            &  & &\\
           &    & IRAS 02439+6143 &0.33 &ML  &N&        &  & &\\
           &    & GSC 04051-01677 &     &B4  &N&        &  & &\\
{\rm S201} &4.0 & IRAS 02593+6016 &56   &O8.5&Y&(23.3,10.1) &  & &+\\
           &    & IRAS 02598+6014 &50   &O9  &N&(9.7,7.9)   &  & &$-$\\
{\rm G139} &4.2 & IRAS 03035+5819 &67   &O8  &Y&($-$24.8,$-$2.9)&+ &+&+\\
           &    & IRAS 03037+5819 &9.6  &B0.5&N&($-$11.8,$-$7.2)&$-$ &$-$&$-$\\
{\rm S217} &2.8 & IRAS 04547+4753 &8.7  &B0.5&Y&(0.0,0.0)   &  & &+\\
           &    & IRAS 04551+4755 &1.76 &B3  &N&        &  & &\\
           &    & LSV 47$^{\circ}$24&   &O9.5&N&        &  & &\\
           &    & BD+47 1079      &     &B8  &N&        &  & &\\
{\rm G206} &0.5 & IRAS 05393-0156 &13   &B0.5&Y&(0.0,0.0)   &  & &+\\
           &    & IRAS 05391-0152 &1.96 &B3  &N&(0.0,0.0)   &  & &$-$\\
{\rm G189} &2.8 & IRAS 06063+2040 &28   &B0  &Y&($-$28.4,5.7) &+ &+&+\\
           &    & ALS 8745        &     &B   &N&        & & &\\
           &    & ALS 8748        &     &B   &N&        & & & \\
           &    & HD 252325       &     &B1  &Y&            & & &\\
 {\rm G213} &1.0 & IRAS 06084-0611 &11   &B0.5&Y&(0.0,0.0)   &+ &$-$&+\\
{\rm G192} &2.5 & IRAS 06099+1800 &61   &O8.5&Y&(11.2,3.6)   &+ &+&-\\
           &    & IRAS 06096+1757 &21.9 &B0.5&Y&(0,0)        &$-$ &$-$&$-$\\
           &    & IRAS 06105+1756 &7.6  &B1  &Y&         & &&\\
           &    & HD253327        &     &B0.5&Y&         & &&\\
           &    & HD253247        &     &B1  &Y&         & & &\\
{\rm S288} &7.2 & IRAS 07061-0414 &78   &O8 &Y&($-$2.5,1.1)   &$-$ &+&$-$\\
           &    & HD 296489       &     &B8  &N&         & & & \\
{\rm G70}  &8.6 & IRAS 19598+3324 &228  &O4  &Y&(0.0,0.0)    &+ &+&+\\
{\rm S127} &11.5& IRAS 21270+5423 &100  &O7  &Y&(21.6,42.8)  &+ &$-$&+\\
           &    &                 &     &    &&($-$24.1,$-$19.8)&$-$ &+&+\\
{\rm S138} &5.7 & IRAS 22308+5812 &84   &O7.5&Y&($-$3.6,$-$2.5)  &+ &$-$&+\\
           &    & GRS105.63-00.34 &     &O7.5&Y&(1.3, $-$2.4)  &+ &$-$&+\\
{\rm S149} &5.5 & IRAS 22542+5815 &48   &O9  &Y&($-$9.0,0.4)   &+ &$-$&+\\
           &    & IRAS 22543+5821 &2.6  &B3  &N&         & & & \\
           &    & IRAS 22546+5814 &4.3  &B2  &N&         & & &\\
           &    & SH2-148         &     &O8  &N&         & & &\\
    \noalign{\smallskip}
    \hline
  \end{tabular}
\end{center}
\renewcommand{\footnoterule}{}{$^a$The distance (D) is
from Kurtz, Churchwell \& Wood (1994) and Wouterloot \& Brand
(1989). The far-infrared luminosity (L) is derived based on the
formula of Casoli, Combes \& Dupraz (1986). The corresponding
spectral type (ST) of the far-infrared luminosity (L) is referred
to the work of Panagia (1973); the spectral type of the massive
stars is taken from SIMBAD database. ML marks intermediate or low
mass star. ${\rm P_{MSX}}$ presents the offset (arcsec) of MSX
sources relative to IRAS sources. ${\rm R_{b}}$, ${\rm R_{r}}$ and
${\rm R_{c}}$ indicate the maximum intensity ratios determined
from the blue-shifted and red-shifted wings of the outflow and
core component, respectively, as shown in the wedges of the
Figs.B.1-12. The plus symbol indicates that the MSX sources are
associated with the maximum ratios (${\rm R_{b}}$, ${\rm R_{r}}$
and ${\rm R_{c}}$). The minus symbol indicates that the MSX
sources are not related to the maximum ratios.}
\end{table*}

 \clearpage
\newpage
\onecolumn

\clearpage
\newpage

\end{document}